\RequirePackage{fix-cm}

\documentclass[smallextended]{svjour3}
\smartqed

\usepackage{makecell}
\usepackage{acronym}

\usepackage{array}

\usepackage[inline]{enumitem}
\usepackage{xcolor}
\usepackage{xspace}
\usepackage{booktabs} % For formal tables
\usepackage{soul}
\usepackage[htt]{hyphenat}
\usepackage{tabularx}
\usepackage{tabularray}
\usepackage{natbib}
\usepackage{amsmath}
\usepackage{amssymb}
\usepackage{url}
\usepackage{placeins}
\newcommand{\rev}[1]{\textcolor{blue}{#1}}

\newcommand{\eg}{\emph{e.g.,}\xspace}
\newcommand{\ie}{\emph{i.e.,}\xspace}
\newcommand{\etal}{\emph{et al.}\xspace}

\newcommand{\secref}[1]{Section~\ref{#1}\xspace}
\newcommand{\figref}[1]{Fig.~\ref{#1}\xspace}
\newcommand{\tabref}[1]{Table~\ref{#1}\xspace}

\newcommand{\RQ}[1]{\textbf{RQ$\pmb{_#1}$}}
\newcommand{\RQe}{\emph{RQ}\xspace}
\newcommand{\RQs}{\emph{RQs}\xspace}

\newcommand{\CH}[1]{\textit{CH${_#1}$}}

\newcommand{\Var}[1]{\textsf{Var$_#1$}}

\usepackage{lipsum}                     % Dummytext
\usepackage{xargs}                      % Use more than one optional parameter in a new commands

\usepackage{graphicx,calc}
\newlength\myheight
\newlength\mydepth
\settototalheight\myheight{Xygp}
\usepackage[]{mdframed}
\usepackage{multirow}

\DeclareRobustCommand{\SEpapers}{\includegraphics[height=2ex]{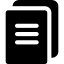}}
\DeclareRobustCommand{\SEcompletepapers}{\includegraphics[height=2ex]{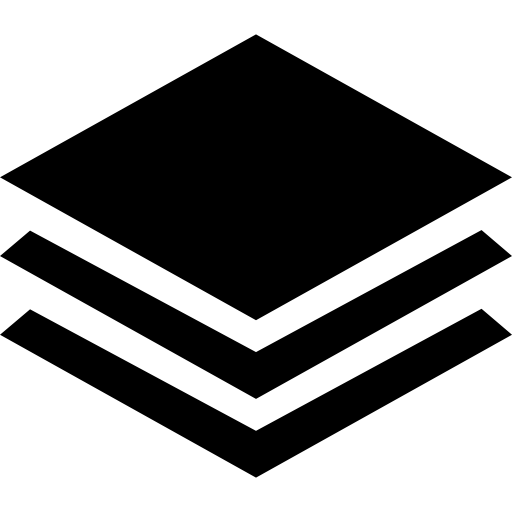}}

\DeclareRobustCommand{\SEauthors}{\includegraphics[height=2ex]{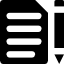}}
\DeclareRobustCommand{\SEresearchers}{\includegraphics[height=2ex]{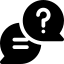}}

\DeclareRobustCommand{\Aone}{\raisebox{-0.2\height}{\includegraphics[height=2.5ex]{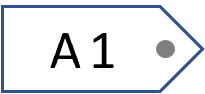}}}
\DeclareRobustCommand{\Atwo}{\raisebox{-0.2\height}{\includegraphics[height=2.5ex]{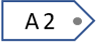}}}
\DeclareRobustCommand{\Athree}{\raisebox{-0.2\height}{\includegraphics[height=2.5ex]{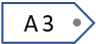}}}
\DeclareRobustCommand{\Afourone}{\raisebox{-0.2\height}{\includegraphics[height=2.5ex]{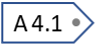}}}
\DeclareRobustCommand{\Afourtwo}{\raisebox{-0.2\height}{\includegraphics[height=2.5ex]{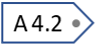}}}
\DeclareRobustCommand{\Afiveone}{\raisebox{-0.2\height}{\includegraphics[height=2.5ex]{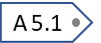}}}
\DeclareRobustCommand{\Afivetwo}{\raisebox{-0.2\height}{\includegraphics[height=2.5ex]{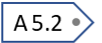}}}

\DeclareRobustCommand{\Bone}{\raisebox{-0.2\height}{\includegraphics[height=2.5ex]{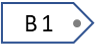}}}
\DeclareRobustCommand{\Btwo}{\raisebox{-0.2\height}{\includegraphics[height=2.5ex]{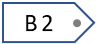}}}
\DeclareRobustCommand{\Bthree}{\raisebox{-0.2\height}{\includegraphics[height=2.5ex]{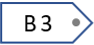}}}
\DeclareRobustCommand{\Bfour}{\raisebox{-0.2\height}{\includegraphics[height=2.5ex]{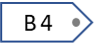}}}
\DeclareRobustCommand{\Bfive}{\raisebox{-0.2\height}{\includegraphics[height=2.5ex]{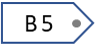}}}

\acrodef{ML}{Machine Learning}
\acrodef{AI}{Artificial Intelligence}
\acrodef{SE}{Software Engineering}
\acrodef{DL}{Deep Learning} 
\acrodef{CsQ+A}[CQ\&A]{Communities of Question and Answers}

\acrodef{ML4SE}[ML4SE]{Machine Learning for Software Engineering} 
\acrodef{SE4ML}{Software Engineering for Machine Learning}
\acrodef{SE4AI}{Software Engineering for Artificial Intelligence} 
\acrodef{IRR}{inter-rater reliability}

\acrodef{LLM}{Large Language Model}

\newcommand{\PreserveBackslash}[1]{\let\temp=\\#1\let\\=\temp}
\newcolumntype{C}[1]{>{\PreserveBackslash\centering}p{#1}}
\newcolumntype{R}[1]{>{\PreserveBackslash\raggedleft}p{#1}}
\newcolumntype{L}[1]{>{\PreserveBackslash\raggedright}p{#1}}

\usepackage{layouts}
\newcommandx{\qs}[1]{\textit{#1}}

\newcommand{\qcl}[1]{(\textit{#1})}

\usepackage{fontawesome}
\usepackage{tasks}
\usepackage{tikz}
\usepackage[usestackEOL]{stackengine}
\usepackage{graphicx}

\makeatletter
\newcommand*{\radiobutton}{\@ifstar{\radiobuttonON}{\radiobuttonOFF}}
\makeatother
\def\radiobuttonON{\raisebox{-1.5pt}{\stackinset{c}{}{c}{.35pt}{$\bullet$}{\scalebox{2}{$\circ$}}}}
\def\radiobuttonOFF{\raisebox{-1.5pt}{\scalebox{2}{$\circ$}}}

\newcommand*{\checkboxicon}{$\square$}

\usepackage[most]{tcolorbox}
\newtcbox{\inlinebox}[1][]{enhanced,
 box align=base,
 nobeforeafter,
 colback=gray!10,
 colframe=black,
 size=small,
 left=0pt,
 right=0pt,
 boxsep=2pt,
 boxrule=0.5pt,
 #1}

\newcommand{\textfieldtext}{~\inlinebox{\small{\textit{Text\phantom{a}Field}}}~}

\newcommand{\textfieldnum}{~\inlinebox{\phantom{T}\#\phantom{T}}~}

\newcommand{\emptyboxOthers}{~\inlinebox{\small{\phantom{This is my}}}~}

\makeatletter
\renewcommand\makeheadbox{\small Preprint. Submitted to \emph{Empirical Software Engineering}.}
\makeatother

\begin{document}

\title{Perspective of Software Engineering Researchers on Machine Learning Practices Regarding Research, Review, and Education}

\titlerunning{Perspective of SE Researchers on ML Practices}

\author{Anamaria Mojica-Hanke \and
        David Nader Palacio \and
        Denys Poshyvanyk \and
        Mario Linares-Vásquez \and
        Steffen Herbold
}

%\authorrunning{Short form of author list} % if too long for running head

\institute{%
Anamaria Mojica-Hanke, 0000-0002-5292-2977 \at
University of Passau, Passau, Germany.
\email{anamaria.mojica-hanke@uni-passau.de}
\and
David N. Palacio, 0000-0001-6166-7595 \at
Microsoft, Redmond, WA, USA
\email{davidnad@microsoft.com}
\and
Denys Poshyvanyk, 0000-0002-5626-7586 \at
William \& Mary, Williamsburg, USA
\email{dposhyvanyk@wm.edu}
\and
Mario Linares-Vásquez, 0000-0003-0161-2888 \at
Universidad de los Andes, Bogota, Colombia
\email{m.linaresv@uniandes.edu.co}
\and
Steffen Herbold, 0000-0001-9765-2803 \at
University of Passau, Passau, Germany.
\email{steffen.herbold@uni-passau.de} 
}

\date{Received: date / Accepted: date}
% The correct dates will be entered by the editor

\maketitle

\begin{abstract}
\textbf{Context:} \ac{ML} significantly impacts \ac{SE}, but studies primarily focus on practitioners, neglecting researchers. This ignores practices and challenges in teaching, researching, or reviewing \ac{ML} applications in \ac{SE}.\\
\textbf{Objective:} This study aims to contribute to knowledge of the synergy between ML and SE from the perspective of SE researchers, by providing information on the practices followed when researching, teaching and reviewing \ac{SE} studies that apply \ac{ML}.\\
\textbf{Method:} We analyzed SE researchers familiar with ML or who authored SE articles using ML, along with the articles themselves. We examine practices, SE tasks addressed with ML, challenges faced, and perspectives of reviewers and educators using open and axial coding and qualitative analysis.\\
\textbf{Results:} We found diverse practices focused on data collection, model training, and evaluation. Some recommended practices (\eg hyperparameter tuning) appeared in less than 20\% of the literature. Common challenges involve data handling, model evaluation (including non-functional properties), and involving human expertise in evaluation. Hands-on activities are common in education, although traditional methods persist. Recent data show a shift from statistical learning towards deep learning and Large Language Models (LLMs), leading to new practices such as prompt engineering.\\
\textbf{Conclusion:} Despite the accepted practices in applying ML to SE, significant gaps remain. By improving guidelines, adopting diverse teaching methods, and emphasizing underrepresented practices, the SE community can bridge these gaps and advance the field.
\keywords{machine learning practices, machine learning challenges}
\end{abstract}
%\linenumbers
\section{Introduction}
\label{sec:introduction}

Software engineering (\textbf{\ac{SE}}) is defined by \textit{IEEE} as \textit{``the application of a systematic, disciplined, quantifiable approach to the development, operation, and maintenance of software''}~\citep{SE_IEEE90}. {Over the last decade, machine learning (\textbf{\ac{ML}}) has become deeply intertwined with this discipline. Industry tools now apply \ac{ML} to code completion~\citep{tabnine, copilot}, code review~\citep{codeguru}, automated testing~\citep{Applitools}, and application performance monitoring~\citep{dynatrace}. \ac{SE} research has followed the same trajectory, with \ac{ML} appearing in studies of code completion~\citep{kim2021code}, software testing~\citep{bertolino2020learning}, vulnerability detection~\citep{kakkar2021combining}, and code representation/embedding~\citep{Chakraborty2022}, among many others. When \ac{ML} is applied to \ac{SE} problems, this is referred to as \ac{ML4SE}~\citep[\eg][]{kotti2023machine, dell2023evaluating}; the converse direction, applying \ac{SE} knowledge to \ac{ML} systems, is referred to as \ac{SE4ML}~\citep[\eg][]{serban2020adoption, martinez2022software}.

A growing body of work studies how \ac{ML} and \ac{SE} interact, and four strands are visible. The first mines question-and-answer platforms and version control systems to surface trending topics~\citep{bangash2019developers} and challenges when using \ac{ML}~\citep[\eg][]{hamidi2021towards, Islam_2019, Alshangiti_2019}, with data analyzed either manually~\cite{hamidi2021towards, Islam_2019} or with \ac{ML} models~\citep{bangash2019developers}. The second analyzes the published literature to characterize how \ac{ML} has been applied in \ac{SE}, including the \ac{SE} tasks addressed~\citep{kotti2023machine, watson2022systematic} and the \ac{ML} or deep learning techniques used~\citep[\eg][]{kotti2023machine, watson2022systematic}. The third draws on the authors' own research or teaching experience to articulate challenges, pitfalls, guidelines, and best practices~\citep[\eg][]{MichaelLones2021, ArpQuiPen_22, biderman2020pitfalls, zinkevich_2021}. The fourth surveys or interviews researchers, developers, and companies to understand best practices~\citep[\eg][]{serban2020adoption, serban2021practices}, \ac{ML} pipeline stages~\citep[\eg][]{amershi2019software}, mismatches~\citep[\eg][]{LewisGrace2021WAIN}, and \ac{ML} challenges~\citep[\eg][]{amershi2019software}.

Across this work, two limitations recur. First, the human subjects studied are almost always industry practitioners; \ac{SE} \textit{researchers} who use \ac{ML} have not been examined as subjects in their own right, even though they produce the published body of \ac{ML4SE} work. Second, prior studies look at one perspective at a time. We argue that to characterize how \ac{ML} practices in \ac{SE} are produced, evaluated, and propagated, three perspectives must be studied \textit{jointly}:

\begin{itemize}
    \item \textbf{Researchers} reveal the \textit{state of practice}: what is actually done when \ac{ML} is applied to \ac{SE} problems, and which challenges arise in doing so. Without this perspective, we cannot say what the community is currently doing.
    \item  \textbf{Reviewers} reveal the \textit{standard of rigor}: the criteria the community uses to decide which \ac{ML4SE} work merits publication. Without this perspective, we cannot say what the community considers good practice.
    \item \textbf{Educators} reveal what is \textit{transmitted to the next generation}: the practices, principles, and pitfalls deemed essential enough to teach. Without this perspective, we cannot say which practices will persist or which gaps may close as the field matures.
\end{itemize}

Each perspective is incomplete on its own. Studying researchers alone would tell us what is done but not whether it meets community standards or what is taught as foundational. Studying reviewers alone would tell us what the community asks for but not whether the work it accepts actually meets that bar. Studying educators alone would tell us what is taught but not whether teaching reflects current practice or current expectations. Examined together, the three perspectives expose the gaps that any one perspective cannot: most importantly, the gap between what researchers do, what reviewers expect, and what educators consider important enough to teach. These gaps are where the field is least mature, and they are not visible from any single vantage point.

 Guided by this motivation, we conduct an \textit{exploratory} study investigating five questions: i) which \ac{ML} practices \ac{SE} researchers apply in their published research articles and report in surveys and interviews; ii) which \ac{ML} practices \ac{SE} researchers themselves consider best practices; iii) which challenges \ac{SE} researchers face when applying \ac{ML}, drawn from researchers not confined to a single company or group; iv) which aspects matter when teaching \ac{ML} in an \ac{SE} context; and v) which aspects reviewers consider when assessing \ac{SE} research that uses \ac{ML}. We additionally analyze articles from 2023--2025 to surface recent trends. The contributions of our research are the following:

\begin{itemize}
    \item A characterization of the \ac{ML} practices used by \ac{SE} researchers, drawn from \textbf{195 research articles, 47 author survey responses, and 14 interviews} with influential \ac{ML4SE} researchers. To our knowledge, this is the first study that combines these data sources for the \ac{SE} research community. \ac{SE} research articles \textbf{commonly use practices related to model training and their functional evaluation} (\eg accuracy), \textbf{data preparation} (\eg split test train), and report on \textbf{how data are collected or reused}, the \textbf{use of baselines}, and \textbf{data cleaning} methods applied; deployment is the notable exception, being only rarely considered.
    \item Evidence of \textbf{a gap between what is considered best practice and what is reported in research articles}. \ac{SE} researchers commonly cite hyperparameter tuning, involvement of human expertise in evaluations, exploratory data analysis, manual validation, consideration of non-functional aspects, and evaluation of models in specific scenarios. Although all of these aspects also appear in research articles, they are not used consistently ($\leq$20\% of research articles).
    \item A characterization of the \textbf{challenges} that \ac{SE} researchers face when applying \ac{ML}, which \textbf{are mainly data-related} (\eg data collection, ground truth, data cleaning, data quality, and information leaks). Evaluation beyond the precision, \eg of non-functional aspects, and the participation of human experts is also seen as a major challenge.
    \item A reviewer-oriented catalog of aspects to consider when reviewing \ac{ML} in \ac{SE} research, covering a wide range of concerns (\eg quality of the process followed, validity of the results). \textbf{Guidelines from related fields cover many aspects that are also considered relevant for reviewing \ac{SE} research.} However, several \textbf{SE-specific aspects (\eg evaluation of non-functional aspects, qualitative assessments, and participation of human experts) are not yet sufficiently covered} by existing guidelines, highlighting the need for complementary guidelines for research at the intersection of \ac{ML} and \ac{SE}.
    \item An educator-oriented characterization showing that the literature and our study both highlight \textbf{hands-on activities as an important method of teaching \ac{ML}} in an \ac{SE} context. Although the literature focuses almost exclusively on such methods, our expert interviews also indicate that traditional teaching methods (\eg courses, text resources) are also considered common. Our experts additionally put a high value on non-functional aspects of \ac{ML} when teaching, \eg human-centered \ac{ML}.
    \item A current-trends analysis identifying the \textbf{use of \acp{LLM} as a recent trend}. However, we did not observe changes to \textbf{other parts of the machine learning pipeline} in recent papers, \ie practices regarding aspects such as data processing or model evaluation \textbf{are not yet affected} by this new technology.
\end{itemize}

The remainder of the article is organized as follows. In Section \ref{sec:related}, we present the related work. This is followed by Section \ref{sec:challenges}, in which we specify the challenges identified in the related work, which leads to a set of research questions presented in Section \ref{sec:rsch_questions}. This is followed by Section \ref{sec:rsch_protocol}, in which we describe our research protocol, including materials, variables, execution plan, and analysis plan. Next, in Section \ref{sec:results}, we present the results, followed by a discussion of the results in Section \ref{sec:discussion}, in which we also present the threats to validity. Finally, we conclude in Section \ref{sec:conclusions}.

% -------------------------------------------- Related work -----------------------------------
\section{Related Work}
\label{sec:related}

In this section, we describe previous papers that are related to our report. Previous works have focused on four main topics: i) \ac{ML} (best) practices  \citep[\eg][]{ArpQuiPen_22, zinkevich_2021}; ii) challenges in \ac{ML} \citep[\eg][]{biderman2020pitfalls}; iii) aspects that should be considered when teaching \ac{ML} \citep[\eg][]{shouman2022experiences, kreuzberger2023machine}; and iv) aspects that should be considered when reviewing \ac{SE} research/tools/papers that use \ac{ML}. %to \ac{SE} students (\eg \cite{}). 
Note that these topics are interrelated, and previous work could focus on more than one previously mentioned topic, \eg literature that discussed challenges and \ac{ML} practices \citep[\eg][]{horneman2020ai, tantithamthavorn2018experience, MichaelLones2021}.  In particular, the first and second topics are interrelated, so they are presented in a single subsection below. 

\subsection{\ac{ML} (Best) Practices and Challenges }

Some of the previous work has focused on ML practices and challenges based on the experience of their \textit{own authors' or experience in a similar environment (\eg a single company)}\footnote{By this, we mean that the practices were retrieved only from this single data source. Some studies used external experts to further validate these results, \eg \cite{ArpQuiPen_22}} while teaching~\citep[\eg][]{MichaelLones2021}, doing research~\citep[\eg][]{MichaelLones2021, biderman2020pitfalls, ArpQuiPen_22}, or applying \ac{ML}~\citep[\eg][]{MichaelLones2021, zinkevich_2021, horneman2020ai, tantithamthavorn2018experience, breck2017ml, ArpQuiPen_22, biderman2020pitfalls}.  Those studies discuss different practices such as \textit{``keep the first model simple and get the infrastructure right''}~\citep{zinkevich_2021}, \textit{``ensure you have a problem that both can and should be solved by AI''}~\citep{horneman2020ai}, \textit{``correlated metrics must be mitigated (\eg removed) in models'' }~\citep{tantithamthavorn2018experience},  and monitor that \textit{``data invariants hold in training and serving inputs''}~\citep{breck2017ml}. In addition, some studies also discuss \ac{ML} challenges such as  \textit{``data snooping''}~\citep{ArpQuiPen_22}, \textit{``label inaccuracy''}~\citep{ArpQuiPen_22, biderman2020pitfalls}, \textit{``not engaging with stakeholders''}~\citep{biderman2020pitfalls}, \textit{``statistical and social bias''}~\citep{biderman2020pitfalls}, and \textit{``data leakage''}~\citep{MichaelLones2021}. This shows not only interest in understanding \ac{ML} practices and challenges, but also interest in different aspects/steps required for building a \ac{ML}-enabled system. This first approach, studies based on authors' experience, allows an in-depth understating of the reasons for identifying practices and challenges since they are based on the authors' experiences. However, experience could be limited to a specific type of context (\eg specific types of industries, projects, or research). \cite{naher2022collab} also study \ac{ML} challenges, but with a focus on collaboration. Using interviews with participants from 28 different companies, they identified challenges, \eg with respect to the interaction between \ac{ML} and non-\ac{ML} experts for requirements, increased complexity for effort estimation due to \ac{ML}, handling and collection of training data, and how \ac{ML} can be integrated into products.

Another group of studies lists a set of practices and challenges mainly from \textit{question-and-answering communities} such as Stack Overflow or \textit{version control and collaboration platforms} such as Github. Most of these studies focus mainly on challenges/difficulties and issues encountered and discussed on those platforms, studying challenges in different \ac{ML} libraries~\citep{Islam_2019}, problems with the documentation of \ac{ML} libraries~\citep{Hashemi_2020}, issues in \ac{ML} libraries~\citep{Han_2020}, and an overview of \ac{ML} challenges~\citep{Alshangiti_2019}. \cite{mojicahanke2023machine} also investigated posts in \textit{question-and-answer systems} with the purpose of finding the best \ac{ML} practices discussed by practitioners in ML-related post questions. 

A third group of studies discusses \ac{ML} practices and challenges based on the analysis of \textit{previous publications}. In particular, \cite{watson2022systematic} present a systematic literature review on the use of \ac{DL} in \ac{SE}, describing the type of \ac{SE} data used, the \ac{SE} tasks for which \ac{DL} was used, the way the data was prepared and the \ac{DL} models used. In addition, \cite{kotti2023machine} conducted a tertiary study that analyzed 83 \ac{ML4SE} secondary studies published between 2009 and 2022. In their analysis, \cite{kotti2023machine} classified the studies according to the following aspects: the \ac{SE} task(s) addressed, the \ac{SE} knowledgeable area covered, and the \ac{ML} techniques mentioned in the studies, obtaining a set of challenges and actions, for the \ac{ML} domain, that could be helpful for the industry and researchers in the \ac{SE} domain. 

\cite{Clemmedsson2018} presents a set of \ac{ML} pitfalls extracted with a literature review; later, the pitfalls were validated with four semi-structured interviews. They concluded that the main three pitfalls relate to the lack of experience, data, and to the preference of short-term solutions over long term solutions in \ac{ML} projects.

\cite{serban2020adoption,serban2021practices} extracted 37 best \ac{ML} practices and six \ac{SE} traditional practices from gray-and-white literature. In which their importance and adoption in developer teams were studied via a survey. 
They distributed the survey using a snowball strategy, starting with their network, and then it was also distributed openly through the channels used by practitioners (\eg Twitter, Medium and HackerNoon). They identified that larger teams tend to adopt more practices~\citep{serban2020adoption}. They also identified that practice adoption for trustworthy \ac{ML} is relatively low, in particular, practices related to assuring the security of ML components~\citep{serban2021practices}.

\cite{LewisGrace2021WAIN} studied mismatches\footnote{Mismatch: ``a problem that occurs in the development, deployment, and operation of an ML-enabled system due to incorrect assumptions made about system elements by different stakeholders (roles): data scientist, software engineer, and operations, that results in a negative consequence''} in ML-enabled systems through interviews followed by a survey. They identified that most mismatches occurred when an incorrect assumption was made about a trained model, and the importance of sharing information about it varies depending on the role. 

\cite{amershi2019software} studied \ac{ML} projects in Microsoft. They conducted semi-structured interviews and an open-ended questionnaire that asked about existing \ac{ML} practices and challenges in the nine stages (see \tabref{tab:amershitable}) to build a  large-\ac{ML}-enabled system. After the analysis of 551 responses to the questionnaire, they identified that traditional challenges and needs in \ac{SE} differ from the ones faced in \ac{ML} applications (\eg \textit{``versioning the data needed for \ac{ML} is much more complex and difficult than other types of software engineering''~\citep{amershi2019software}}). On the topic of education, they found that while ``education and training'' were negatively correlated with personal \ac{AI} experience (\ie people with less \ac{AI} experience found education and training to be more important than people with more \ac{AI} experience), ``educating others'' was positively correlated with \ac{AI} experience.

\begin{table}[t]
\centering
\caption{Machine Learning pipeline stages by \cite{amershi2019software}.}
\begin{tabularx}{\textwidth}{>{\raggedright\arraybackslash}p{0.2\textwidth}>{\raggedright\arraybackslash}p{0.75\textwidth}}
\toprule
\textbf{ML Pipeline Stage} & \textbf{Description of the ML Pipeline Stages by \cite{amershi2019software}} \\ 
\midrule
Model Requirements & Designers decide which features are feasible to implement with machine learning and which can be useful for a given existing product or for a new one. \\
Data Collection  & Teams look for and integrate available datasets (\eg internal or open source) or collect their own. \\
Data Cleaning    & Involves removing inaccurate or noisy records from the dataset, a common activity to all forms of data science. \\
Data Labeling    & Assigns ground truth labels to each record. \\
Feature Engineering & Refers to all activities that are performed to extract and select informative features for machine learning models. \\
Model Training & The chosen models (using the selected features) are trained and tuned on the clean, collected data and their respective labels. \\
Model Evaluation & The engineers evaluate the output model on tested or safeguard datasets using pre-defined metrics. \\
Model Deployment & The inference code of the model is deployed. \\
Model Monitoring & The deployed model is monitored for possible errors during real-world execution. \\
\bottomrule
\end{tabularx}
\label{tab:amershitable}
\end{table}

In general, while previous work made exhaustive efforts to identify challenges and practices when using \ac{ML}, there is no study that uses \ac{ML4SE} researchers and researchers as subjects. We close this gap by studying how this large and impactful research community deals with \ac{ML}.

% ------------------------------   RESEARCH  -------------------------------------------

\subsection{Aspects Considered when Reviewing \ac{ML} in a \ac{SE} context}
 
When reviewing \ac{SE} research, general guidelines for reviewers are provided by the journals (\eg guidelines for Transactions on Software Engineering and Methodology (TOSEM)~\citep{tosem_guidelines} and conferences (\eg guidelines for the International Conference of Software Engineering (ICSE)~\citep{Pollock_Di_Penta_2023} in which general instructions are presented to the reviewers, such as searching and evaluating for novelty, relevance, significance, (technical) soundness, quality of writing/presentation, and an appropriate state-of-the-art. Similar criteria are also established in the \ac{ML} community~\citep[\eg][]{ICML_guidelines, jmlr_guidelines}, showing the importance of those general aspects when reviewing research. However, often those concepts are not clear~\citep[\eg][]{ralph2021empirical}, and are not focused on the particular challenges, practices, and pitfalls that could be present when applying \ac{ML} in \ac{SE}.

When searching for more concrete studies that analyze or describe guidelines when reviewing \ac{SE} studies, it is possible to find some studies whose main goal is to provide guidelines for \ac{SE} empirical research~\citep{ralph2021empirical, arshad2021towards, chatterjee2022empirical,kapoor2024reforms}. From which two of them \citep{arshad2021towards, chatterjee2022empirical} are the continuation of the main study \citep{ralph2021empirical} and will be described together.

\cite{ralph2021empirical} present  \textit{``a brief public document that communicates expectations for a specific kind of study (\eg a questionnaire survey)''}. In this study, beyond general standards, standards for 18 different types of empirical studies are included, such as Action Research, Grounded Theory, Repository Mining, and Data Science. In these standards, \ac{ML} is included as a tool for different types of study such as data science and Repository Mining. However, when \ac{ML} is identified as a tool in Repository Mining studies, the reader is redirected to the Data Science guidelines. In these guidelines, attributes, patterns, and anti-patterns are presented when  \ac{ML} is used as a data-centric tool to analyze \ac{SE} phenomena or artifacts. However, \ac{ML4SE} can also interact in broader aspects than in analyzing the phenomena in \ac{SE}. In addition, since the practices apply to data-centric analysis methods in general, the practices could be broad and not focused on punctual problems that \ac{ML} presents \eg biasing a model or leaking data.

Regarding the complementary studies of the previous study \citep{ralph2021empirical} , the first complementary study \citep{arshad2021towards} is the formalization of a tool that allows the creation of a checklist of standards when writing or reviewing empirical studies. The second complementary study \citep{chatterjee2022empirical} presents a tutorial in which the standards for a specific type of empirical study (Mining Software) are described. And the same comparisons made with the main study apply to them.

\cite{kapoor2024reforms} in addition to the set of guidelines for reviewing \ac{ML} studies, provides a recommendation on how to conduct and report research using \ac{ML}. The recommendations are a consensus of 19 researchers in multiple fields and cover topics ranging from defining the goal and motivation of the use of \ac{ML} methods to how to present generalizability and limitations of the research, also analyzing aspects such as data preprocessing, quality and leakage; model design; computational reproducibility; and metrics and uncertainty.

% ------------------------------   TEACHING --------------------------------------------
\subsection{Aspects Considered when Teaching \ac{ML} in a \ac{SE} context}

We have identified previous works that have studied relevant aspects, practices, or challenges while teaching \ac{ML} at different levels of education, including kindergarten to $12^{th}$ grade (K-12)~\citep[\eg][]{marques2020teaching, tedre2021teaching, ziglar2022teaching, elhashemy2024adapting}, undergraduate~\citep[\eg][]{Huang_2018, Chinmay_2021, acquaviva2022teaching} and master levels~\citep[\eg][]{acquaviva2022teaching, van2008teaching, shouman2022experiences}. However, the education levels from kindergarten to $12^{th}$ (K-12)  are beyond our scope. Therefore, in the following, we will discuss only previous work related to higher education. 

Several papers consider the teaching of \ac{ML} at a university level to non-computer science students. For Electrical Engineering students, \cite{Huang_2018} and \cite{Chinmay_2021} identify the importance of real-world examples, hands-on examples, and projects, but also the importance of general interest of students in \ac{ML} \citep{Huang_2018}. \cite{acquaviva2022teaching} used their experience in teaching \ac{ML} for the Physical Sciences to students to formulate best practices. They also stress the importance of hands-on examples. However, they also highlight that high-quality materials are crucial and find that teaching across-disciplines concepts (\eg interpretability) is challenging. When teaching students with mixed backgrounds, \cite{shouman2022experiences} also observed the importance of hands-on examples, but also emphasized the positive impact of live code during lectures. \cite{van2008teaching} go one step further with hands-on examples using Embodied Intelligence~\citep[\eg][]{brooks1991intelligence, pfeifer2006body} method to teach \ac{ML} (\ie the concepts of Q-Learning and neural networks) using Lego Mindstorm NXT. The tangible machines and the positive association with the tool (Lego) motivated the students. They concluded that \textit{``the students with less technology affinity successfully completed the course, while the students with more technology affinity excelled in solving advanced problems.''}

Regarding studies more focused on the relationship between \ac{SE} and \ac{AI} / \ac{ML} in an educational setting,  some studies analyze different aspects, \eg the usage of generative tools for \ac{SE} \citep{yabaku2024university}, the integration of tools \ac{AI} in \ac{SE} education \citep{vierhauser2024towards, daun2023chatgpt}. Other full studies have studied teaching \ac{AI} with a special focus on software engineers/computer scientists at the graduate or undergraduate level \citep[\eg][]{lanubile2023teaching, lanubile2023training, kastner2020teaching, mashkoor2023teaching, heck2021lessons, acuna2023developing}. Since we are interested in aspects that should be considered when teaching \ac{AI} to a \ac{SE} audience, we will focus on the last group of studies.

\cite{acuna2023developing} discusses the experience of developing a course that introduces data science to undergraduate and graduate students, covering \ac{ML} techniques. For the course, they map the Software Engineering Body
of Knowledge (SWEBOK) v3 \citep{SWEBOK} areas to specific topics in their data science course. They conclude that this alignment between SWEBOK and data science domains is feasible and provides a context for SE students to learn concepts.

\cite{lanubile2023teaching} and \cite{lanubile2023training} present lessons learned from a hands-on course that covers the end-to-end ML component life cycle (\ie from model building to production deployment), targeted to students that already have experience in \ac{ML} and \ac{SE}. Regarding the  teaching methodology, they identified that the most helpful methodology for learning for the students was the project-based nature of the course and the team work. In addition, they identified that some of the most useful practices for the students were code versioning and experiment tracking.

\cite{mashkoor2023teaching} designed a graduate-level course for students with mixed backgrounds (\ie students from \ac{AI} or \ac{SE}), with the goal of teaching \textit{``how to employ state-of-the-art SE practices to engineer AI-intensive systems''}. This course included lecture-based instruction, collaborative learning, and group projects. The lessons learned presented after the course was executed include that having interdisciplinary groups work, as well as the hands-on approach to real-world problems and peer learning had a positive effect. In addition, they present that both background groups learned new tools and platforms. They also describe that giving constant and constructive feedback also helps in the learning experience.

\cite{kastner2020teaching} created a course for higher education, but with a different focus from the previously presented related work, \ac{SE4AI}, in which they were interested in teaching \textit{``how software engineering techniques can be used to build better systems with or around \ac{AI} components''}. After teaching the course, they formulate several recommendations, including finding challenging  scenarios for students to seriously consider the efficiency of the system. The use of practical examples was useful and real-life use/applications of AI should be considered. There is a lack of tools for emerging tasks in \ac{AI} and also for data management in an educational setting. Note that although this study has the opposite focus SE4AI, in their course they teach concepts and present challenges from both areas (\ie \ac{SE} and \ac{AI}), therefore it is relevant for our study.

Although most of the previous work on identifying key aspects in teaching \ac{ML} has executed specific case studies for teaching \ac{ML} in higher education \citep[\eg][]{Huang_2018, Chinmay_2021, acquaviva2022teaching, van2008teaching, kapoor2024reforms, kastner2020teaching, lanubile2023teaching, acuna2023developing} or presents lessons learned based on their own experience during years of teaching~\citep[\eg][]{acquaviva2022teaching}, our approach rely on the experiences of educators who have been teaching \ac{ML} in different institutions for multiple years. This gives us a broader perspective that is not focused on experiences from a single institution or a pair of institutions (\eg two universities). 

\section{Research Challenges}
\label{sec:challenges}

In this section, we present the challenges (\textit{\textbf{CHs}}) identified in the related work for each of the topics. Our report is aimed at overcoming four key challenges that software researchers face while working on Machine Learning:

\begin{itemize}
    \item \CH{1}: \textbf{Omitting software researchers perspective from best ML practices.} We are unaware of the current (best) \ac{ML} practices that software engineering researchers consider, report, or use when applying \ac{ML} in \ac{SE}.  This is due to the fact that the current state-of-the-art does not focus on \ac{SE} researchers as the main subjects of study, or if it does, it is not based on a diverse curated set of researchers.  In particular: 
    
    \begin{itemize} 
        \item \CH{{1.1}}: We disregard the \ac{ML} practices that are used in \ac{SE} research studies when  \ac{ML} is used as part of them. 
        
        \item \CH{{1.2}}: We have an unclear concept of what a ``professional procedure is considered to be correct or most effective''~\citep{b_practice_def} (best practice) when applying \ac{ML} in \ac{SE} research. 
        
        \item \CH{{1.3}}: We do not have a clear idea of what is considered a \ac{ML} practice by \ac{SE} researchers. 

    \end{itemize}

    \item \CH{2}: \textbf{We do not have a clear picture of the challenges \ac{SE} researchers face when applying \ac{ML} in \ac{SE}} based on a curated and diverse set of researchers.
    
    \item \CH{3}: \textbf{Review of software research focused on the applicability.} of ML, we are not familiar with the relevant aspects that should be taken into account when reviewing research centered on \ac{SE} topics that use \ac{ML}.
    
    \item \CH{4}: \textbf{Teaching Machine Learning for Software Engineering.} We ignore the relevant aspects that should be taken into account when teaching \ac{ML} in the specific context of computer science or software engineering programs, \eg when teaching about ML4SE or SE4ML. 
    
\end{itemize}

\section{Research Questions}
\label{sec:rsch_questions}

The purpose of this study is to understand different perspectives on \ac{ML} practices and challenges. In order to achieve these goals and align with the previous challenges, we %will
investigate the following four research questions:

\begin{itemize}
    \item \RQ{1}\textbf{:}\textit{What are the \ac{ML} practices that the \ac{SE} researchers use and declare when using \ac{ML}? 
    }
    \begin{itemize} 
        \item \RQ{{1.1}}\textbf{:} What are the \ac{ML} practices used in \ac{SE} research articles? %  Survey
        \item \RQ{{1.2}}\textbf{:} What are the best ML practices declared by the authors of \ac{SE} research articles? 
        \item \RQ{{1.3}}\textbf{:} What are the \ac{ML} practices declared by \ac{SE} researchers? % W&M interview
    \end{itemize}
    
    \item \RQ{2}\textbf{:} \textit{
    What are the difficulties/challenges faced by researchers when applying \ac{ML} for \ac{SE} research?
    } 
    \item \RQ{3}\textbf{:} \textit{
    What are the significant aspects that should be considered when reviewing \ac{SE} research that has used \ac{ML}? 
    }
    \item \RQ{4}\textbf{:} \textit{
    What are the significant aspects that should be considered when teaching \ac{ML} from a perspective of \ac{SE}?
    }
    \item RQ{5}\textbf{:} \textit{
    What are the recent trends when applying ML for \ac{SE} research?
    }
\end{itemize}

With this set of questions, as mentioned previously,  we want to better understand the different aspects that should be taken into account when using \ac{ML}. In particular, \ac{ML} practices and challenges faced by \ac{SE} researchers, reviewers, and educators.  In this sense, the study focuses mainly on \ac{ML4SE}. This is due to the practices and challenges that belong to the \ac{ML} domain and are being applied in \ac{SE}. However, our study follows a \ac{SE4ML} approach as the process of identifying and coding the practices needed to answer the questions comes from the \ac{SE} domain.

\section{Research Protocol}
\label{sec:rsch_protocol}

In this section, we define the materials, variables, execution plan, and analysis plan of our research protocol. A general overview of the following steps described in this section is depicted in  \figref{fig:execution_plan}.

\begin{figure}[h]
    \centering
    \includegraphics[width=\textwidth]{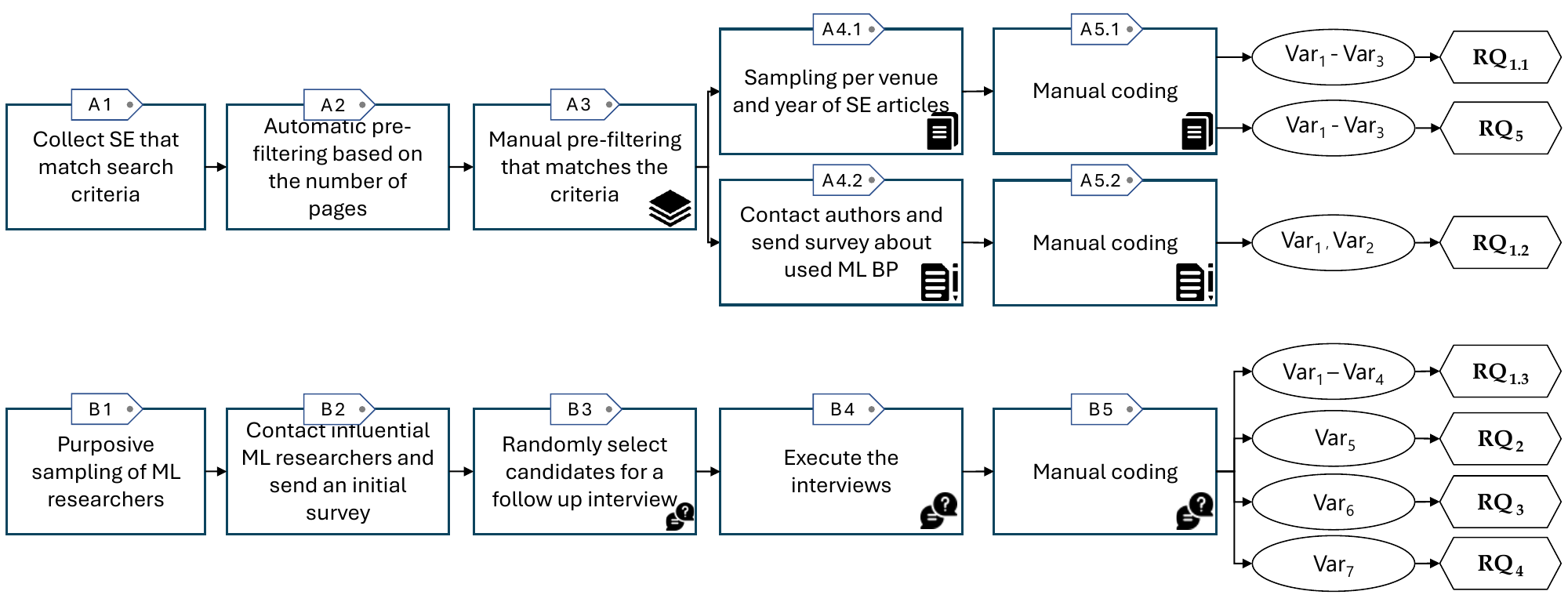}
    \caption{Research Protocol steps and their relation with the Variables (\textsf{Var}) and the Research Questions (\RQs).}
    \label{fig:execution_plan}
\end{figure}

\subsection{Materials}
\label{subsec:materials}
Our study is based mainly on \ac{SE} research articles published in  $A^*$ conferences from 2011 to 2022 and \ac{SE} researchers using \ac{ML}, which we used to identify suitable subjects.

%-------------------------------------------------------------SUBJECTS----------------------------------------------------------------------------
\subsubsection{Subjects}
\label{sec:subjects}

We have three groups of subjects in our study:
\begin{enumerate}
    \item[\SEpapers]  \ac{SE} research articles.
    \item[\SEauthors]  Authors of at least one paper in the complete set of the above-mentioned \ac{SE} articles (\SEcompletepapers).
    \item[\SEresearchers]   \ac{SE} researchers that use \ac{ML}.  
\end{enumerate}

%----------------------------------------------- SB: PAPERS -----------------------------------------------------------
\SEpapers \textbf{\textit{\ac{SE} research articles.}}  We search papers that match the query ``machine learning'' in the ACM digital library and IEEE Explore engines in the period between 2011 and 2022 from the three top $A^*$ SE conferences in the CORE ranking~\citep{CORE_INC_2023}, \ie \textit{ASE}, \textit{FSE}, and \textit{ICSE} (see \figref{fig:execution_plan}, \textit{step}  \Aone). Our rationale for this narrow focus on few, highly ranked conferences is that the selection of research articles for these venues is extremely competitive and the rigor of the review process is well documented, meaning that published research articles should follow what the community deems to be good practices. We note that this does not mean, nor do we intend to imply, that this may not be the case for other venues (\eg high-impact journals or topical conferences). For this, we execute the following queries while varying the \texttt{venue} and the \texttt{year} (\eg \texttt{Venue}: ICSE, \texttt{Year}: 2011):

\begin{itemize}
    \item \texttt{\textbf{ACM Digital Library:}
    \texttt{Venue}: ICSE '11: Proceedings Of The 33rd International Conference On Software Engineering. \textit{Search criteria}: [\textit{All}: "machine learning"] AND [\textit{E-Publication Date}: (01/01/2011 TO 12/31/2011)]}
    \item \texttt{\textbf{IEEE Explore:} \texttt{Venue}: 2011 33rd International Conference on Software Engineering (ICSE). \textit{Search criteria}: ``machine learning''}
\end{itemize}

In addition to this, we execute additional filtering to reduce the number of possible false positives (\ie research articles that mention the keywords ``machine learning'' but do not use it) to analyze. We keep \ac{SE} research articles that meet the following criteria: 
\begin{itemize}
    \item Research articles with at least 10 pages long.
    \item Research articles that are not tutorials or keynotes.
    \item Research articles in which at least one of the \ac{ML} pipeline stages is mentioned/described.
    \item Research articles related to software engineering and \ac{SE} for \ac{ML}.
\end{itemize}

This second filtering is performed, as previously described, to reduce the number of possible false positives that are analyzed in the following steps of the analysis plan. The first aforementioned criterion is executed automatically with a \textit{Python} script (see \figref{fig:execution_plan}, \textit{step}~\Atwo), which reads the metadata of the articles and determines the length of each article. The rest of the criteria in this pre-filtering is executed by at least two people per article, and \textit{Krippendorff's alpha}~\citep{krippendorff2004content}  is computed as a coefficient of inter-rater reliability (IRR); see \figref{fig:execution_plan}, \textit{step}~\Athree. Once we have this \textit{complete sample of pre-filtered research articles} (\SEcompletepapers), due to the popularity of the \ac{ML} topic and in order to have a sample that can be manually analyzed, one-fifth of the research articles is be selected with stratified sampling by venue (\eg \textit{FSE}) and year (\eg \textit{2011}); see \figref{fig:execution_plan}, \textit{step}~\Afourone.

We executed a search for over 10 years (2011\textendash2022), executing the queries previously described, obtaining a total of 1398 articles. From this set, we executed the filter of automatic pre-filtering, giving us a set of 1,015 in which we executed the manual round of pre-filtering,  obtaining a group of 580
``pre-filtered'' research articles. For this pre-filtering process, we had an IRR score of  0.668, which is an agreement level that allows us to keep the data and draw tentative conclusions~\citep{krippendorff2004content}. To mitigate the risk of noise within our research articles, we also discussed all disagreements between raters to come to a commonly agreed judgment regarding inclusion. With stratified sampling,  we obtained a total of 117 articles.\footnote{one-fifth of 580 is 116, but in order to keep the strata/groups' proportions and doing the corresponding rounding we sampled 117 articles}  

To further broaden the perspective and understand recent trends, we also collected a secondary set of \ac{SE} articles with data for 2023\textendash2025. For these three most recent years, we considered the three conferences above (\ie ICSE, ASE, FSE), as well as three major software engineering journal: the IEEE Transactions on Software Engineering (TSE), the ACM Transactions on Software Engineering and Methodology (TOSEM), and Empirical Software Engineering (EMSE). The choice of these journals was motivated by the journal-first track of \textit{ICSE}, which allows submissions from these journals. Due to the recent rise in software engineering papers with ML aspects, we did not sample a fixed percentage. Instead, we select five articles per year and venue randomly (\ie 85 articles in total).\footnote{At the time the articles were downloaded, the ASE'25 articles were not yet publicly available} Articles that do not match the inclusion criteria are dropped and we resample additional articles until we have five for each venue.

%----------------------------------------------- SB: SURVEY -----------------------------------------------------------

\SEauthors \textbf{\textit{Authors of at least one research article in} the complete set of pre-filtered research articles in the \ac{SE} articles~(}\SEpapers\textbf{)}.  To identify our first group of \ac{SE} researchers that uses \ac{ML} for \ac{SE} tasks, we sample from the complete list of authors the first and last authors of the complete set of articles identified in the previous subject group (\figref{fig:execution_plan}, \textit{step}~\Athree).  This is based on the assumption that the first author is the one who contributes the most to a research article. Therefore, they have a better understanding of the research article. In addition, the selection of the last author is based on the assumption that, in many cases, the last author is the principal investigator.  After sampling the required authors, we extract the authors' contact information (name and email) using a \textit{Python} script to read the metadata of the research articles. Subsequently, two people look for i) mismatches between names and emails (\ie the email matches with another author's name); ii) cases where there are different emails per a single author; and iii) empty names or emails (\ie emails/names are not able to be extracted with the script). In the cases where the same author has more than one email associated, we choose the more recent one (by using the publication date). Additionally, in the case where the authors'emails are not present in the research article, we search for the authors' public website.  After having the list of \ac{SE} researchers, we send them a survey asking about possible \ac{ML} best practices that they used in the articles previously identified in the complete set of \ac{SE} research articles (\figref{fig:execution_plan}, \textit{step}~\Afourtwo\!), a step explained in a follow-up section (\secref{sec:explan}).

In the complete set of \ac{SE} research articles (\SEcompletepapers), we identified 2,792 authors, of which 1,734 were unique,\footnote{If an author appeared multiple times with different names, it is considered as a different author. It can also happen the opposite, two different authors with the same name were considered the same author.} and on average, each author has written 1.6 articles. Then, we extracted only the first and last authors, which gave us a sample of 829 authors, from which we could identify an email address for 769, which we contacted and sent the survey. Of these authors, 58 responded.

%----------------------------------------------- SB: Interviews -----------------------------------------------------------

\SEresearchers \textbf{\textit{\ac{SE} researchers that use \ac{ML}}}. This second group of \ac{SE} researchers that use \ac{ML} in their studies is selected with purposive sampling~\citep{patton2014qualitative} (\figref{fig:execution_plan}, \textit{step}~\Bone\!). The participants are academic contacts of the authors, who are influential researchers in the field of \acf{ML4SE}. For this second group of researchers, we send a survey asking for demographic information (\figref{fig:execution_plan}, \textit{step}~\Btwo\!), and from those who answer, a random sub-sample of a maximum of 20 is selected (\figref{fig:execution_plan}, \textit{step}~\Bthree\!). Then, a post-hoc interview is conducted with this sub-sample, in which we ask questions related to machine-learning best practices, challenges, and their perspective not only as researchers but also as reviewers and educators. This step, \figref{fig:execution_plan}~-~\textit{step}~\Bfour, is explained in a follow-up section (\secref{sec:explan}). 

We contacted a total of 38 influential ML4SE researchers,\footnote{A total of 40 researchers were contacted via email. However, two of the emails were invalid. The researchers were all authors of multiple research articles at highly ranked venues (\eg TOSEM, TSE, ICSE, FSE, ASE).} from whom 19 responded to the survey. Since the number of respondents was less than 20,\footnote{20 was the initial number established and submitted as part of the protocol} we could not randomly select  20 participants to interview. Therefore, the 19 researchers were contacted to conduct the follow-up interviews. From this, we were able to interview 14 of them.\footnote{Of these, 11 were professors or senior researchers in industrial research labs at the time of the interview, and 3 were PhD students in their final year with exceptional track records.}

%-------------------------------------------------------------VARIABLES----------------------------------------------------------------------------
\subsection{Variables}
\label{sec:variables}

We study seven main \textsf{variables}, depicted in \tabref{tab:variables}.  This table also shows the relationship between each variable and each group of subjects previously defined, and as can be seen, not all \textsf{variables} are studied (\checkmark) in every group of subjects. For the \textsf{Practice} variable (\Var{1}), we identify three sub-variables: \Var{{1.1}}~\textsf{Input}\textemdash the data or information that was processed as part of a practice, \Var{{1.2}}~\textsf{Technique} \textemdash the technique that was applied to process the input, and \Var{{1.3}}~\textsf{Purpose/output} \textemdash the purpose or goal behind the application of a practice. This separation allows us a more fine-grained analysis and better understanding of not only the techniques, but also the motivations behind it as well as for which inputs best practices are applied. The variables are chosen to provide the required information for the research questions (see Section~\ref{sec:analysys_plan}).

\begin{table}[]

\begin{tabular}{@{}L{1cm}L{2cm}L{5.2cm}L{0.5cm}L{0.5cm}L{0.5cm}@{}}
\toprule
\textbf{ID} & \textbf{Name}                           & \textbf{Description}                                                                                                                & \SEpapers & \SEauthors & \SEresearchers \\ \midrule
\Var{1}     & \textsf{Practice}       & \textit{The actual application or use of a method}~\citep{practice_def} used in a ML pipeline.                                                                  & \checkmark              & \checkmark              & \checkmark              \\ [0.2cm]
\Var{{1.1}} & \textsf{Input}           & The input required for the \textit{method} in order to be executed.  For example, the data (including the type) that is needed to train a Random Forest (method). The input can be elements/results of previous steps in the \ac{ML} pipeline, \eg cleaned-data, trained model.                                                                  & \checkmark              & \checkmark              & \checkmark              \\ [0.2cm]
\Var{{1.2}} & \textsf{Technique}       & \textit{A way of doing something}~\citep{technique_def}. In particular, it is a way of executing a possible action in the \ac{ML} pipeline. In other words, the \textit{method} that is being applied.                                                                 & \checkmark              & \checkmark              & \checkmark              \\ [0.2cm]
\Var{{1.3}} & \textsf{Purpose/output}  & The reason for applying certain techniques which will lead to an output. In our terms, this variable represents the reason (which includes the output) for applying a \textsf{Technique} to a specific \textsf{Input}.                                                                  & \checkmark              & \checkmark              & \checkmark              \\ [0.2cm]
\Var{2}     & \textsf{ML pipeline stages}  & ML stages that are associated to a \textsf{practice}. The possible stages are based on the \ac{ML} pipeline proposed by \cite{amershi2019software} & \checkmark              & \checkmark              & \checkmark              \\ [0.2cm]
\Var{3}     & \textsf{SE tasks }        & Software engineering tasks that are addressed using ML.                                                                             & \checkmark              & \textemdash              & \checkmark              \\ [0.2cm]
\Var{4}     & \textsf{Quality attributes}  & Properties that a ML-enabled system could have.                                                                                     & \textemdash              & \textemdash              & \checkmark              \\ [0.2cm]
\Var{5}     & \textsf{ML challenges}   & Challenges faced when developing and researching ML including possible difficulties faced when assuring the \textsf{Quality attributes} in a ML system.                                                                                & \textemdash              & \textemdash              & \checkmark              \\ [0.2cm]
\Var{6}     & \textsf{Reviewer's perspective} & Important aspects that should be considered when reviewing research articles that use ML.                                          & \textemdash              & \textemdash              & \checkmark              \\ [0.2cm]
\Var{7}     & \textsf{Educator's perspective} & Aspects and tools that should be considered when teaching ML.                                                                       & \textemdash              & \textemdash              & \checkmark              \\
\bottomrule
\end{tabular}%
\caption{Analyzed variables, including the associated group(s) of subject(s). The group of subjects are represented by the icons, previously defined. \SEpapers\ : \textbf{\textit{\ac{SE} research articles}}, \SEauthors\ : \textbf{\textit{Authors of \ac{SE} articles}},  \SEresearchers\ : \textbf{\textit{\ac{SE} researchers that use \ac{ML}}}.}
\label{tab:variables}
\end{table}
%-------------------------------------------------------------EXecution (ex) PLAN----------------------------------------------------------------------------

\subsection{Execution Plan}
\label{sec:explan}
Once the subjects are identified, as well as the variables studied for each of them (see \secref{sec:subjects} and \secref{sec:variables}), we collect the data needed for a subsequent step of manually coding the values for all variables. For this last step, we use procedures from the grounded Theory~\citep{strauss1998basics}, focusing only on open coding, where concepts and their properties are identified; and axial coding, where connections between the codes are identified. We do not execute selective coding, where a core category is identified, since our goal is not to create a theory. This aligns with \cite{strauss1998basics}, where it is described that a researcher can use only some of its procedures to meet the research goals. Since we use the codes generated by this manual coding as input for further qualitative analysis, we describe the manual coding as part of the execution plan and not as part of the analysis. In the following paragraphs, we describe how the manual coding is executed, taking into account the different subjects previously described. 

%------------------------------------------------ ex: ARTICLES ------------------------------------------------------------------------

\subsubsection{Manual Coding for the \ac{SE} articles}
\label{sec:explanART}

This inductive coding~\citep{leach2022transformativist} process uses as input the sub-sample of articles from 2011-2022~(\SEpapers) and is conducted by a group of seven people (see \figref{fig:execution_plan}, \textit{step}~\Afiveone). First, the set of articles are distributed between the coders, always guaranteeing that each article is assigned to two taggers. Then, each group of two authors independently code all the possible variables (see \secref{sec:variables}) in each of the assigned articles. Subsequently, a third independent author reviews the assigned codes and performs a merge between the two groups of the identified codes. The independent author resolves conflicts in wording (\eg by using fixed syntactic patterns by starting the techniques with a verb) without changing the meaning of the codes. In this case, the merging process is executed with the main purpose of having a complete picture of the possible codes for the variables, \eg \textsf{Practice}, since one of the coders could miss to identify a code, rather than contradict the other coder. However, since having contradicting codes is possible, the third independent author also reviews the input article and conveys a decision on the codes. This merging process is executed by three of the authors (\ie $mr_{a1}$, $mr_{a2}$, and $mr_{a3}$). In which one of these authors ($mr_{a1}$) executes the merging process in all the labeling processes in which she was not involved. The rest of the labeled articles are merged by the other two authors (\ie $mer_{a2}$ and $mer_{a3}$), and they merge the labels in which they were not involved (\eg $mer_{a2}$ merges the labeling in which $mer_{a1}$ and $mer_{a3}$ are involved).

After the three involved authors execute the merging process, we  have discussion rounds between two authors for conveying to an initial set of codes that represent the information extracted in the previous labeling process $mr_{a}$. Once the initial codes are established, we execute consecutive rounds of axial coding~\citep{corbin1990basics, strauss1998basics} to have categories that group the initial codes with similar or related meanings. This coding process enables a subsequent analysis of the variables of interest. Analysis that is explained in \secref{sec:analysys_plan}. In this process, we performed the labeling for 117  \ac{SE} articles.

The papers from the years 2023\textendash2025 were coded after the full analysis for the results from 2011-2022 was already finished. The intent was to observe if our results generalize or which new aspects became relevant in the last years. Due to this, we followed a slightly different coding procedure. First, we used a mixture of open coding and closed coding, by giving all raters access to the full code book that we created for the earlier batch of papers. Second, we used a single rater per article. This is facilitated by the aforementioned change, since the closed coding has less potential for disagreements and because the expectations on the labels were already established by the earlier batch of papers. While this introduces a certain risk with respect to the data quality, we believe this risk is acceptable given that this data is not for our primary analysis, but rather only used to understand generalization to later years (see Section~\ref{sec:plan_trends}).

%------------------------------------------------ ex: Surveys ------------------------------------------------------------------------

\subsubsection{Survey to the Authors of \ac{SE} Articles and Manual coding of the Responses}
\label{sec:explanSURV}
Once the targeted authors are identified~(~\SEauthors~, see \secref{sec:subjects}), we distribute a survey\footnote{For this survey, we did not collect any personal information. An ethics approval was  obtained for the Ethics Committee of the University of Passau on 2024-05-10.}\footnote{The question and instructions on how to answer it can be found in the Appendix.} with a single open question (\figref{fig:execution_plan}, \textit{step}~\Afourtwo\!), in which we ask them the following question:

\begin{center}
``\textit{What machine learning best practices have you used in your software engineering research articles, and how often did you apply those practices [Always, Frequent, Sometimes, Never]}''.  
\end{center}

The main purpose of this open question is to obtain insights about ``what is considered a best \ac{ML} practice'' by \ac{SE} researchers. This implies that we do not go through the articles written by the authors and extract \ac{ML} practices and ask the authors which ones of the identified practices are considered best practices, nor give them an example of what is a good practice, or answer what we refer to as an  \ac{ML} best practice. While doing this, we also avoid biasing the respondents on what we consider an \ac{ML} best practice. 

To extract the codes from the responses, first, the answers are be skimmed to spot not useful answers (\ie empty responses or responses that do not relate to \ac{ML} or that only point to papers). Then, for the remaining answers, in a first approximation of identifying individual practices, we split the responses by line breaks (\ie enter, \textbackslash n),\footnote{Based on the indications given on the survey, each line should represent a practice.}  taking into account of not removing context from each practice. Next, for each line of the valid responses, two authors execute an open coding~\citep{corbin1990basics, strauss1998basics} and axial coding~\citep{corbin1990basics, strauss1998basics} procedure (\figref{fig:execution_plan}, \textit{step}~\Afivetwo\!), allowing the identification of codes and categories for the \textsf{Practice} sub-variables. In the case that the practices are not divided by line breaks, this separation is also done by the coders. Since no structure is be given to the authors about what is a best practice, it could happen that the identified practices only contain one of the \textsf{Practice} sub-variables, and the other sub-variables will be empty. 

The survey is sent to the 769 selected authors, from which we received 58 answers. In the first skimming process, we filtered 11 answers as i) they were not related to \ac{ML} (six responses), ii) they only referred us to external research articles (two answers), iii) they related to \ac{ML} but did not address the question posed (two answers), and iv) the research article that was a false positive, not filtered in the pre-filtering step (one answer). In particular, for the two answers that were discarded due to the authors referring us to external research articles, we did not continue with an analysis of the articles since, as previously mentioned, we wanted to gain insights about what is considered a best practice (including its structure), and the process of extracting the practices ourselves would break this goal.  In addition, during the period in which we conducted the survey, we also received three emails asking for clarifications about what we considered an \ac{ML} best practice. In these cases, we replied that we could not answer the question since we would bias the answer. 

%------------------------------------------------ ex: Interviews ------------------------------------------------------------------------

\subsubsection{Interview to the \ac{SE} researchers and Manual Coding of the Responses}
Once the targeted researchers are identified, after executing the demographic survey\footnote{Since for this survey we collect personal data, as well as the follow-up interview. This protocol was submitted to an ethics committee.} and the random sample of the researchers (~\SEresearchers~, see \secref{sec:subjects}) follow-up interviews\footnote{The questions of the demographic survey and the script for the interviews can be found in the Appendix.} are executed; see \figref{fig:execution_plan}, \textit{step}~\Bfour.  In the interviews, we ask different questions related to \ac{ML} best practices, challenges, and perspectives as a reviewer or educator. These interviews are the input for our third process of inductive coding~\citep{leach2022transformativist}; see \figref{fig:execution_plan}, \textit{step}~\Bfive.

The interview questions are structured into four groups. The first group of questions covers best practices from multiple perspectives, to understand what the researchers consider as best practices and what/why consider to be important. In particular, this first group of questions was an extension of \textit{components of learning} and \textit{learning principles} theory by \cite{ML_principles_Mostafa}. The second group aims to understand how \ac{ML4SE} is already impacting education. The third group is about the perspective of the reviewer to understand if there are special concerns when reviewing research on ML for SE. After covering these general aspects, the fourth group follows the structure of the ML pipeline by \cite{amershi2019software} to understand how the different phases of the ML workflow are executed by the different researchers. Finally, we finish the interview by asking potential follow-up questions. 

In order to be able to execute an inductive coding process, we i) record the interviews and ii) generate transcripts of the interviews. We schedule and conduct the meeting via Zoom, a platform that allow us to record the interviews. Before recording the meeting, the required permission to record is requested from the participants. Then, in order to have transcripts that can be coded, we use \textit{Whisper}~\citep{radford2023robust}.

For this manual coding, two authors independently code the same 20\% of interviews to create a codebook (coding all the variables related to this group of subjects, see \secref{sec:subjects}). This practice has been suggested and used in previous studies ~\citep{elder1993working, mayer2021now}. In this process, the two coders meet to discuss their codes to harmonize the code book created up to that point. Since we found after this initial coding that a single coder was sufficient since the coder identified all information also identified by the other coder in the first 3 interviews (21\% of the data), the single coder proceeded to code the remaining interviews. Afterward, a second coder double-checked all the results by reading the interviews and checking for the correctness of the assigned codes. Thus, we still had two coders to ensure the reliability of the data but avoided the additional effort of harmonizing the wording of codes.

As previously mentioned, the demographic survey was sent to 38 \ac{ML} influential researchers, from which 19 responded, and we were able to interview 14 \ac{SE} influential researchers who use \ac{ML}. The interviews were conducted online with a video conference. As a result of this process, we have 14 recorded interviews; for each interview, we have an audio file (.m4a), a video file (.mp4), and a chat file (.txt).  The audio files were used as input to the  \textit{Whisper}~\citep{radford2023robust} tool to generate the required transcript files.
  
%--------------------------------------------------------------- Analysis Plan  ------------------------------------------------------------------------

\subsection{Analysis Plan} 
\label{sec:analysys_plan}
We conduct a qualitative analysis taking as a basis the output of the inductive coding processes previously mentioned. This analysis aim to answer our \RQs. In particular, when executing the qualitative analysis, we use only the relevant \textsf{variables} and subjects for each research question (see \figref{fig:execution_plan}).

\subsubsection{\RQ{1} What are the ML practices that are used and declared by SE researchers when using ML?} 
To answer this research question, we use the identified codes and categories for the \textsf{Practice} variable in the manual coding with the aim of understanding what is being considered \ac{ML} (best) practices, from three angles:

\begin{enumerate}[label=\roman*)]
    \item Used practices in \ac{SE} research (\RQ{{1.1}}), by identifying what practices are reported in the \ac{SE} studies (\SEpapers).
    \item What \ac{SE} researchers think is a best \ac{ML} practice (\RQ{{1.2}}), by analyzing not only the codes and categories but also the existence or not of the different components (sub-variables) of a practice.
    \item \ac{SE} researchers that use \ac{ML} (\RQ{{1.3}}), by analyzing the obtained codes and categories in the interviews for the \textsf{Practice} variable. 
\end{enumerate}

In addition, for all the sub-research questions, the variable \textsf{ML pipeline stages} complements the perspectives while indicating the stage(s) where the practices are applied. The \textsf{SE tasks} variable provides an additional context of when the practices are being used for \RQ{{1.1}} and \RQ{{1.3}}. Furthermore, since for the third group of subjects (~\SEresearchers~), we have the opportunity to go deeper and obtain more details, we complement also their perspective (\RQ{{1.3}}) with the codes and categories obtained for the \textsf{Quality attributes} and \textsf{ML principles}. When analyzing the codes and categories, counts of all codes are reported, as well as interesting quotes from the articles, surveys, and interviews. 

\subsubsection{\RQ{2} What are the difficulties/challenges faced by researchers when applying ML for SE research?} 
To answer this research question, we use the identified codes and categories for the variables \textsf{ML challenges} and \textsf{Challenges to ensure quality attributes}. Those two variables are relevant for this \RQe in order to identify the general challenges (\textsf{ML challenges}) that can be found when using \ac{ML} either for pure research or more application related or to identify challenges to obtain a desired property (\textsf{Challenges to ensure quality attributes}) in an \ac{ML}-enabled system. When analyzing the codes and categories, counts of all codes are reported, as well as interesting quotes from the interviews. 

\subsubsection{\RQ{3} What are the important aspects that should be considered when reviewing research that has used \ac{ML} in \ac{SE}?} 
For answering this research question, we use the identified codes and categories for the \textsf{Reviewer’s perspective} variable. With these codes, we are able to analyze possible relevant elements and characteristics that should be considered when reviewing a study that uses \ac{ML} in the \ac{SE} community. When analyzing the codes and categories, counts of all codes are reported, as well as interesting quotes from the interviews.

\subsubsection{\RQ{4} What are the important aspects that should be considered when teaching \ac{ML} from a perspective of \ac{SE}?} 
To answer this research question, we  use the identified codes and categories for the \textsf{Educator’s perspective} variable. With these codes, we will be able to analyze possible relevant tools and characteristics that should be considered when teaching \ac{ML} in an \ac{SE} university context, since we interview \ac{SE} educators who teach at the university level. When analyzing the codes and categories, counts of all codes are reported, as well as interesting quotes from the interviews. 

\subsubsection{\RQ{5} What are the recent trends when applying ML for \ac{SE} research?}
\label{sec:plan_trends}

We answer this question by studying the differences between the data collected during different time frames. Similar to \RQ{1.1}, we will analyze the codes and categories we identified for coding the data for 2023\textendash2025 and compare these results to the data from 2021\textendash2022. We qualitatively describe the differences in the trends, taking into account the differences in sample size (110 articles from 2021\textendash2022 and 85 from 2023\textendash2025) and time frame covered (11 years vs. 3 years).

%------------------------------------------------ RESULTS -------------------------------------------------------------------------------------

\section{Results}
\label{sec:results}

During the open-coding of the different subjects, \ie the \SEpapers \textit{\ac{SE}} \textit{research articles}, the \SEauthors\ survey responses from \textit{authors of \ac{SE} articles}, and the transcripts from the \SEresearchers\ \textit{interviews} of the \textit{\ac{SE} researchers that use \ac{ML}}, some of the collected data needed to be dropped to ensure a consistent quality of our data (see \tabref{tab:subjectsnumber}).
\begin{itemize}
    \item 11 survey responses were dropped because they did not respond to the question asked (see \ref{sec:explanSURV}).
    \item 7 research articles were dropped because they were false positives, \eg because they only mentioned ML but did not study it. 
\end{itemize}
The following results are with reference to the data after removing these invalid data points, \ie the percentages and totals do not consider the dropped data.  All results are published online in our replication kit.

\subsection{General Statistics}

The coding process yielded 4,107 initial codes extracted from the three subjects in the eight variables (see \tabref{tab:codes}). Subsequently, we harmonized the codes (\eg removing differences in wording), resulting in a total of 3,057 codes. Although some variables retained the initial number of codes throughout the process (\ie \Var{3}-\Var{7}), since different codes represented different semantic aspects, others (\ie \Var{{1.1}}- \Var{2}) exhibited a reduction in the number of codes. The subsequent axial coding leads to a strong reduction in the counts, \ie many codes belong to the same category, \eg different variants of data splitting for evaluation. This strong reduction is visible across all variables.

As can be seen in \tabref{tab:codes}, during the coding process, the variables \Var{{1.1}} - \Var{{1.3}} have the highest number of different codes. This is in part because these variables were coded in all three subjects and the expressiveness of the natural language (\ie an idea can be expressed in multiple ways). However, this also already highlights the many different practices that SE researchers apply. In contrast, the variables \ac{ML} pipeline stage (\Var{2}) encodes a specific, small set of concepts and, therefore, less codes.

In addition, during open-coding, we observed that the ~\SEauthors\ \textit{surveys} had a peculiarity that is not shared with the other two subjects.  The practices variables (\Var{{1.1}} -\Var{{1.3}}) during the open-coding of the ~\SEpapers~\textit{research articles} and \SEresearchers~\textit{interviews} were coded with the full context of the article or the \textit{interview}'s script, respectively. For the surveys, this was not possible, due to the short answer that lacked a broader context (\eg SVM, Adversarial ML, or Hyperparameter tuning). The mean length of useful characters (\ie text without punctuations and numbers) of a \textit{survey line} (practice) is 53.7 characters, which drops to 44.9 characters when we omit the indicated frequency (\eg [Always]). This is an artifact of our use of a free-text field, where we believe that most respondents answered in a short manner to save time. Due to this, we discuss the results for the surveys independently of the results of the research articles and interviews, even though we report their numbers together. 

Another aspect that we consider during the analysis is the smaller sample size of the interviews: since we have only 14 data points, we cannot expect that all research topics are covered in equal breadth as within the research articles. Similarly, single responses from interviews have a larger impact on the overall ratio of responses coded in a certain manner than for articles: each \textit{interview} makes up 7\%, while an article only accounts for 0.9\% of the data. We account for this when analyzing how common the codes for practices are across these data sources.

%------------------------------------------------ Subjects counts -------------------------------------------------------------------------------------
\begin{table}[]
\centering
\caption{Number of subjects available for the study and actual number of subjects coded.}
\begin{tabular}{@{}lrrr@{}}
\toprule
Subject                                              &  \makecell[c]{Initial\\Subjects} & \makecell[c]{Coded\\Subjects} & \makecell[c]{Avg.\\Length/duration}\\ \midrule
\SEpapers\  \ac{SE} research articles (2011-2022)    & 117              & 110            &  11.95 Pgs.\\
\SEpapers\  \ac{SE} research articles (2023-2025)    & -                & 85             & 23.05 Pgs. \\
\SEauthors\  Authors of \ac{SE} articles             & 58               & 47             &  259.06 chars\\
\SEresearchers\ \ac{SE} researchers using \ac{ML} & 14               & 14             & 48.6 min.\\ \bottomrule
\end{tabular}
\label{tab:subjectsnumber}
\end{table}

%----------------------------------------- Code counts -------------------------------------------------------------------------------------
\begin{table}[t]
    \centering
    \caption{Number of codes and categories in the open coding process (\ie from initial codes to categories).}
    \begin{tabular}{@{}lrrr@{}}
    \toprule
       Variable & Initial Codes & After Harmonizing & After Axial Coding \\
       \midrule
        \Var{{1.1}} \textsf{Input}                 & 597   & 365   & 49  \\
        \Var{{1.2}} \textsf{Technique}             & 1,843 & 1,454 & 176 \\
        \Var{{1.3}} \textsf{Purpose/output}        & 1,293 & 866   & 72  \\
        \Var{2} \textsf{\ac{ML} pipeline stage}    & 14    & 10    & 10  \\
        \Var{3}  \textsf{\ac{SE} tasks}            & 116   & 116   & 10  \\
        \Var{4}  \textsf{Quality attributes}       & 36    & 36    & 9   \\
        \Var{5} \textsf{\ac{ML} Challenges}        & 89    & 89    & 16  \\
        \Var{6} \textsf{Reviewer's perspective}    & 71    & 71    & 21  \\
        \Var{7} \textsf{Educator's perspective}    & 42    & 42    & 14  \\   
    \end{tabular}
    \label{tab:codes}
\end{table}

%------------------------------------------------ CODES per variable -------------------------------------------------------------------------------------
In the following subsections, we will present the statistics per category in the \SEpapers~\textit{research articles} and \SEresearchers~\textit{interviews}. For the variables \Var{{1.1}} \textendash \Var{2}, the statistics that are  presented are normalized counts per category per subjects (\eg data appears 100\% in \textit{interviews}, means that the data category appeared in all 14 interviews). For the rest of the variables (\Var{3} \textendash \Var{7}), we present the normalized number of appearances in general (\ie counts/total of subjects). The reason of this, is because the first three variables are asked for all the three subjects, therefore are coded, for the rest of the variables, except for variable \Var{2}, are only present for the ~\SEresearchers~\textit{interviews} (see \tabref{tab:variables}). 

Due to the high number of categories and the resulting size of the result tables and figures, all supporting data are included in the Appendix of the paper. In addition, the most notable findings for the \Var{{1.1}}\textendash \Var{{1.3}} variables are shown in Tables \ref{tab:inputInterest}, \ref{tab:techniqueInterest}, and \ref{tab:outputinterest}. 

%------------------ C.p.V: Input  -------------------------------------------------------------------------------------

\subsection{Results for \RQ{1}: What are the ML Practices that are Used and Declared by SE Researchers when Using ML?} 

We now report the results for the ML practices, \ie the inputs, techniques, and purposes that we obtained from our subjects. Then we proceed to report additional data to put this in context in terms of the ML stages that were considered, the \ac{SE} tasks, and the quality attributes considered.

\subsubsection{\Var{{1.1}}: Input}

The inputs used in ML evaluation practices, as expected, center on data artifacts in both subjects, though practitioners and researchers differ notably in how they specify them. While every \textit{interview} referenced model(s) and data as inputs in broad terms, \textit{research articles} decompose these into fine-grained categories such as numeric data or deep learning models. Common across both subjects are inputs directly tied to evaluation output, \ie data and results.

Another identified difference between the two studied subjects emerges around two categories that appear frequently in \textit{interviews} but are nearly absent from \textit{research articles}: unverified labels and the problem to be solved. \rev{This shows differences not only} in the abstraction level, but also in how different concerns of the research are discussed.   For example, in  I6 references ``\textit{sanity checking, manual inspection, a couple of data}'' and I9 describes reasoning about whether ``\textit{we have [...] that we need to solve the problem we're looking to solve}'', illustrating how input validity is actively questioned during evaluation.

Beyond these common categories, both subjects contain a long tail of domain-specific inputs,  such as time-series incident, provided statistics of an existent data set, and categorical data in \textit{research articles} (less than 3\%); and metadata, unstructured, and multi-modal data in \textit{interviews} (less than 8\%). Shared across both subjects, a variety of inputs such as image data, screenshots, embeddings, model activations, tabular data, and biometric data appear less than 15\% for both subjects.

\begin{table}[ht]
\centering
\caption{Notable codes and the ratio of occurrence for \Var{{1.1}} \textsf{Input}. Results are selected due to the impact indicated by high occurrence frequencies, or because of strong differences between the two data sources. The full results can be found in figures~\ref{fig:input_axial1} and \ref{fig:input_axial2}  in the appendix.}
\begin{tabular}{@{}lrr@{}}
\toprule
\textbf{Code}                          & \SEresearchers\ \textbf{Interviews} & \SEpapers\ \textbf{Articles}\\ \midrule
data (ground truth) + results (output) & 78\%                & 85\%            \\
results (metrics)                      & 64\%                & 50\%            \\
data (code data)                       & 92\%                & 48\%            \\
data (textual data)                    & 41\%                & 35\%            \\
data                                   & 100\%               & 40\%            \\
model                                  & 100\%               & 31\%            \\
data (unverified labels)               & 42\%                & 5\%             \\
problem                                & 50\%                & 1\%             \\ \bottomrule
\end{tabular}
\label{tab:inputInterest}
\end{table}

%------------------ C.p.V: Technique  -------------------------------------------------------------------------------------
\subsubsection{ \Var{{1.2}}: Technique of Practices}

\begin{table}[]
\caption{Notable codes and the ratio of occurrence for \Var{{1.2}} \textsf{Technique}. Results are selected due to the impact indicated by high occurrence frequencies, or because of strong differences between the two data sources. The full results can be found in figures~\ref{fig:technique_axial1}-\ref{fig:technique_axial5}  in the appendix.}
\centering
%\small
\begin{tabularx}{\textwidth}{@{}>{\raggedright\arraybackslash}p{6,5cm}>{\raggedleft\arraybackslash}p{1.9cm}>{\raggedleft\arraybackslash}p{2.5cm}@{}}
\toprule
\textbf{Code} &\SEresearchers\! \textbf{Interviews} & \SEpapers\!\textbf{Articles}\\ \midrule

computing accuracy metrics & 28\% & 65\%  \\
use statistical learning models & 28\% & 42\% \\
use deep learning models & 57\% & 40\% \\
usages collaborative knowledge sources & 42\% & 36\% \\
split data  & 42\% & 30\% \\
use existing datasets & 71\% & 30\% \\
compare with baseline model(s) & 21\% & 19\% \\
compare with baseline model(s), baselines non ml & - & 25\% \\
compare with baseline model(s), baselines deep learning & - & 21\% \\
compare with baseline model(s), baseline statistical learning & - & 17\% \\
use embeddings & 14\% & 1\% \\
use embeddings, word2vec & - & 11\% \\
use bag-of-words/n-gram embeddings & - & 10\% \\
use embeddings, contextual & 7\% & 7\% \\
collect own data & 64\% & 12\% \\
select/filter data based on quality & 42\% & 3\% \\
hyperparameter tuning & 78\% & 20\% \\
\bottomrule
\end{tabularx}
\label{tab:techniqueInterest}
\end{table}

The \textit{research articles} commonly compute accuracy metrics, and use statistical learning and deep learning models. Nearly a third also report on how they re-used data and/or the splits they made. Overall, these are normal and unsurprising practices for the development of \ac{ML4SE} approaches. When own data is sourced, a common practice is to use collaborative knowledge sources \rev{\eg GitHub, StackOverflow}. All of these aspects are also frequently discussed by the interview participants.

In general, the \textit{research articles} were more specific than \textit{interviews}, especially with respect to some processes that are executed in the \ac{ML} pipeline. Notably, the research articles provided more details about comparisons with baselines and the types of baselines (\rev{non-ML}, statistical learning, and deep learning) that were used, while the interviewees more generally described comparisons with baselines as important. Similar fine-grained aspects about the techniques used can also be seen in other aspects, \eg the features that were used for the machine learning which are described in depth in the \textit{research articles} (\eg type of textual embeddings), but only generally discussed in the interviews (e.g., I8: ``\textit{We just try to engineer features that could, in our opinion, get something that was useful to predict the outcome variable. For example, with readability prediction, the main goal was to engineer and test some specific features to predict readability. In that case, they were textual features.}''). 

A clear distinction between the \textit{interviews} and \textit{research articles} visible with respect to own data collections. While this is mentioned in over two thirds of the interviews as important, less than one third of the the \textit{research articles} collected own data. The two main drivers for own data collection we identified in the interviews were the desire for high-quality data (\eg I6: ``\textit{I used to collect data sets for which the dependent variable I want to predict is somehow reliable}''), but also the lack of data for a given problem (\eg I10: ``\textit{but sometimes it happens that we do not have a data.}''). While this is a large gap, this can possibly be explained by the fact that while researchers sometimes face problems with data quality or availability, this is not the case for all papers, thereby explaining the discrepancy.

For other notable differences between the \textit{interviews} and \textit{research articles}, there is no such explanation. While hyperparameter tuning is mentioned in in almost all interviews, only one-fifth of the articles used this technique. There are also similar, but smaller gaps with respect to techniques that should ensure high quality and soundness of results, \eg involvement of humans (incl. manual validation and qualitative analysis of results), data exploration, and removing duplicates. The statements of the \SEresearchers\ in this regard are often very definitive regarding the importance (\eg I6: ``\textit{I think some form of data cleaning is absolutely a must}''). Similarly, \textit{interviews} frequently highlight aspects like relating to analyzing the problem and related categories like model selecting models or evaluation methods, while the rationale for decisions is often not explained within the \textit{research articles}. However, this does not automatically mean that authors of \textit{research articles} do not use these techniques, but it could also mean that they are implicitly assumed and not explicitly reported (\eg the importance of the problem). Since our coding process only considered explicit statements, this would explain this difference in contrast to other aspects that are usually specifically described in \textit{research articles} (\eg data collection, training process, and model evaluation).

We believe that similar considerations can also be made for other categories, although the differences are less pronounced. The usage of concrete SE tools (\eg code parsers) or existing tools (\eg TensorFlow) is often not part of the text of \textit{research articles}, but rather of replication packages, which were not considered in our study. Moreover, the \textit{research articles} we studied rather focus on what is being done without explaining why alternative solutions were not considered, explaining differences regarding not using certain techniques. Similarly, when data are re-used, \textit{research articles} often implicitly assume a high quality of data sets without specifically stating this. 

In addition, we identified 28 techniques that were not part of any of the \textit{research articles} but were mentioned in the \textit{interviews}. These techniques were mostly on a relatively high level of abstraction (\eg select models, conduct sanity checks, use machine learning) and were mentioned in at most 35\% of the interviews. However, there are also interesting corner cases, \eg considering ethical aspects, using A/B testing, and using model monitoring. Beyond this, we identified a long tail of corner cases  that appear in less than 20\% for both subjects: data manipulation (\eg tokenize by words), how models are evaluated (\eg ablation study, changing the model or the data used), the usage of visualizations, usage of formal concepts (\eg use formal data representations or formal proofs), specific types of data prepossessing (\eg graph prepossessing), specific types of metric used (\eg model-specific or cost-efficiency metrics), and interpreting the model (\eg use explainable AI). 

%------------------ C.p.V: Purpose  -------------------------------------------------------------------------------------

\subsubsection{\Var{{1.3}}: Purpose}

\begin{table}[]
\centering
\caption{Notable codes and the ratio of occurrence for \Var{{1.3}} \textsf{Output/purpose}. Results are selected due to the impact indicated by high occurrence frequencies, or because of strong differences between the two data sources. The full results can be found in figures~\ref{fig:output1}-\ref{fig:output3}  in the appendix.}
\begin{tabularx}{\textwidth}{@{}lrr@{}} 
\toprule
\textbf{Code} & \SEresearchers\! \textbf{Interviews} & \SEpapers\! \textbf{Articles}\\ \midrule
to evaluate model(s) for a functional attribute & 85\% & 86\% \\
to collect data & 92\% & 62\%\\
to compare against others & 14\% & 47\%\\
to embed and encode data & 28\% & 28\% \\
to evaluate model(s) for a non-functional attribute & 21\% & 23\%\\
to split the data into different sets & 28\% & 22\%\\
to understand the requirements and possible solutions & 100\% & 4\%\\
to evaluate model(s) with human judges & 64\% & 4\%\\
to evaluate model(s) in specific scenarios & 50\% & 3\%\\
to ensure statistical significance & 7\% & 15\%\\
\bottomrule
\end{tabularx}
\label{tab:outputinterest}
\end{table}

Across both subjects, a dominant purpose is the evaluation of functional attributes of the model, most notably accuracy, which appears in the vast majority of both \textit{interviews} and \textit{research articles}, contrasting with the low attention given to the evaluation of non-functional aspects. Data collection is also shared as a prominent purpose, though with greater prominence in \textit{interviews} than in \textit{research articles}. Both subjects further share a set of moderately frequent purposes, each appearing in roughly 17\%--28\% of both, mostly related to data preparation (\ie embedding data, removing noisy data, and splitting data into subsets), as well as optimizing the model, consistent with patterns identified in other variables.

Beyond this shared focus, the two subjects diverge notably. \textit{Research articles} concentrate on comparison with others, reflecting a benchmarking norm, whereas \textit{interviews} instead emphasize understanding the requirements and possible solutions, a purpose rarely mentioned in \textit{research articles}, likely because published work implicitly assumes problem relevance by targeting known use cases. This is illustrated by I6 discussing ``\textit{the reason why we want to use machine learning}'', I4 identifying ``\textit{a particular pain point}'', and I10 noting that ``\textit{if, say, there are a couple of papers [...], the project has some importance to it}''.

A similar pattern holds for data quality purposes, aligning with previous findings in other variables. Labeling, specifically labels for instances and to validate and correct labels, appears in at least half of \textit{interviews} but less than 18\% of \textit{research articles}, consistent with practitioners working with existing datasets rather than collecting new (\eg I7: ``\textit{You reflect on the data. You look for errors. You try to patch bad labels}''). Practitioners also place considerably greater emphasis on human involvement in evaluation and evaluation under specific conditions, both appearing in over half of \textit{interviews} but less than one-tenth of \textit{research articles}.

This divergence extends further to purposes that appear predominantly or exclusively in \textit{interviews}:extracting and building features, handling bias, understanding the data, and ensuring data quality each appear in 70\%--95\% of \textit{interviews} but under 25\% of \textit{research articles}; and reproducibility and understanding the model in roughly half of \textit{interviews} but only around 20\% of \textit{research articles}. Purposes such as monitoring models, handling concept drift, and deploying a model appear exclusively or nearly exclusively in \textit{interviews}, pointing to life-cycle concerns outside the scope of most published research. Finally, statistical significance receives little attention in either subject - present in only 15\% of \textit{research articles} and a single \textit{interview}, which is notable given its centrality to evaluation methodology.

\subsubsection{\Var{2}: \ac{ML} Stages}

%------------------ C.p.V: SE STage  -------------------------------------------------------------------------------------

For this variable, we cannot conduct a meaningful analysis between the \textit{research articles} and \textit{interviews}. The reason for this is that the interviews were designed to mention each stage described by \cite{amershi2019software} explicitly. The only cases where there is no data for a stage in an \textit{interview} were when the interviewees explicitly mentioned that they do not do \textit{Feature Engineering} (1 \textit{interview}), 
Data Labeling (2 \textit{interviews}), \textit{Model Deployment} (4 \textit{interviews}), or \textit{Model Monitoring} model (11 \textit{interviews}).

For the \textit{research articles}, the most notable pattern is the near-complete absence of life-cycle-related stages: \textit{Model Monitoring} does not appear at all, and \textit{Model Deployment} is present in only in 2,7\% of articles while \textit{Model Requirements} appears in only 13,6\%. This aligns well with prior results. In contrast \textit{Model Evaluation} and \textit{Model Training} are common and present in more than 95\% of the \textit{research articles}. We also manually checked the \textit{research articles} that did not conduct training or evaluation. In three cases, the \textit{research articles} did not train a model and only compared existing models for a specific purpose (\eg testing whether the models were reproducible). In four cases, the training phase was mentioned, but no clear practices were obtained due to the short description of it or because it was not clear to the taggers what exactly was done. For \textit{Model Evaluation}, two \textit{research articles} were about approaches for deep learning models, and the evaluation was the evaluation of the tool that was not used \ac{ML} and not the evaluation of deep learning models. For two other \textit{research articles}, the evaluation phase was mentioned in the \textit{research articles}. However, the taggers did not identify any explicit practices.

When coding the data for the \textit{research articles}, we deviated from the definition of the \textit{Data Cleaning} stage from \citep{amershi2019software} to include techniques regarding the understanding of the data, under the assumption that understanding of the data is a precursor for cleaning. We also decided to add generic data transformations to \textit{Data Cleaning} (\eg normalization) instead of to \textit{Feature Engineering}. This was done to ensure that \textit{Feature Engineering} is only about the conscious modeling of features for specific aspects and not just transformations of existing information. Again, the overall frequencies align well with the frequencies we observe for specific techniques in the previous sections. 

Finally, we added the stage \textit{Other} to cover \ac{SE} and research considerations that cannot be associated with any specific stage, \eg ethical considerations, whether ML is required to solve a problem, transparency on how research is done, and how to make results more reproducible.

%------------------ C.p.V: SE tasks  -------------------------------------------------------------------------------------
\subsubsection{\Var{3}: \ac{SE} Tasks}

\iffalse
\begin{table}[ht!]
\centering
\caption{Codes identified for \Var{{3}} \textsf{SE tasks} and their frequency in the 117 \textit{research articles}.}
\begin{tabular}{@{}lr|lr@{}}
\toprule
\textbf{Codes} & \multicolumn{1}{l|}{\textbf{Frequency}} & \textbf{Codes} & \multicolumn{1}{l}{\textbf{Frequency}} \\ \midrule
Testing                 & 44  & Requirements and Specifications & 4  \\
Implementation          & 26  & Implementation ML               & 2  \\
Management              & 18  & Maintenance                     & 2  \\
Testing ML              & 12  & Team/developer aspects          & 2  \\
Software Representation & 6  & Other                            & 1  \\ \bottomrule
\end{tabular}
\label{tab:SEtasks}
\end{table}
\fi

While the SE tasks are not directly related to our {\RQe}s, they help contextualize for which research topics our \textit{research articles} results are representative. Since the small sample size of \textit{interviews} precludes a meaningful distribution, we report counts directly. The axial coding grouped the SE tasks based on the phase in the software engineering process.

The \textit{research articles} span a range of SE phases, but with a pronounced concentration in testing-related topics: testing alone accounts for 40\% of articles, covering tasks such as ``bug localization'', ``test prioritization'', and ``developers' perception of software quality'', and a further 10.9\% (12 articles) directly address testing of ML (\eg ``deep learning testing'', ``XAI for ML testing''). Together, testing-related articles represent more than half of the set of sampled \textit{research articles}. Implementation (\eg ``code source competitive generation'', ``automated error feedback'', and ``comment generation'') and management (\eg ``incident linking'' and ``bug report classification'') follow at a distance, with the remaining articles distributed sparsely ($\leq$ 5\%) across software representation, requirements, maintenance, and team aspects%(See~\tabref{tab:SEtasks}).
.

%------------------ C.p.V: QA   -------------------------------------------------------------------------------------
\subsubsection{\Var{4}: Quality Attributes}

With this variable, we take a closer look at the quality considerations discussed during the interviews. As expected, considerations about functional aspects (incl. accuracy) were common (71\%). However, there was also a strong focus on non-functional properties: reliability, security, privacy, usability, explainability, interpretability, and fairness were all mentioned by at least 21\% of the experts we interviewed. Resource utilization and process metrics were also mentioned, but by at most 2 participants. 

\subsubsection{Insights from the Survey}

As we explained above, the survey responses were generally very short and, therefore, yielded limited insights, which precluded a detailed comparison of frequencies. Nevertheless, the results show us clearly what the first things are that come to SE researchers' minds, when they are asked about machine learning best practices, without biasing them through any additional information. The data for \Var{{1.1}} (\textsf{Input}) shows that such practices are about \textit{data}, \textit{models}, or \textit{results}. The techniques are more insightful. Here, \textit{cross validation}, \textit{hyperparameter tuning}, and \textit{using deep learning} are the most common (20\%-30\%). These are followed by aspects like \textit{running multiple experiments}, \textit{considering specific scenarios for evaluations}, \textit{splitting data}, \textit{using statistical learning}, and \textit{handling data imbalance}. Although the percentages are lower, the trend that these are the most common aligns well with the interviews, supporting their representativeness, although the survey was intentionally brief and did not even define the meaning of the practices.

%------------------ C.p.V: Challenges  ------------------------------------------------------------------------------------

\subsection{Results for \RQ{2}: What are the Difficulties/Challenges Faced by Researchers when Applying ML for SE Research?}

For the second research question, we have only one variable to consider, \ie \Var{5} (\textsf{Challenges}), collected using the \textit{interviews}. We identified six common challenges. The most frequently mentioned is getting the right data (64\%), whether in terms of sufficient quality (\eg I12: ``data sets in the domain where I work are junk'') or the cost associated with the collection, often due to human effort (\eg I12: ``\textit{a human study [is] expensive soil}''). Almost as common is the challenge of knowing how to evaluate the model and getting a correct ground truth (57\%). Thus, not only are data a problem in general, but ground truth data poses an additional layer of difficulty. Additionally, how to best evaluate models even if data is available is seen as a problem (\eg due to the weak correctness assumption) and because it is difficult to involve human expertise in model evaluations. The interviewees also noted that having enough computational resources is increasingly a pressing concern (50\%).

Three further challenges were each mentioned in 42\% of interviews. Conducting studies involving humans is considered difficult due to subjectivity, recruitment costs, and finding a representative sample (\eg I3: \textit{if you're working on a generative code model and code that people are actually going to work on}''). Interpretability and trustworthiness reflect that the power of \ac{ML} comes at the cost of limited transparency, difficulty understanding why a model behaves as it does, makes it hard to trust in certain domains (\ie medical domain) and hard to explain to users (\eg I8: \textit{How do I explain something to someone else in a way that they understand it?}''). Reproducibility and deployability are also a challenge, as the randomness of training processes and the difficulties in sharing models, data, and deployments remain largely unresolved.

Beyond these, model selection (28\%), including considerations of model size, fine-tuning strategies, and balancing computational cost, is closely followed by a broader concern about the lack of tooling (21\%) that would otherwise facilitate this, \eg data management and the implementation of specialized models. Finally, a number of challenges were mentioned in at most 14\% of interviews, including fairness, model governance, security, privacy, response time efficiency, lack of knowledge, and challenges intrinsic to the problem itself.
%------------------ C.p.V: Reviewers  -------------------------------------------------------------------------------------

\subsection{Results for \RQ{3}: What are the Important Aspects that Should be Considered when Reviewing Research that has Used \ac{ML} in \ac{SE}?} 

The data for the third research question was collected during the \textit{interviews} and encoded in \Var{6} (\textsf{Reviewers' Perspective}). Two aspects stand out as most prominent, each appearing in 50\% of the interviews. First, reviewers are attentive to the gap/lack in coverage some areas, such as the usage of multi-modal data, debugging, and licensing issues. Second, reviewers carefully analyze the metrics used, particularly their alignment with the problem being solved. I9 shows this concern well ``\textit{And I think it's hard to let go of of these bad habits that we've built as a community for using these metrics that aren't necessarily aligned.}''. Closely related is a broader concern about model evaluation (42\%), encompassing aspects such as ``recognizing the importance of qualitative analysis'' and ``the inclusion of necessary baselines''. 

About one-third of the interviewees (35\%) highlight \textit{fairness}, \textit{methodology} (incl. how justifications and experiment execution are reported), and whether models are usable and suitable for practice.  A similar percentage (28\%) raise the importance of looking for errors that impact data integrity, model selection (\eg insufficient hyperparameter exploration), and the reproducibility and replicability of the results, as well as the analysis of the results (\eg result tables and whether contributions are clearly stated). 

Less frequently mentioned concerns include how model comparisons are conducted and whether limitations are adequately discussed (14\%). Several concerns appeared in only a single interview, such as errors in the feature engineering process, in the general process and methodology, and in describing limitations, as well as unclear usability, lack of guidelines, and lack of ML expertise. The latter is illustrated by I3: ``\textit{concepts for machine learning that maybe aren't [...] quite as well known in the software engineering research community, because that reduces your pool of qualified reviewers even further.}'' I3 also notes that ``\textit{the community has come a long way in terms of the overall collective knowledge that people have on machine learning techniques}'', suggesting this challenge may diminish over time.

%------------------ C.p.V: EDUCATORS  -------------------------------------------------------------------------------------

\subsection{Results for \RQ{4}: What are the Important Aspects that Should be Considered when Teaching \ac{ML} from a Perspective of \ac{SE}?} 

We report the results of this research question based on our insights on \Var{7} (\textsf{Educators' Perspective}) from the \textit{interviews}. The most common teaching strategies combine \textit{experiential learning} (64\%) with learning from \textit{textual sources} (50\%) such as books, blogs, and articles, showing a trend of students gaining hands-on experience while also building theoretical foundation (I8: ``\textit{You need to get your hands dirty}''). Other approaches such as \textit{classical} and \textit{non-classical courses} (incl. online courses), and \textit{peer learning} are also mentioned, though less frequently (below 30\%).

Regarding content, \textit{quality attributes} and \textit{possible issues with ML} are considered important by 28\% of interviewees (\eg I6: ``\textit{I actually take great care of letting students understand what are the common issues when designing a machine learning pipeline}''). A notable finding is that 21\% emphasize that ML should be human-centered, incorporating the socio-technical impact of the technology, the relationship between model performance and user trust, and the use of human evaluation of model outputs, aspects that go beyond technical correctness.

%------------------ C.p.V: TRENDS  -----------------------------------------------------------------------------------------

\subsection{Results for \RQ{5}: What are recent trends when applying ML for \ac{SE} research?}

When comparing the data from more recent conferences and journals (2023--2025) to the earlier data (2011--2022), the most important finding is that most practices remain stable. From the perspective of the types of practices used, how \ac{ML} is processed and evaluated for SE research is largely unchanged.

The most notable shift is towards deep learning models and \acp{LLM}. Please note that we explicitly distinguish between deep learning models (incl. those using the transformer architecture), and \acp{LLM}, which we define as fully trained (and usually instruction tuned models) that are used by prompting, without further training.\footnote{We acknowledge that further training, \eg own alignment could also happen for \acp{LLM}. However, since we have not observed such practices, we do not make this distinction in this work.} This shift is visible across all three variables: \Var{{1.1}} \textsf{Input}, \Var{{1.2}} \textsf{Technique}, and \Var{{1.3}} \textsf{Purpose}. For example, 32.5\% of the conference papers from 2023-2025 report techniques that \textit{use \acp{LLM}}, while the use of the use of statistical learning models dropped to 17.5\%. Similarly, the use of \ac{ML} for classification declined (down to 20\% at conferences) while generation tasks increased (up to 27.5\%). New practices around prompt engineering and using prompts also emerged, appearing in $\approx$10\% of papers, roughly one third of those reporting \acp{LLM} use. The shift is further reflected in \Var{2}, where model training appears in only 80.0\% of recent papers compared to 93.6\% in earlier work. A similar but weaker trend is visible in journal articles, which is expected given their longer review and revision cycles.

A further difference between venues concerns the soundness of results. Journals show higher rates of statistical methods (28.9\% vs. 10.0\% at conferences) and manual inspection of results (13.3\% vs. 2.5\%). Researchers also make greater use of visualizations in journals (15.6\% vs. 2.5\%), likely reflecting the additional space they afford.

Finally, we observe a trend towards greater diversification in \Var{3} \textsf{SE Tasks}, maintenance tasks have become as common as implementation and testing. For the other \ac{SE} tasks, we observe no differences.

%------------------ C.p.V: Discussion  -------------------------------------------------------------------------------------
\section{Discussion}
\label{sec:discussion}

We now discuss our results in a broader context. First, we interpret the differences we found between our subjects, \ie \SEpapers \textit{research articles}, \SEresearchers \textit{interviews}, and the \SEauthors \textit{surveys}. Then, we compare what we found with the state of the art regarding best practices, challenges, reviewing, and education. 

\subsection{Comparison Between ML Practices Identified in Research Articles, Interviews, and Surveys}

Our results show that both analyzing \textit{research articles} and interviewing researchers can yield a large amount of best practices on how to use machine learning in software engineering. Our results are complementary rather than redundant: \textit{research articles} provide granular, reproducible accounts of specific pipeline stages, such as model training (\eg model type used), model evaluation (\eg evaluation performance), feature engineering and data handling (\eg data splitting strategies). While \textit{interviews} reflect more abstract concerns that may motivate those practices, including that rarely appear in \textit{research articles}, in stages such as model requirements (\eg problem understanding) and model monitoring. In the less structured survey, we found similar information as within the interviews, though the level of abstraction is again higher. We attribute this to the fact that participants in interviews are more committed in terms of time they invest in comparison to participants in online surveys.

A key finding from this comparison is the gap between what interviewees consider important and what \textit{research articles} actually report. Practices related to achieving high-quality results, such as hyperparameter tuning and exploratory data analysis, and practices that relate to how results should be assessed, such as human involvement in evaluation, and handling of non-functional properties, appear consistently in \textit{interviews} but are underrepresented in the \textit{research articles}. Notably, these same aspects were mentioned by the interviewees not only when discussing \ac{ML} practices, but also when discussing them from the perspective of the reviewer. We believe that this shows that the work at the intersection of \ac{ML} is still maturing, \ie \textit{our community (SE community) is aware of some practices that should be used and aspects that should be considered, but they are not yet being used consistently}.  %\newline maybe too strong -_> removed This convergence across perspectives suggest that the SE community is aware of what constitutes rigorous ML practice - but has not yet converged on applying it. 

\subsection{Comparison to the State of the Art}

%------------------ Discussion: challenges  -------------------------------------------------------------------------------------
% Arpteg

\subsubsection{\RQ{1} Practices} 

Prior work on \ac{ML} practices varies considerably in the granularity of the practices described, from short actionable rules \citep{serban2020adoption, serban2021practices} to lengthy\footnote{the practice is immersed in a paragraph that presents a guide or a recommendation} pipeline guides \citep{amershi2019software, ArpQuiPen_22}, or analyses of specific dimensions such as inputs or techniques \citep[\eg][]{watson2022systematic}. In general, studies that do consider multiple dimensions tend to classify practices into broad groups \citep{serban2020adoption, ArpQuiPen_22, Arpteg}. \textit{These differences make a direct one-to-one comparison difficult; nevertheless, a general comparison is possible and reveals both overlaps and gaps with prior work}, as we discuss below.

For practice, only two previous studies specifically consider the input dimension~\citep{watson2022systematic, serban2020adoption}. Our finding that code data is the most common input is consistent with \cite{watson2022systematic}, and aligns with \cite{serban2020adoption}, assuming their textual data category includes code. However, our analysis extends prior work by identifying inputs not previously considered, such as results and models, specific data quality aspects (\eg (un)verified labels), and a wider range of domain-specific data types (\eg images, screenshots), showing the the input dimension is more diverse then previously considered in the SE literature.

Beyond the dimension-specific comparisons, our most frequent findings on both subjects are well covered by prior work in general.  For example, the use of metrics for classification tasks aligns with \cite{watson2022systematic}, who find a similar proportion in their sample of research articles (74\%). The use of baselines is also studied in prior work, either in a general way \citep[\eg][]{MichaelLones2021, ArpQuiPen_22, zinkevich_2021} or distinguishing between types based on their origin \cite{watson2022systematic}. 

Interestingly, the practices frequent in \textit{interviews} but underrepresented in \textit{research articles}, were also discussed in prior work as recommended practices, and in some cases also identified as pitfalls if not followed. This is the case for hyperparameter tuning \citep[\eg practices and pitfalls:][]{MichaelLones2021, serban2020adoption, watson2022systematic, ArpQuiPen_22, amershi2019software, biderman2020pitfalls}, human expertise in evaluation \citep[\eg  practices:][]{MichaelLones2021, amershi2019software}, understanding the problem \citep[\eg practices:][]{biderman2020pitfalls, MichaelLones2021, ArpQuiPen_22, zinkevich_2021},  manual label validation \citep[\eg practices:][]{serban2020adoption, Clemmedsson2018, ArpQuiPen_22, serban2021practices}, and exploratory data analysis \citep[\eg practices:][]{watson2022systematic, MichaelLones2021, biderman2020pitfalls}. In each case, prior work highlights different aspects, such as the importance of human involvement and expertise in evaluation, and the rigor of the processes involved in training a model. For lower frequency practices, the literature also partially covers aspects like preprocessing techniques (\eg tokenization, embeddings \citep{watson2022systematic}) and data quality \citep{MichaelLones2021, serban2020adoption}, though our results provide more detail.

Finally, there are blind spots on both sides (\ie our study and prior work). Our work does not cover practices related to deployment \citep{serban2020adoption, ArpQuiPen_22, breck2017ml}, teamwork \citep{serban2020adoption, amershi2019software}, or how to conduct research \citep{MichaelLones2021}, mainly because these topics are not usually discussed in \textit{research articles} nor were they central to our \textit{interview} script. Conversely, our study presents a higher level of detail for some practices that are also present in previous work, for example for data manipulation aspects~\citep{watson2022systematic, MichaelLones2021}, we have it specified in practices such as collecting own data, using existing datasets, cleaning data, filtering data, and feature selection. \textit{This higher level of detail allows us also to distinguish between practices for statistical \ac{ML} and deep learning, a distinction that, to our knowledge, no prior work has made directly}.

%------------------ Discussion: challenges  -------------------------------------------------------------------------------------
\subsubsection{\RQ{2} Challenges} 
\label{sec:rq2-discussion}

Our findings on challenges largely align with prior work, with data-related concerns being prominent in both, encompassing data collection and quantity \citep{biderman2020pitfalls, amershi2019software}, leakage and snooping \citep{ArpQuiPen_22, MichaelLones2021}, and data manipulation \citep{Islam_2019, Alshangiti_2019, tantithamthavorn2018experience, Clemmedsson2018}.  However, challenges specific to operational environments, such as concerns related to data versioning \citep{LewisGrace2021WAIN} or the inclusion of new data in production \citep{kreuzberger2023machine}, are covered by studies focused on ML in industry but not captured by our study. Most other identified challenges are also covered in a similar manner in the literature, though not as broadly as the data concerns, including pitfalls in model interpretation \citep{tantithamthavorn2018experience}, reproducibility and model deployment \citep{tantithamthavorn2018experience, Alshangiti_2019}, model selection \citep{MichaelLones2021, ArpQuiPen_22, tantithamthavorn2018experience, Islam_2019}, lack of technology or knowledge \citep{Islam_2019, amershi2019software, kreuzberger2023machine, Clemmedsson2018}, security \citep{ArpQuiPen_22}, fairness \citep{amershi2019software, naher2022collab}, privacy \citep{Arpteg} and problem intrinsic challenges \citep{ArpQuiPen_22, MichaelLones2021, biderman2020pitfalls}. \textit{The broad coverage across the identified categories indicates that our interviews were representative and that our experts were knowledgeable about the challenges of applying \ac{ML}}.

Moreover, our experts went beyond prior work in three aspects. First, for ground truth and evaluation, while previous studies focus on how to execute evaluation \citep{tantithamthavorn2018experience}, comparison with baselines \citep{ArpQuiPen_22, biderman2020pitfalls}, and metric selection \citep{biderman2020pitfalls}, our interviews additionally identified the difficulty of building a reliable ground truth and the complexity of testing ML models due to their approximate nature. Second, for resource-related challenges, prior work takes a more operational perspective on resource availability across the ML life-cycle \citep{LewisGrace2021WAIN, kreuzberger2023machine, Arpteg}, whereas our experts also highlight challenges specific to an academic research environment, such as access to computational resources and keeping up with the pace of technological innovation. Third, our \textit{interviews} encompasses a human-centered perspective that is neglected by prior studies, particularly the challenges of finding representative participants, dealing with the subjectivity of human evaluations, and managing the associated recruitment costs.} %  this is the merge of the three middle paragraphs the quotations found in those paragraphs were removed

However, we note that our focus on SE researchers introduces some blind spots. Challenges specific to the operation of ML systems in industry~\citep[\eg][]{kreuzberger2023machine, LewisGrace2021WAIN, tantithamthavorn2018experience, Arpteg, serban2020adoption} are not covered by our study. Similarly, challenges arising from differences between communities~\citep{kotti2023machine}, \eg SE and \ac{ML} community, are also outside of our scope.

%------------------ Discussion: reviewers  -------------------------------------------------------------------------------------

\subsubsection{\RQ{3} Reviewer's perspective} 

Our results overlap primarily with common recommendations from the literature, covering aspects such as process and methodology, data sources and manipulation, data integrity, model selection, metrics, and study reproducibility \citep[\eg][]{ralph2021empirical, kapoor2024reforms}. Interestingly, the literature sometimes directly recommends practices alongside recommendations~\ie reviewing criteria \citep{kapoor2024reforms}, suggesting that the boundary between best practices and review guidelines is often blurred and hard to distinguished.

Analyzing by specific categories, we identified that our experts go further in two areas (\ie metrics and evaluation). For metrics, while prior work emphasizes justifying the suitability of a metric for the problem \citep{kapoor2024reforms, ralph2021empirical}, our experts additionally highlight the importance of using metrics recognized by the community and considering multiple quality attributes. For the broader category, evaluation, our experts stress the value of qualitative and quantitative analysis of results, the usage of use cases to evaluate the model, and the explicit consideration of fairness, aspects not yet covered by existing guidelines \citep{kapoor2024reforms, ralph2021empirical}. We also note blind spots in our \textit{interviews}, regarding data selection and data description \citep{ralph2021empirical, kapoor2024reforms}, topics that were not commonly raised by the interviewees.

Finally, our experts raised general observations that may be relevant for the \ac{SE} community to consider, though not directly actionable. These include uneven coverage of research areas (\eg licensing issues related to models or reverse engineering), lack of guidelines (\eg for doing prompting engineering), and concerns about insufficient \ac{ML} expertise among reviewers in niche areas. The last point connects, at least partially, to a recent editorial framing which intersection aspects between ML and SE research are appropriate for SE venues \citep{Uchitel2024-editorial}.

%\textcolor{orange}{the missing aspects are the ones noted before, specific details inside each category}

%------------------ Discussion: Educators  -------------------------------------------------------------------------------------

\subsubsection{\RQ{4} Educators' perspectives} 

Our findings on the educators' perspective align with prior studies, but also complement them at the content and method level. Regarding teaching methods, experiential learning is the dominant approach in both our \textit{interviews} and prior work \citep{mashkoor2023teaching, lanubile2023teaching, Huang_2018, kastner2020teaching, shouman2022experiences, acuna2023developing, acquaviva2022teaching, Chinmay_2021}, with an emphasis on hands-on activities such as developing projects, confirming consensus with our study. Using real-life examples, while frequently mentioned in the literature \citep{Huang_2018, mashkoor2023teaching, lanubile2023teaching, kastner2020teaching, Chinmay_2021}, was less specifically highlighted by our experts, possibly because it was discussed in the context of challenges rather than as a separate teaching consideration (see Section~\ref{sec:rq2-discussion}). Furthermore, peer learning, common in the literature \citep{kastner2020teaching, lanubile2023teaching, mashkoor2023teaching, shouman2022experiences}, was mentioned by only two of our interviewees. Conversely, other teaching methods such as classical courses, non-classical courses, and text-based learning resources were frequently mentioned in our interviews but largely absent from the literature. This may be because the literature focuses on novel teaching concepts that can replace or supplement text-based courses.

The most notable divergence concerns content emphasis. While the literature focuses on state-of-the-art tools and data management \citep{kastner2020teaching, lanubile2023teaching, mashkoor2023teaching, van2008teaching, shouman2022experiences}, our experts place greater weight on quality attributes, common pitfalls, and the human-centered dimensions of ML. In contrast, we did not cover granular aspects such as providing feedback to students, or  encouraging interest in \ac{ML} tasks, topics that were addressed by prior work \citep{kastner2020teaching, Huang_2018, mashkoor2023teaching}. This suggests that, the literature should possibly explore more aspects of ML education. Furthermore, ML education from an SE perspective is still evolving, so educators may not be aware of all the new concepts and ideas that are being developed. This is in contrast to other teaching aspects that are more stable (\eg software processes).

\subsection{\RQ{5} Recent trends}

Our results for recent changes are -- while perhaps unspectacular -- very important: while observe the expected rise in \ac{LLM} use and techniques related to this (\eg prompt engineering), it seems like this only affects one part the pipeline, \ie the model training (and to some degree the engineering of textual and code features) is now sometimes replaced with \ac{LLM}-based approaches. Our data indicates that other parts of the \ac{ML} pipeline do not seem to be affected, with respect to the practices: model requirements, data collection, data cleaning, data labeling, model evaluation, model deployment and model monitoring remain mostly the same so far.

\subsection{Implications}

The implications can be divided into those relating to the practices themselves and their evolution and those relating to the research that studies these practices. For the former, the key implications can be summarized as follows:
\begin{description}
    \item[\textbf{IM1}:] There has been a recent shift towards \acp{LLM} as models, but we have not observed changes to practices for the other stages of the machine learning pipeline.
    \item[\textbf{IM2:}] Some best practices are often mentioned, but less often followed, indicating a gap between what is said to be done in research and how research is actually conducted.
    \item[\textbf{IM3:}] Dealing with data (incl. data collection, ground truth data, data cleaning, ensuring data quality, and preventing information leaks) is most commonly seen as the main challenge for \ac{ML} in SE. However,  thorough evaluations that consider not only functional aspects (\eg accuracy), but also non-functional aspects and human involvement in the evaluation are also seen as difficult. 
    \item [\textbf{IM4}:] Hands-on teaching with experimental learning is the most common approach, though our interviews also reveal that many people are still using traditional teaching methods such as courses and text resources. 
\end{description}
Our multi-source study, which  looked at both \textit{research articles} and \textit{interviews},  also yielded interesting insights into different methods employed when studying practices: 
\begin{description}
    \item[\textbf{IM5}:] The data source matters when studying practices. Fine-grained details are best studied using detailed artifacts, such as \textit{research articles}. Broad perspectives are best studied by interacting with stakeholders, \eg through interviews.
    \item[\textbf{IM6}:] Multi-source studies in practices can effectively reveal differences between the desired state-of-the-art and the actual state of practice.
\end{description}

\subsection{Threats to Validity}

We report threats to the validity of our work following the classification of \cite{Cook1979} suggested for software engineering by \cite{Wohlin2012}. In addition, we discuss the reliability as suggested by \cite{Runeson2009}. 

\subsubsection{Construct Validity}

The first major threat to the construct of our work is the assumption that research articles, interviews, and surveys are suitable sources for eliciting information about \ac{ML} best practices from SE researchers. For research articles, we believe that such a threat is negligible, as research articles have to declare what is done in their methods sections. For interviews and surveys, this depends on the subject selection. In our interviews, we target known experts through a purposive sample, \ie researchers who frequently participate in program committees of high-level conferences, who are (partially) on editorial boards of highly-ranked journals, and who published themselves at these venues on topics in the intersection of ML and SE. Consequently, we believe our assumption that these are good sources of information on best practices is reasonable. For the surveys, subjects were collected from research articles in influential venues, \ie from researchers who managed to apply practices in such a manner that their work passed the rigorous peer-review processes. Again, we believe that this is a suitable indicator that these are suitable sources to mitigate this threat.

The second major threat to our construct is how we collect data from these sources. We structured data collection through well-defined variables to guide the data collection process to mitigate threats with respect to the information that is collected. For the collection of data from the research articles, all taggers were provided with spreadsheets that contained detailed guidance regarding each variable, \ie how it should be collected and coded. To further reduce the risk of missing information, all research articles were considered by two authors. For the interviews, we followed a detailed script that ensured complete coverage of both the \ac{ML} pipeline and our variables. The construct of the survey through a single question has the threat that it is too simple to elicit detailed knowledge about \ac{ML} practices. This threat also manifested in our data and led to a limited usefulness of the survey data. However, any additional structuring that would still allow for a quick completion of the survey (\eg check-boxes for pre-defined practices) would bias the results. Additional structuring of the survey, \eg through multiple free-text fields, one for each pipeline stage, would have biased the results as some participants may have added practices that they never used to not leave fields empty and also likely have reduced the rate of responses, due to the additional effort. Finally, we decided to give none of the human subjects a definition for what constitutes a (best) practice. This introduces the risk that aspects unrelated to practices are described. However, based on our data, this did not happen. 

%Our construct underwent rigorous peer review prior to the execution of the study through pre-registration, further mitigating the above risk through the early involvement of external expertise.

\subsubsection{Internal Validity}
A possible concern with the internal validity of our results are misunderstandings during the interviews, \eg due to misleading questions by the interviewer introducing bias or a wrong understanding of a question by a participant. We mitigate the first threat by using a question catalog for the interviews. While we cannot rule out misunderstanding on the side of the participants, we did not pick up on inconsistencies (\eg caused by an inappropriate change of topics) when reading the transcripts of the interviews. We do not see other major concerns about the internal validity of our work. The results we report are derived directly from the data. The conclusion that there is a gap between what is said in the interviews and what we observe in the data is directly supported by this.

A separate concern with the internal validity is due to the large number of codes and categories, which may be due to an inappropriate level of abstraction. While we cannot conclusively rule this out, our data provides several indications that this is not the case. Notably, the number of codes for \Var{2}\textendash\Var{7} have only 21 or less categories after axial coding. These variables encode already fairly abstract concepts, like the pipeline stages, SE tasks, our \ac{ML} challenges, which makes this lower number of categories reasonable. The variables \Var{{1.1}}, \Var{{1.2}}, and \Var{{1.3}} rather describe the input, technique, and purpose/output associated with practices. Considering that we considered the whole machine learning pipeline and 195 papers that were published over the course of 14 years, with additional data from fourteen interviews and a short survey of 47 authors asking about important practices, we do not believe that the numbers are unreasonably large. For example, having 176 unique categories for the \Var{{1.2}} (\textsf{Technique}) is not unreasonable, as there are already ten \ac{ML} pipeline stages, which all require different techniques. Consequently, we do not believe that the large number indicates inappropriate interpretations, but rather that this highlights the diversity of practices used by the research community. We acknowledge that this diversity is a problem to concisely report and discuss the results in their entirety. We focused on the most common practices, differences between data sources, and aspects that seem to be corner cases and reported the full results in the Appendix to allow readers to dive deeper into our data.

\subsubsection{External Validity}

While we want to identify the different perspectives about \ac{ML}, including challenges and practices when applying \ac{ML}, our selection of subjects may introduce bias due to the studied time frame for the \ac{SE} articles and the authors of these \ac{SE} research articles, the focus on few, highly-ranked SE venues, and the purposive sample for \ac{SE} influential researchers. Our approach may exclude relevant aspects that are not considered in the studies analyzed during the established time frame, the perspective of \ac{SE} researchers that were not first or last authors in the studied research articles or with different level of experience than the ones selected in the purposive sample, we may also miss practices that are not present in English literature. Therefore, we cannot claim that we identified all practices nor that results regarding the prevalence generalize to all contexts. We mitigate this threat by considering a large scope, \ie 110 research articles published between 2011-2022 at top SE conferences, incl. the authors of these research articles and the deep expertise of the interview participants. Additional analysis of 90 publications from 2023-2025 at top SE conferences and journals further broadens this scope with a current perspective. We further note that our comparison with related work showed large overlaps, supporting the expertise of our subjects. 

We cannot rule out that the interviews of our subjects were sufficient to fully cover important aspects from the community. On the hand, our sample size may not be sufficient for saturation, leading to missing aspects. The strong agreement on core concerns indicated by topics often being mentioned in over half of the interviews, indicates that at least these are stable. On the other hand, our focus on SE researchers that use \ac{ML} may mean we miss relevant parts of the SE community, \eg the perspective of non-ML experts on reviewing \ac{ML} papers or of teaching faculty for education.

\subsubsection{Reliability}

Our study heavily relied on human expertise, which threatens the reliability of our work, \ie it is possible that the involvement of different taggers would lead to different results. To mitigate these threats, we provided documentation to everybody involved regarding what exactly to do. Additionally, we build in quality checks through the data collection and analysis pipeline to reduce the reliance on the expertise of a single person: i) all research articles that were candidates for inclusion were checked by two researchers and disagreements were discussed; ii) all research articles were tagged by two independent researchers; iii) the merging of the data from the two researchers was conducted by a third researcher, who also double checked the correctness in case of diverging codes (missing on one side or contradictions); iv) all harmonization of codes and all axial categorizations where checked by at least two authors; and v) all data for the interviews was double checked by an author, that was neither involved in the interviews, nor in the coding of the interviews. Additionally, the protocol we followed to ensure this reliability underwent rigorous peer review prior to the execution of the study through pre-registration, with the goal of identifying possible weaknesses in our procedure early. 

Furthermore, the background of our interview subjects might also affect the reliability of our results. We cannot provide detailed demographics to avoid the possibility of deanonymization. Instead, we report broad data about their expertise here with respect to the relationship of our other subjects we studied, \ie the research articles. Our interview subjects were (co-)authors of 151 papers at FSE, ICSE, and ASE and 82 papers at TSE, TOSEM, and EMSE in the sampling frame relevant for the interviews, \ie 2011-2022. All participants are either senior academics in faculty positions; or senior PhD students that have multiple first-author publications in selective, high-quality venues (journals or conferences) and are involved in teaching at their host institutions both by supporting their advisors (creation of materials and/or course organization) and conducting exercises/TA sessions. We further analyzed our data and found that three of the SE articles we coded were authored by one of the interviewees, \ie less than 3\% of the data might be biased due to this, leading to a negligible risk.

\section{Conclusion}
\label{sec:conclusions}

Within this study, we increase our understanding of what practices are used when \ac{ML} is applied in an SE context. We found that SE researchers use a wide range of practices, covering almost all the best practices identified in the previous literature. The most common were practices directly related to the core of a machine learning pipeline, \ie training a model and the computation of accuracy metrics, as well as aspects related to collecting and splitting data. We also found a gap regarding the prevalence of some techniques that are deemed ubiquitously important in both the literature as well as in our expert interviews. Especially hyperparameter tuning is always mentioned as a best practice but was only used in 20\% of the research articles we considered. Similarly, human involvement in evaluation, manual data validation, or exploratory data analysis are typically mentioned as best practices, but not always used. While there is a recent trend towards using \acp{LLM} and new practices evolve to use these models, the impact of this is restricted to the model training, other practices for other parts of the machine learning pipeline are not affected, yet.

When it comes to challenges, SE researchers consider it especially challenging to deal with data (collection, quality, processing), but evaluation is also a commonly mentioned challenge, especially involving humans. We found that these considerations are very similar to reviewing ML/SE work, where our expert interviews aligned well with already existing guidelines. However, we identified several aspects that either require updates to existing guidelines or additional guidelines for this context, \eg regarding the evaluation of non-functional quality attributes. When SE researchers are involved in teaching activities that involve ML, they typically rely on experimental, hands-on work, which aligns well with the literature. However, traditional teaching methods, like courses and text sources are also mentioned as relevant methods. The focus of SE researchers seems to go beyond what is currently considered in the literature, as human involvement and non-functional properties are also seen as important during teaching. 

Our work opens many possible venues for future work. On the one hand, we demonstrated that studying a large sample of research articles augmented with expert interviews is a good method to collect a broad selection of practices, with the exception being practices regarding deployment, because this is typically not considered by researchers. Future work could investigate why deployments are not studied more often and ideally develop countermeasures to ensure that more research addresses this important activity. Similar considerations can also be made for techniques that are deemed important, but not often used. We need to better ensure that hyperparameters are tuned and determine why we do not involve human subjects more often in the evaluation of our work, even though this is deemed critical. Similarly, we should develop concrete practices that incorporate the assessment of non-functional quality attributes in the evaluation of \ac{ML} models, such that this becomes as common as measuring accuracy. Review guidelines should be updated to incorporate that and mitigate concerns regarding lack of expertise. Finally, we want to highlight the stunningly low presence of statistical methods in many research articles. They were also not often mentioned in interviews and are missing from many guidelines and sets of best practices. We need to investigate why this is the case and improve guidelines, the state of practice, and the sets of best practices, to ensure that our research is based on sound methods.

\section*{Declarations}

\noindent\textbf{Funding} This study was conducted solely based on the funding of the involved academic institutions without any additional third-party funding. 

\vspace{5px}

\noindent\textbf{Ethical Approval} An ethics approval was  obtained for the Ethics Committee of the University of Passau on 2024-05-10.

\vspace{5px}

\noindent\textbf{Informed Consent} Informed consent was obtained from all participants. The consent form was part of the ethics approval. 

\vspace{5px}

\noindent\textbf{Author Contributions} All authors were involved in the planning of the work. AMH organized the data collection process and conducted the sampling of publications. DNP and DP conducted the interviews. All authors were involved in the coding of the data. AMH conducted the qualitative analysis of the results. AMH drafted the manuscript. All authors discussed the results and contributed to the final manuscript. 
\vspace{5px}

\noindent\textbf{Data Availability} All data is available online: \url{https://github.com/aieng-lab/Replication-kit-Perspective-of-SE-Researchers-on-ML-Practices}. A long-term archive on Zenodo will be created for a final manuscript of this paper. 

\vspace{5px}

\noindent\textbf{Conflict of Interest}  The authors have no competing interests to declare that are relevant to the content of this article.
        
\vspace{5px}
        
\noindent\textbf{Clinical Trial Number} Not applicable.

%\begin{acknowledgements}
%If you'd like to thank anyone, place your comments here
%and remove the percent signs.
%\end{acknowledgements}

% Authors must disclose all relationships or interests that 
% could have direct or potential influence or impart bias on 
% the work: 
%
% \section*{Conflict of interest}
%
% The authors declare that they have no conflict of interest.

% BibTeX users please use one of
\bibliographystyle{spbasic}      % basic style, author-year citations
\bibliography{1_local_bib, 1_tools_bib}

%\bibliographystyle{spmpsci}      % mathematics and physical sciences
%\bibliographystyle{spphys}       % APS-like style for physics
%\bibliography{}   % name your BibTeX data base

\appendix

% !TEX root = main.tex

\section{Supplementary material for the Execution Plan with the SE Authors }

In this section of the Appendix, we provide the relevant information for the supplementary material used to survey the authors of the identified Software Engineering (\ac{SE}) research papers using Machine Learning (\ac{ML}). We sent emails to the authors asking them to answer the following question based on a subset of the identified papers in which they were authors, which was presented in the email.

\textbf{Question:}
\textit{What machine learning best practices have you used in your software engineering research papers, and how often did you apply those practices (Always, Frequent, Sometimes, Never)?}

\textbf{Instruction given bellow the question:}

\textit{Please write each practice in a new line:}

\textit{- Good practice [how often]}

%--------------------------------------------------------------------

\section{Supplementary material for the Execution Plan with SE Researchers}
In this section of the Appendix, we provide the relevant information for the supplementary material used to survey and interview the Software Engineering (\ac{SE}) researchers that have used Machine Learning (\ac{ML}).

%--------------------------------------------------------------------

\subsection{Demographic Survey Questions}
We collected and report demographic data about our interview participants. This data is not reported within this article, but can be accessed in our online appendix.
For simplicity, we present each question following the conventions below: multiple choice question with a single answer (\radiobutton), multiple choice question with multiple answers (\checkboxicon), open question with a text field for answering (\textfieldtext), questions with a numeric field with accepted answers between 0 and 100 (\textfieldnum). In addition, if clarifications or instructions were provided for a question, they will be presented after the question in italics between parentheses.

\begin{enumerate} [label=\roman*)]

   \item \qs{How many years of experience in software development do you have?} \textfieldnum

    \item \qs{How many papers have you published related to ML?} \qcl{ML: Machine Learning}\textfieldnum 
    
    \item \qs{How many papers have you published related to SE without involving ML?} \qcl{ML: Machine Learning, SE: Software Engineering}\textfieldnum
    
    \item \qs{How many papers have you published related to ML4SE?} \qcl{ML: Machine Learning, SE: Software Engineering, ML4SE: ML for SE}\textfieldnum
    
    \item \qs{How many papers have you published related to SE4ML?} \qcl{ML: Machine Learning, SE: Software Engineering, SE4ML: SE for ML}\textfieldnum
    
    \item \qs{What venues (journals/conferences) do you tend/prefer to publish in? } \qcl{Please, separate each venue with a comma~(~,~)}\textfieldtext
    
    \item \qs{How would you describe your primary role?} 
    \begin{tasks}[style=itemize](4)
        \task*(2)[\radiobutton] Researcher (academics)
        \task*(2)[\radiobutton] Researcher (industrial)
        \task[\radiobutton] Applied Scientist
        \task[\radiobutton] Tester
        \task[\radiobutton] Project Manager
        \task[\radiobutton] DevOps Engineer
        \task[\radiobutton] Programmer
        \task[\radiobutton] Project Lead
        \task[\radiobutton] IT Manager
        \task[\radiobutton] Other \emptyboxOthers
    \end{tasks}
    
    \item \qs{What is your highest level of education?} 
    \begin{tasks}[style=itemize](4)
        \task*(2)[\radiobutton] Did not graduate from high school
        \task[\radiobutton] Some college
        \task[\radiobutton] High school
        \task[\radiobutton] Bachelors' degree
        \task[\radiobutton] Master's degree
        \task[\radiobutton] Doctoral degree
        \task[\radiobutton] Other \emptyboxOthers
    \end{tasks}

    \item \qs{Which programming languages have you frequently used in past project(s)?} \qcl{If more than an additional language is added, please split them with ``,''} 
    \begin{tasks}[style=itemize, label=\checkboxicon](4)
        \task C/C++
        \task C\#
        \task Java
        \task Javascript
        \task Julia
        \task Python
        \task R
        \task Rust
        \task Dart
        \task Kotlin
        \task Other~\emptyboxOthers
    \end{tasks}
    
    \item \qs{Which types of Learning approaches have you used?}
    \begin{tasks}[style=itemize, label=\checkboxicon](3)
        \task Supervised
        \task Semi-supervised
        \task Reinforcement learning
        \task Unsupervised
        \task Other~\emptyboxOthers
    \end{tasks}
    
    \item \qs{What types of ML tasks have you used?}
    \begin{tasks}[style=itemize, label=\checkboxicon](3)
        \task Regression
        \task Clustering
        \task Machine translation
        \task Object detection
        \task Transcription
        \task Anomaly detection
        \task Other~\emptyboxOthers
    \end{tasks}
    
    \item \qs{What types of systems have you developed involving ML?}
    \begin{tasks}[style=itemize, label=\checkboxicon](3)
        \task Operating systems
        \task Server side
        \task Mobile applications
        \task Middleware
        \task Desktop applications
        \task Web applications
        \task Databases
        \task*(2) Recommender systems
        \task*(2) Development tools (compilers, prog.languages, etc.)
        \task Other~\emptyboxOthers
    \end{tasks}
    
    \item \qs{Have you primarily worked on open-source or closed-source projects?} 
    \begin{tasks}[style=itemize](4)
        \task[\radiobutton] Open-source only
        \task[\radiobutton] Closed-source only
        \task[\radiobutton] Open and closed-source
    \end{tasks}
    
    \item \qs{For which domains (\eg banking or healthcare) have you developed applications/systems?} \qcl{List them next separated by coma~(~,~)}\textfieldtext
    
    \item \qs{Do you have a background in machine learning?} \qcl{Formal (college classes, degree, certification) or informal (self-learning or other training)?} 
    \begin{tasks}[style=itemize](4)
        \task[\radiobutton] Yes, formal
        \task[\radiobutton] Yes, informal
        \task[\radiobutton] No
    \end{tasks}
    
    \item \qs{Do you have a background in software engineering research?} \qcl{Formal (college classes, degree, certification) or informal (self-learning or other training)?}
    \begin{tasks}[style=itemize](4)
        \task[\radiobutton] Yes, formal
        \task[\radiobutton] Yes, informal
        \task[\radiobutton] No
    \end{tasks}
    
    \item \qs{When researching software engineering, do you keep track of your approaches/ models + dataset?, Which versioning systems have you used (\eg CML, DVC, Git)?} 
    \begin{tasks}[style=itemize](2)
        \task[\radiobutton] Yes~\emptyboxOthers
        \task[\radiobutton] No
    \end{tasks}
        
    \item \qs{How often did you release or help release a new major version of ML models over the past two years? Please give your best estimate.} 
    \begin{tasks}[style=itemize](5)
        \task[\radiobutton] Never
        \task[\radiobutton] Annually
        \task[\radiobutton] Quarterly
        \task[\radiobutton] Monthly
        \task[\radiobutton] More frequently
    \end{tasks}
    
    \item \qs{Which country/countries are you and/or your organization based in?}\textfieldtext

\end{enumerate}

%--------------------------------------------------------------------
\subsection{Interview Questions}

\begin{enumerate} [label=\roman*)]
    \item Best Practices and Protocols
    \begin{itemize} 
        \item What do you consider as a best practice?
        \item Which process, guidelines, or pipelines you follow when employing ML in SE?
        \item What do you consider are the “must” practices or protocols you use or implement in your research/papers?
        \item Which are the most recurrent challenges when using ML for SE in any of the ML workflow stages?
        \item Which practices do you use from each ML pipeline stage?
        \item How did you employ the components of learning to guide your ML model design? [Abu-Mustafa, 2012]
        \item Do you take into consideration any of the learning principles (\eg data snooping, Occam’s razor, or sampling bias)? [Abu-Mustafa, 2012]
        \item Which quality attributes are important for SE systems that use ML? >> \textbf{ANSW1}
        \item Do you have any particular difficulties/challenges when building an ML-enabled system to ensure \textbf{ANSW1} attributes? 
    \end{itemize}
    
    \item Education
    \begin{itemize}
        \item What educational resources have you employed to perform ML4SE? (\eg tools, languages, books, etc..)
        \item How do you learn ML4SE?
        \item How do you educate your students to enable ML4SE?
        \item How do you promote \textbf{ANSW1} when teaching about ML4SE?
    \end{itemize}
    
    \item Reviewer’s Perspective
    \begin{itemize}
        \item What issues have you observed when reviewing papers?
        \item What ML4SE areas are not being covered in conference reviews?
        \item As a reviewer, have you seen that the \textbf{ANSW1} attributes are being addressed?
    \end{itemize}
    
    \item ML Workflow Stages (\textit{Amershi}~\etal)
    \begin{itemize}
        \item Model Requirements
        \begin{itemize}
            \item What was the rationale of selecting or proposing an ML model you have utilized in your papers? 
        \end{itemize}

        \item Data Collection
        \begin{itemize}
            \item Do you collect your own data, use previous datasets, or both? How do you select data/datasets?
            \item Which data collection strategies or protocols you employ in your studies?
            \item How do you promote \textbf{ANSW1} when collecting data for ML-enabled systems?
        \end{itemize}
        
        \item Data Cleaning
        \begin{itemize}
            \item Which pipelines do you employ to address data exploration?
            \item How do you perform data cleaning?
            \item How do you handle exploratory analysis in your studies? 
            \item How important are exploratory analyses for data cleaning?
            \item How do you promote \textbf{ANSW1} when cleaning data in ML-enabled systems? 
        \end{itemize}

        \item Data Labeling (if applicable)
        \begin{itemize}
            \item Which data labeling protocols do you use for your supervised tasks?
            \item How do you promote \textbf{ANSW1} when labeling data for ML-enabled systems? 
        \end{itemize}

        \item Feature Engineering
        \begin{itemize}
            \item Which feature engineering methods you have employed? 
            \item How do you promote \textbf{ANSW1} when selecting and building features in ML-enabled systems? 
        \end{itemize}

        \item Model Training
        \begin{itemize}
            \item What frameworks have you employed for model training?
            \item How do you parameterize your models?
            \item How do you promote \textbf{ANSW1} when training models in ML-enabled systems?
        \end{itemize}

        \item Model Evaluation
        \begin{itemize}
            \item How do you evaluate or validate your ML models?
            \item Do you employ any measurements to control for bias?
            \item Do you use any interpretability technique to guide your evaluation?
            \item How are measurements aligned to study RQs, goals, or business objectives?
            \item How do you promote \textbf{ANSW1} when evaluating ML-enabled systems?
        \end{itemize}

        \item Model Deployment
        \begin{itemize}
            \item How do you do promote \textbf{ANSW1} when deploying ML-enabled systems?
        \end{itemize}

        \item Model Monitoring
        \begin{itemize}
            \item Do you employ any measurements to handle concept drift?
            \item How do you monitor models during operation (\eg user studies, industry projects, production)?
            \item How do you promote \textbf{ANSW1} when monitoring ML-enabled systems? 
        \end{itemize}
        
    \end{itemize}
 
    \item Potential Follow-up Questions 

\end{enumerate}

%----------------------------------------------------------------------

\newpage
\section{Supplementary Material for the Results of the Open Coding, 2011-2022}

\subsection{Codes associated to the ML practices}
\FloatBarrier

\begin{figure}[h]
    \centering
    \includegraphics[width=\textwidth]{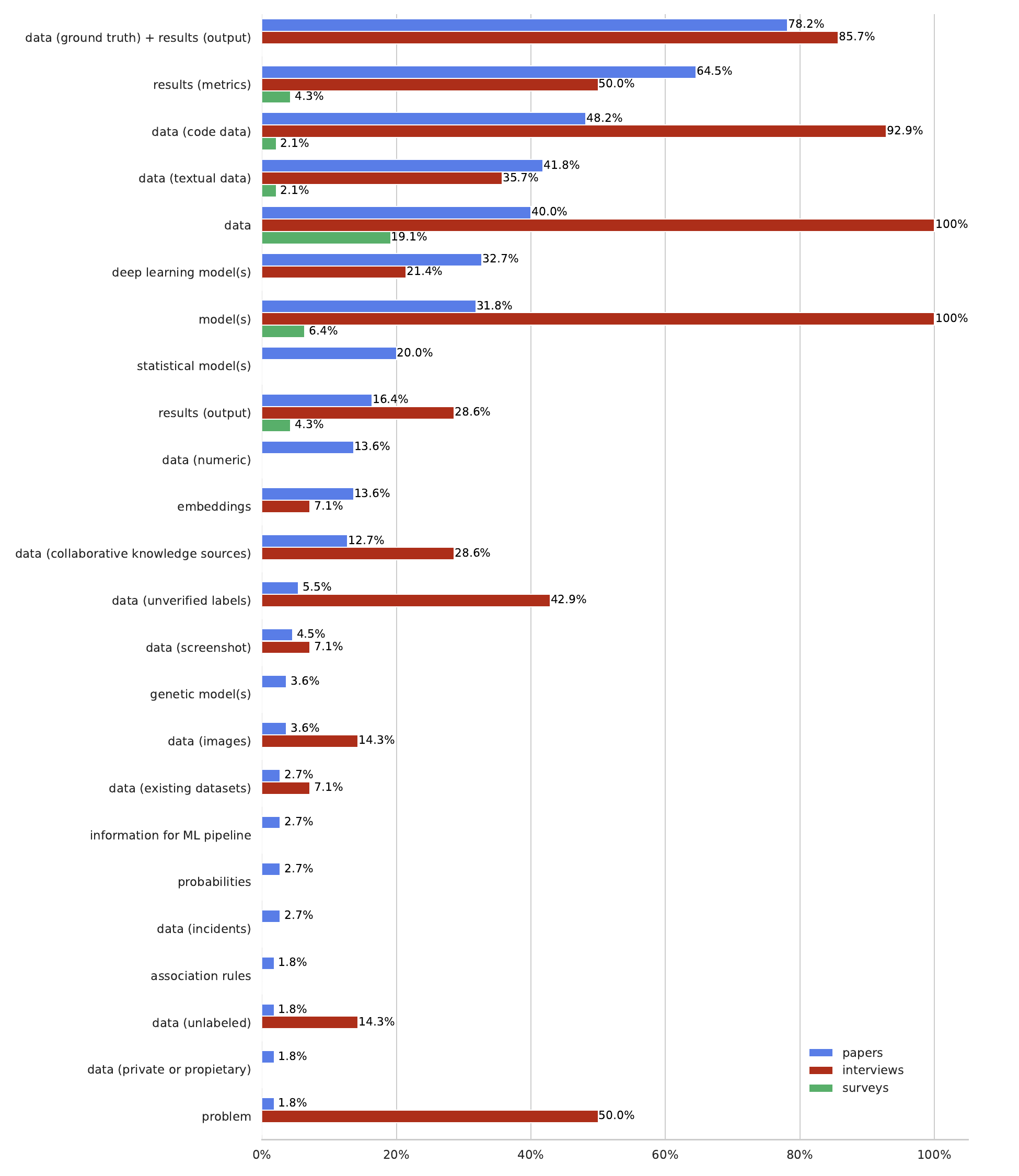}
    \caption{Percentage of subjects (\ie papers, interviewees, and surveys) in which each \textit{input category} appears. The figure displays the first half of the \textit{input codes}, organized by the subject \textit{papers} appearance.}
    \label{fig:input_axial1}
\end{figure}

\begin{figure}[h]
    \centering
    \includegraphics[width=\textwidth]{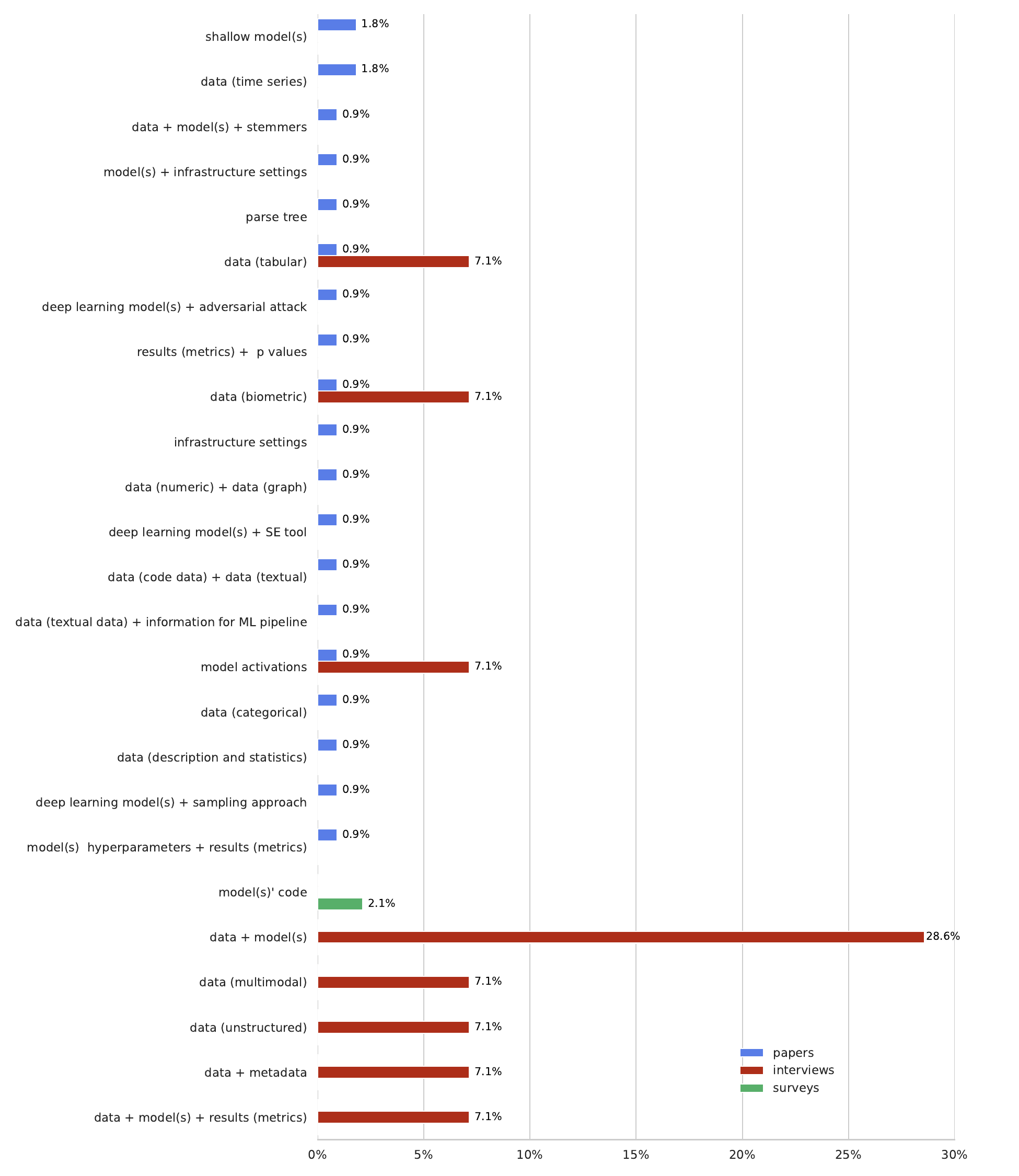}
    \caption{Percentage of subjects (\ie papers, interviewees, and surveys) in which each \textit{input category} appears. The figure displays the second half of the \textit{input codes}, organized by the subject \textit{papers} appearance.}
    \label{fig:input_axial2}
\end{figure}

% ------------------------------------------------------ Techniques --------------------------------------

\begin{figure}[h]
    \centering
    \includegraphics[width=\textwidth]{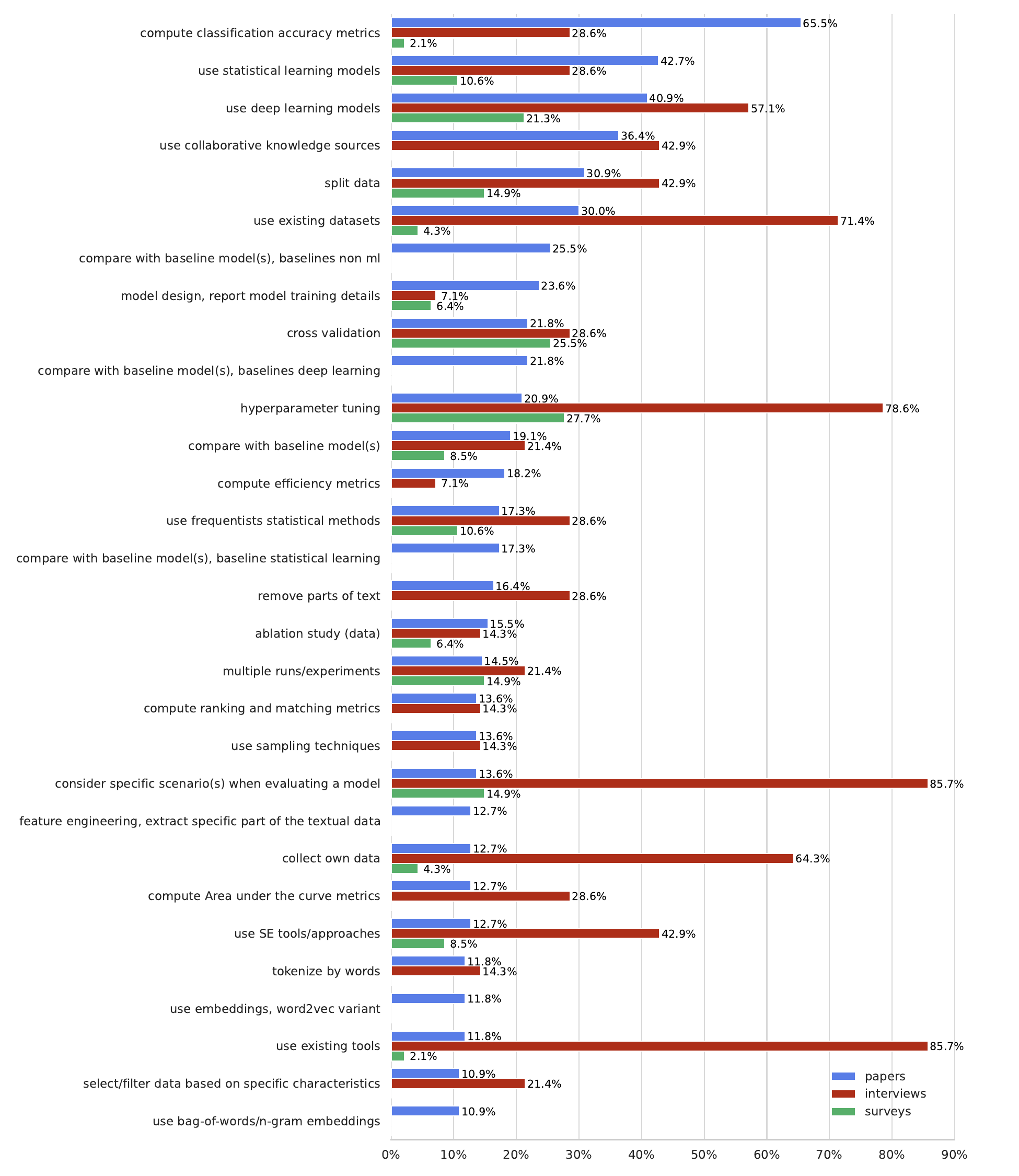}
    \caption{Percentage of subjects (\ie papers, interviewees, and surveys) in which each \textit{technique category} appears. The figure displays approximately the first sixth  of the \textit{technique codes}, organized by the subject \textit{papers} appearance.}
    \label{fig:technique_axial1}
\end{figure}

\begin{figure}[h]
    \centering
    \includegraphics[width=\textwidth]{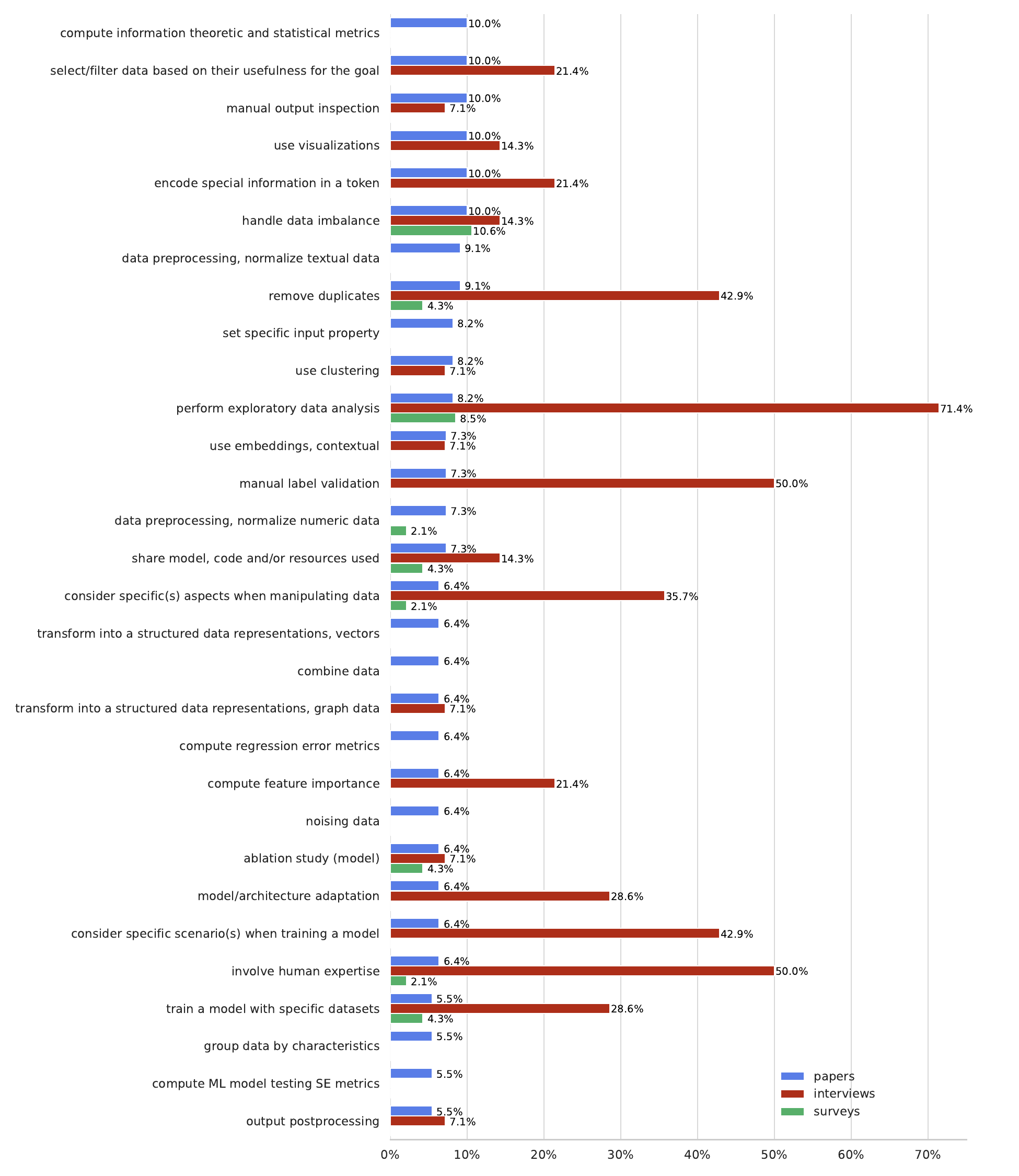}
    \caption{Percentage of subjects (\ie papers, interviewees, and surveys) in which each \textit{technique category} appears. The figure displays approximately the second sixth of the \textit{technique codes},  organized by the subject \textit{papers} appearance.}
    \label{fig:technique_axial2}
\end{figure}

\begin{figure}[h]
    \centering
    \includegraphics[width=\textwidth]{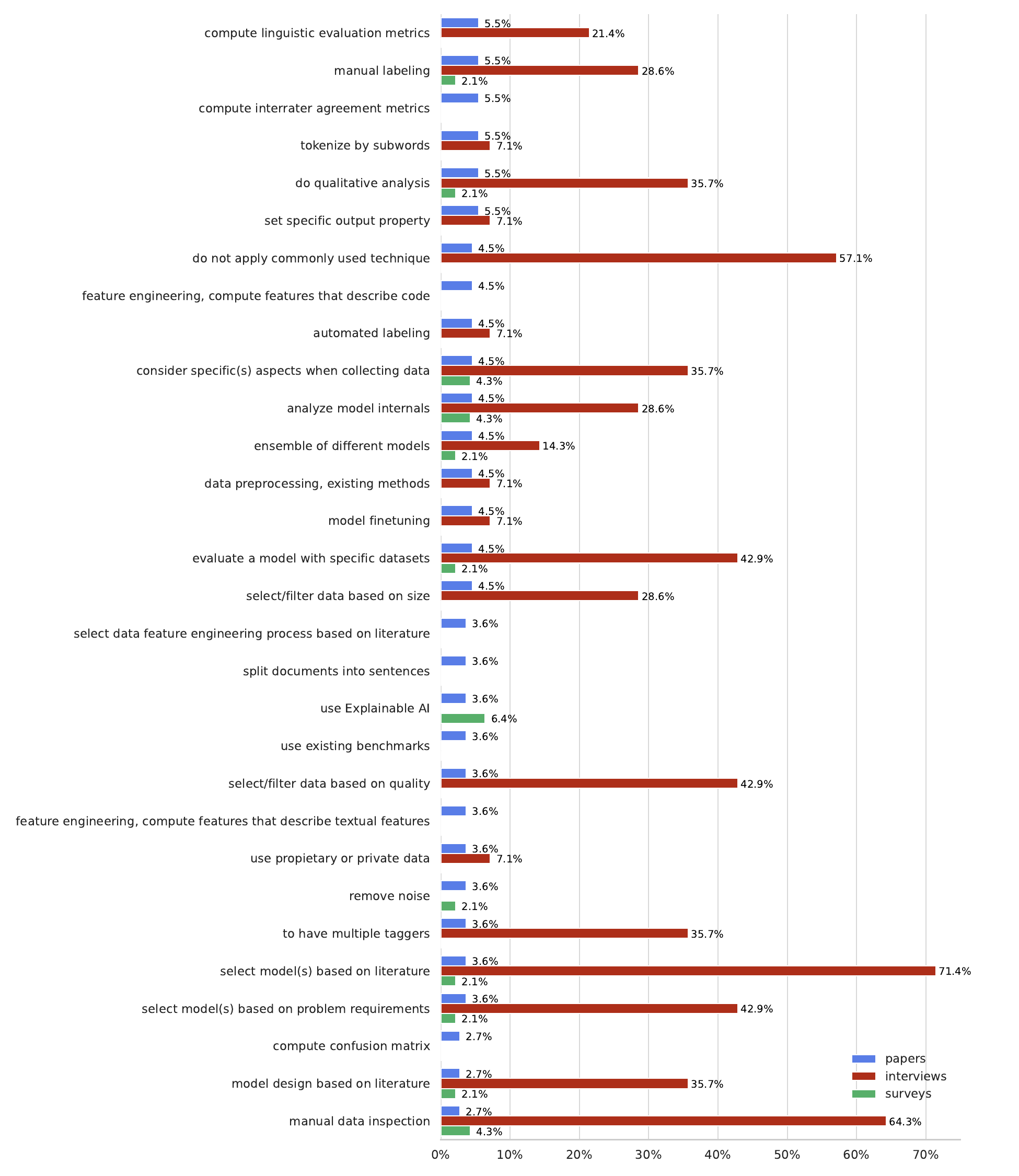}
    \caption{Percentage of subjects (\ie papers, interviewees, and surveys) in which each \textit{technique category} appears. The figure displays approximately the third sixth of the \textit{technique codes}, organized by the subject \textit{papers} appearance.}
    \label{fig:technique_axial3}
\end{figure}

\begin{figure}[h]
    \centering
    \includegraphics[width=\textwidth]{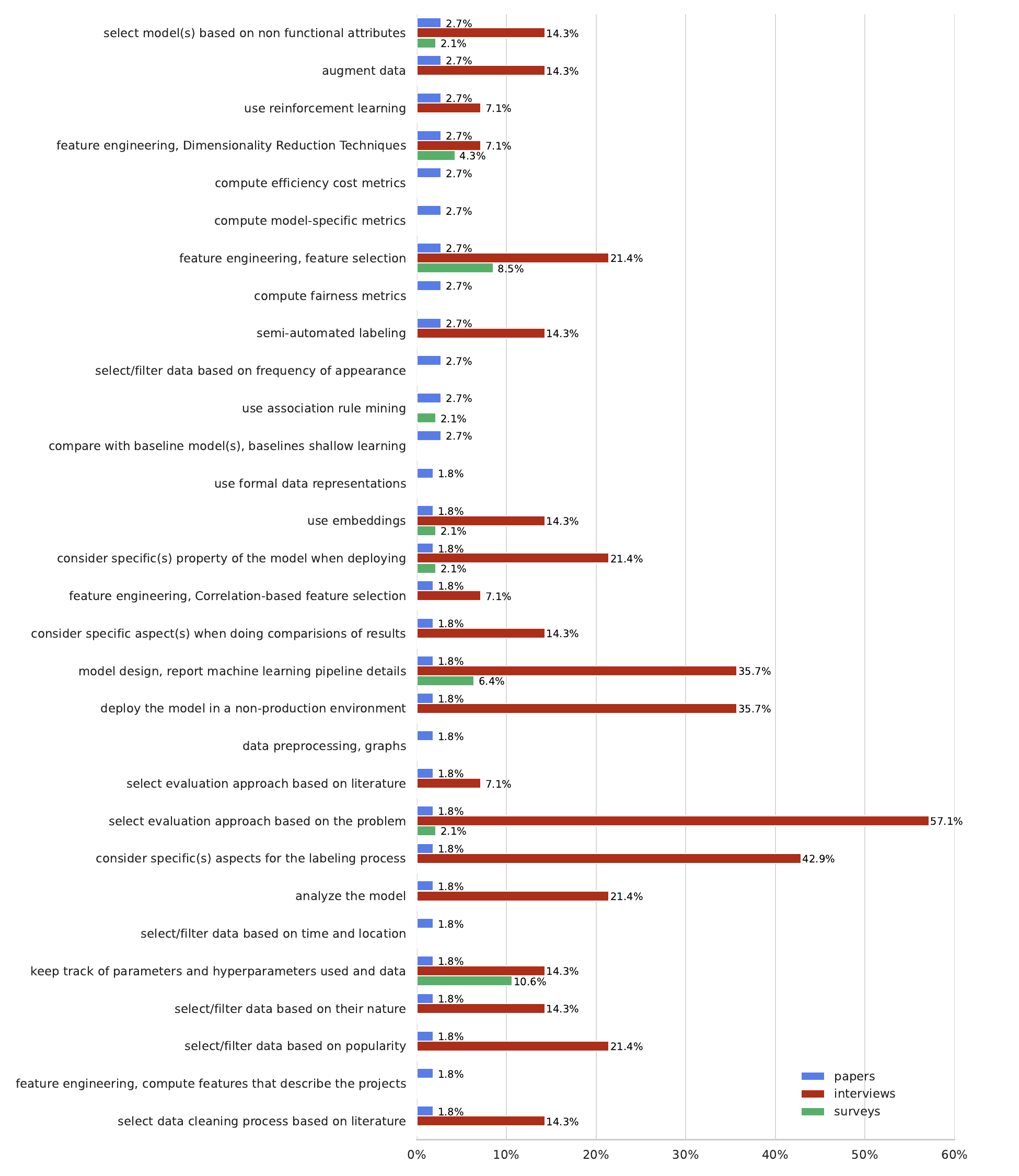}
    \caption{Percentage of subjects (\ie papers, interviewees, and surveys) in which each \textit{technique category} appears. The figure displays approximately the fourth sixth of the \textit{technique codes},  organized by the subject \textit{papers} appearance.}
    \label{fig:technique_axial4}
\end{figure}

\begin{figure}[h]
    \centering
    \includegraphics[width=\textwidth]{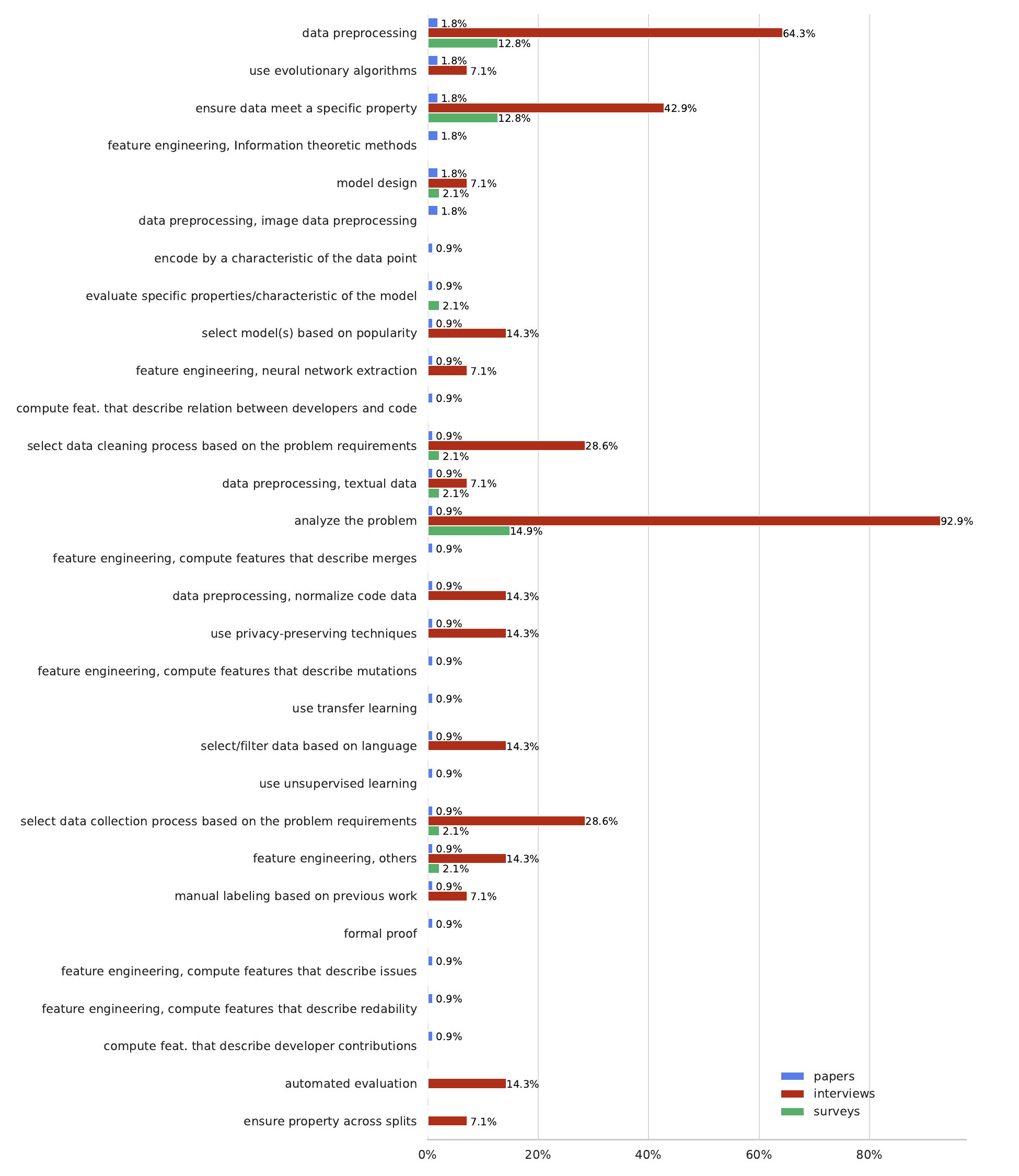}
    \caption{Percentage of subjects (\ie papers, interviewees, and surveys) in which each \textit{technique category} appears. The figure displays approximately the fifth sixth of the \textit{technique codes},  organized by the subject \textit{papers} appearance.}
    \label{fig:technique_axial5}
\end{figure}

\begin{figure}[h]
    \centering
    \includegraphics[width=\textwidth]{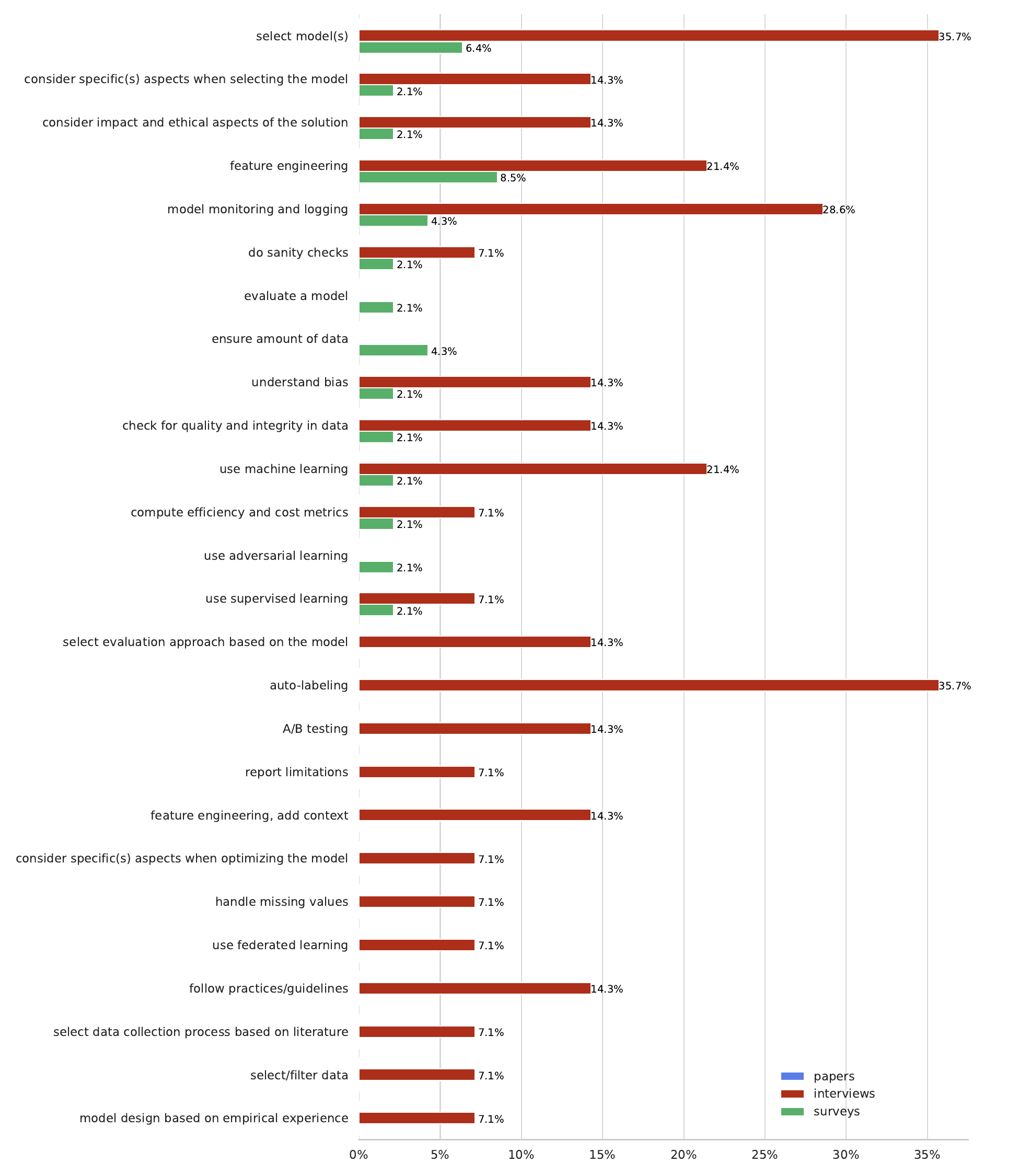}
    \caption{Percentage of subjects (\ie papers, interviewees, and surveys) in which each \textit{technique category} appears. The figure displays approximately the last sixth of the \textit{technique codes},  organized by the subject \textit{papers} appearance.}
    \label{fig:technique_axial6}
\end{figure}

% ------------------------------------------------------ Purpose --------------------------------------

\begin{figure}[h]
    \centering
    \includegraphics[width=\textwidth]{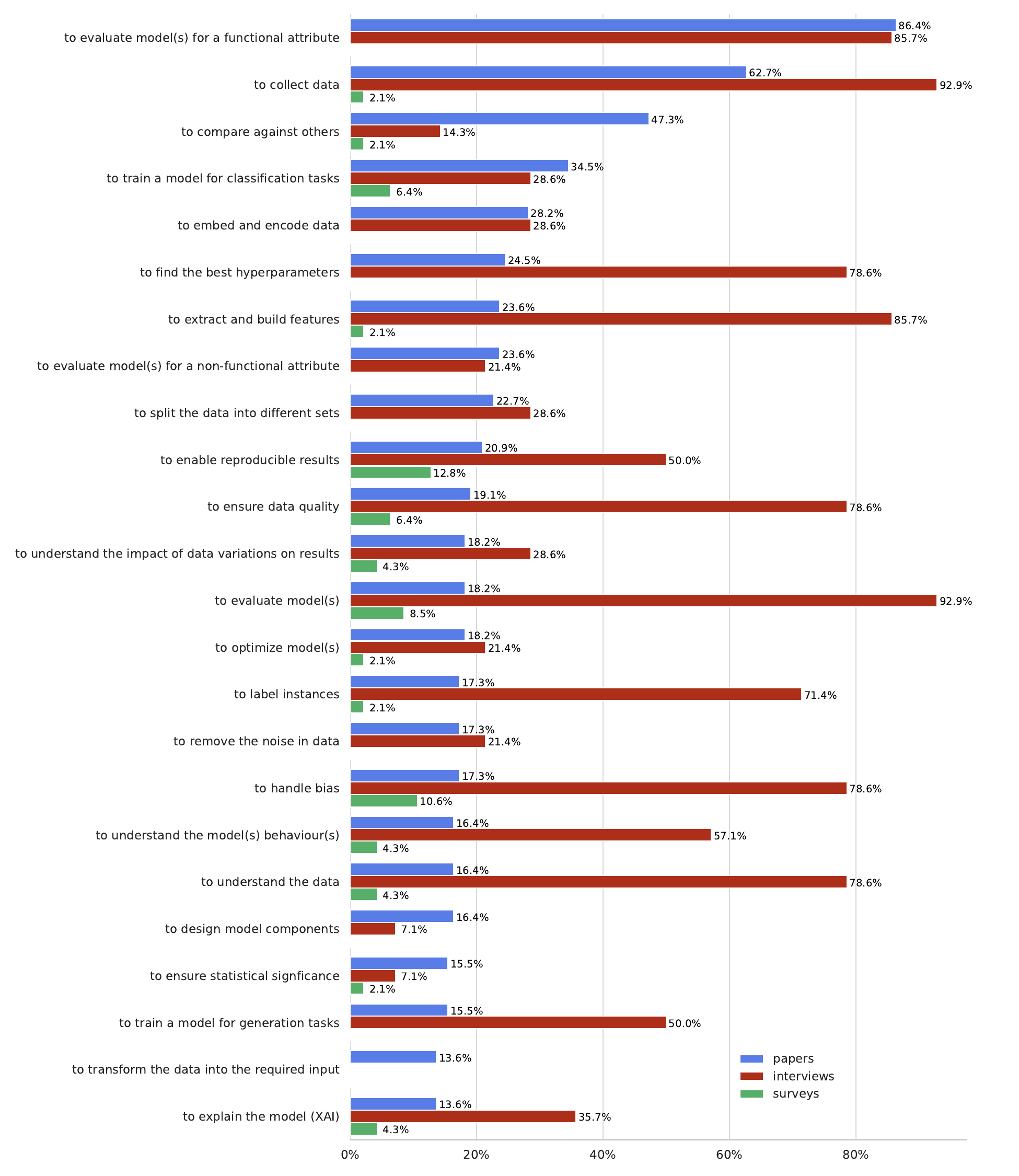}
    \caption{Percentage of subjects (\ie papers, interviewees, and surveys) in which each \textit{purpose/output category} appears. The figure displays approximately the first third of the \textit{purpose/output codes}, organized by the subject \textit{papers} appearance.}
    \label{fig:output1}
\end{figure}

\begin{figure}[h]
    \centering
    \includegraphics[width=\textwidth]{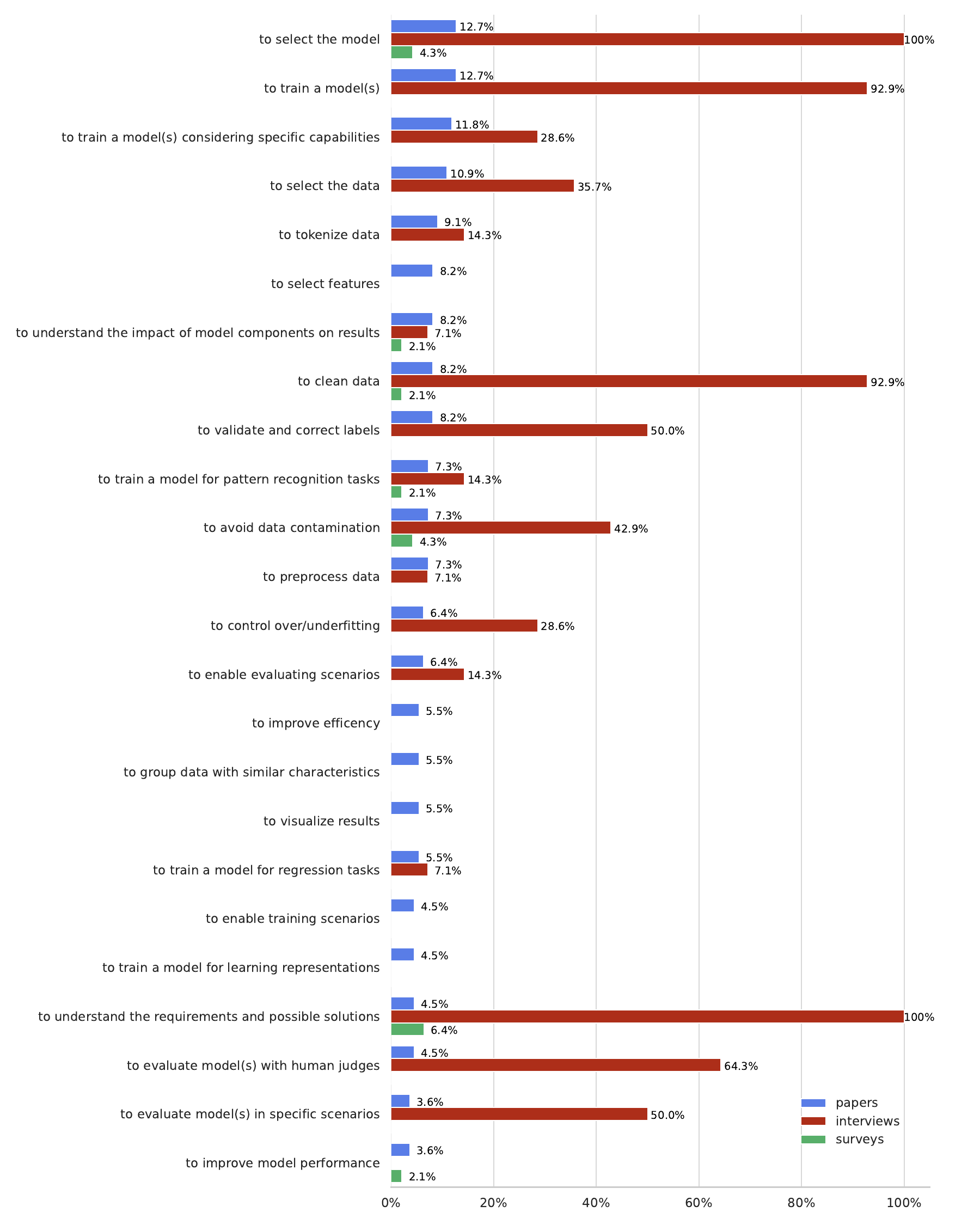}
    \caption{Percentage of subjects (\ie papers, interviewees, and surveys) in which each \textit{purpose/output category} appears. The figure displays approximately the second third of the \textit{purpose/output codes},  organized by the subject \textit{papers} appearance.}
    \label{fig:output2}
\end{figure}

\begin{figure}[h]
    \centering
    \includegraphics[width=\textwidth]{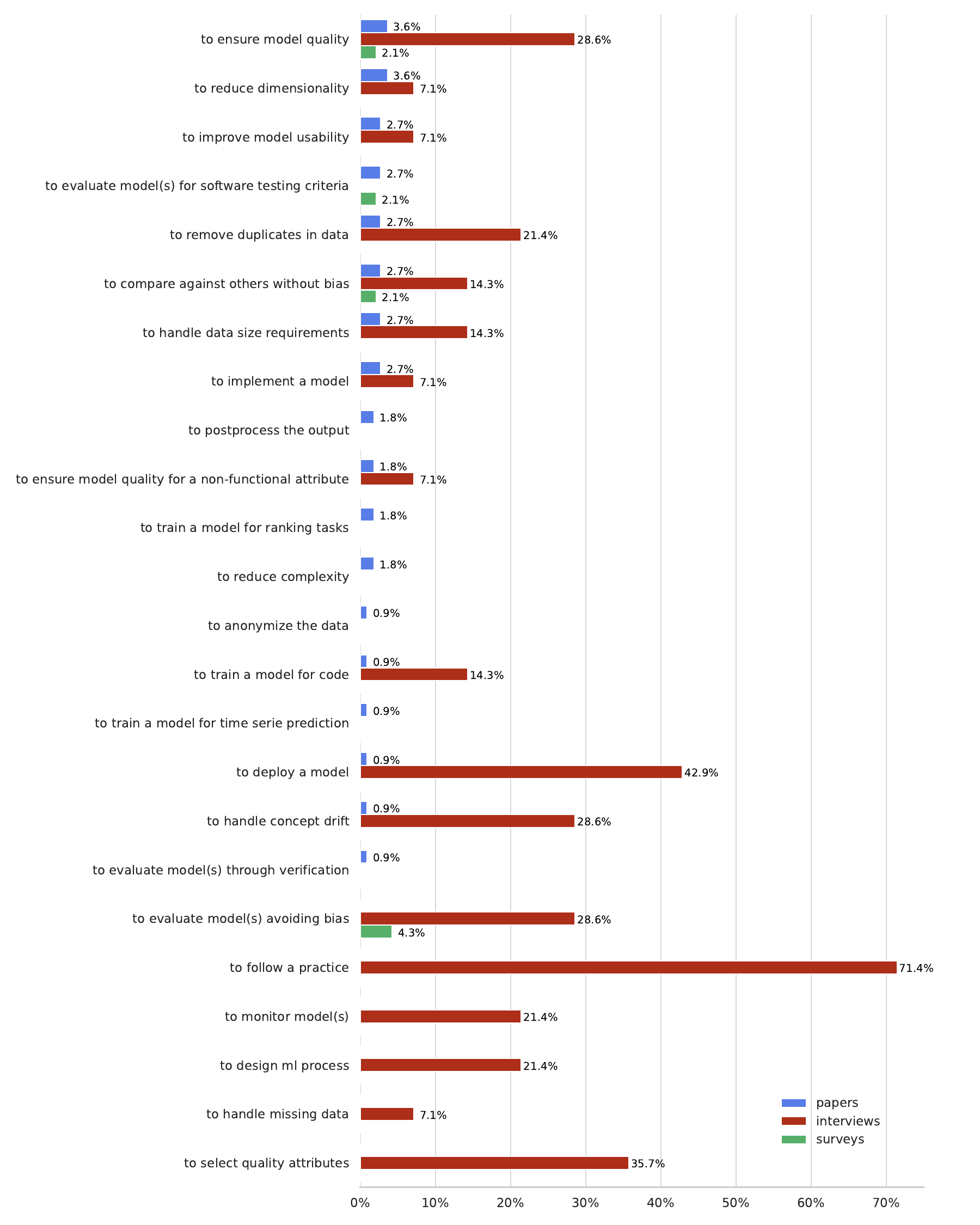}
    \caption{Percentage of subjects (\ie papers, interviewees, and surveys) in which each \textit{purpose/output category} appears. The figure displays approximately the last third of the \textit{purpose/output codes},  organized by the subject \textit{papers} appearance.}
    \label{fig:output3}
\end{figure}

\FloatBarrier
\newpage

\subsection{Categories associated to the ML Stage}

\begin{figure}[h]
    \centering
    \includegraphics[width=\textwidth]{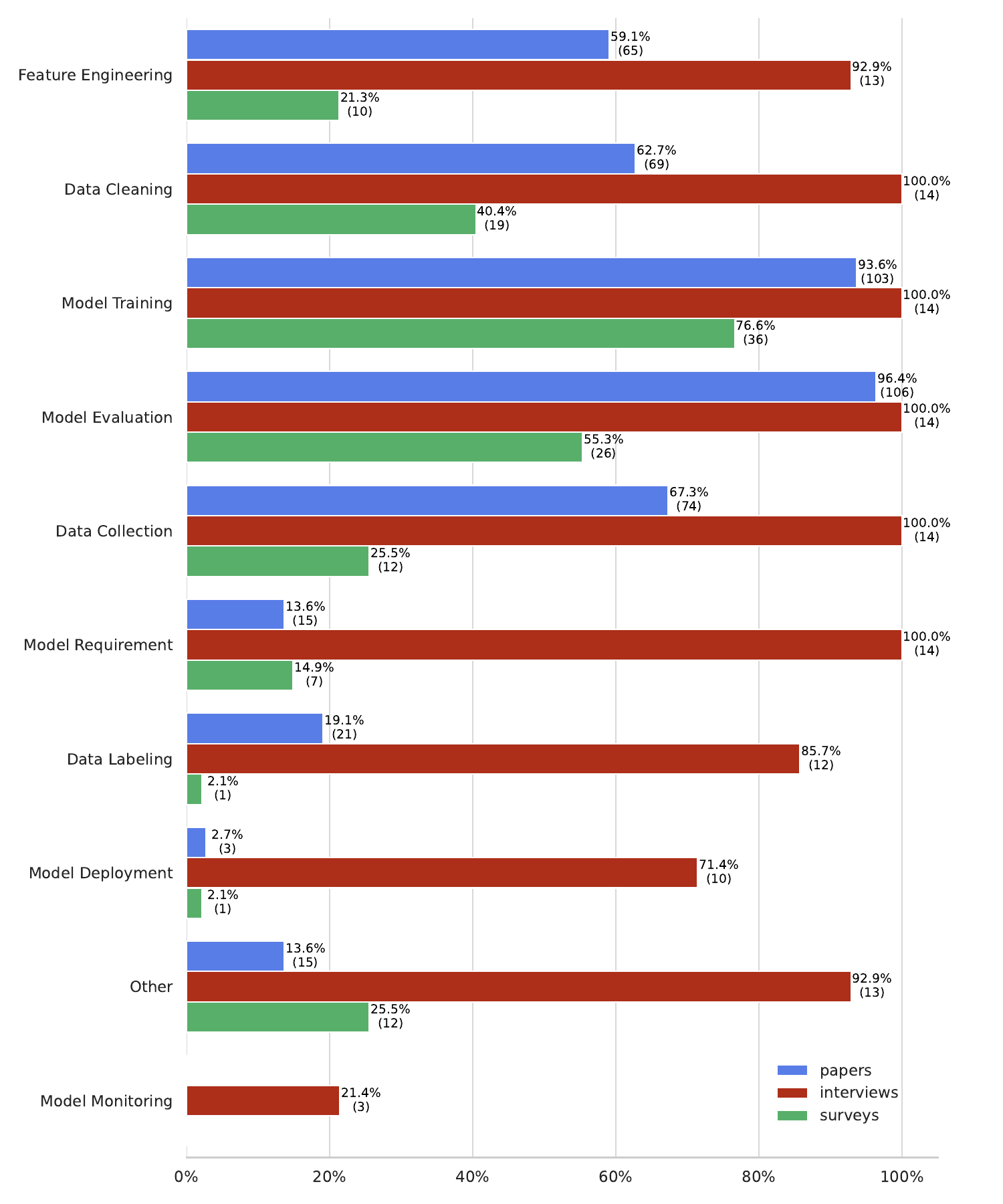}
    \caption{Percentage of subjects (\ie papers, interviewees, and surveys) in which each \textit{ML stage category} appears.}
    \label{fig:stages}
\end{figure}

\FloatBarrier
\newpage
\subsection{Categories associated to the SE Tasks}

\begin{figure}[h]
    \centering
    \includegraphics[width=\textwidth]{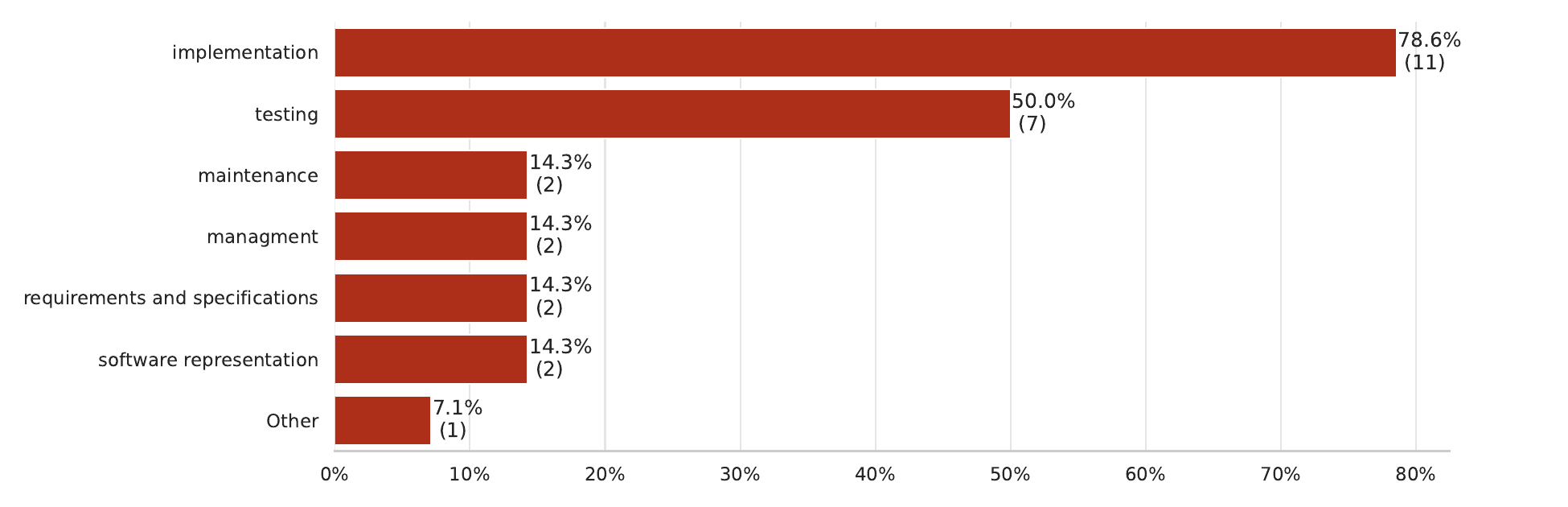}
    \caption{Percentage of interviews in which each \textit{SE task category} appears.}
    \label{fig:SEtasksI}
\end{figure}

\begin{figure}[h]
    \centering
    \includegraphics[width=\textwidth]{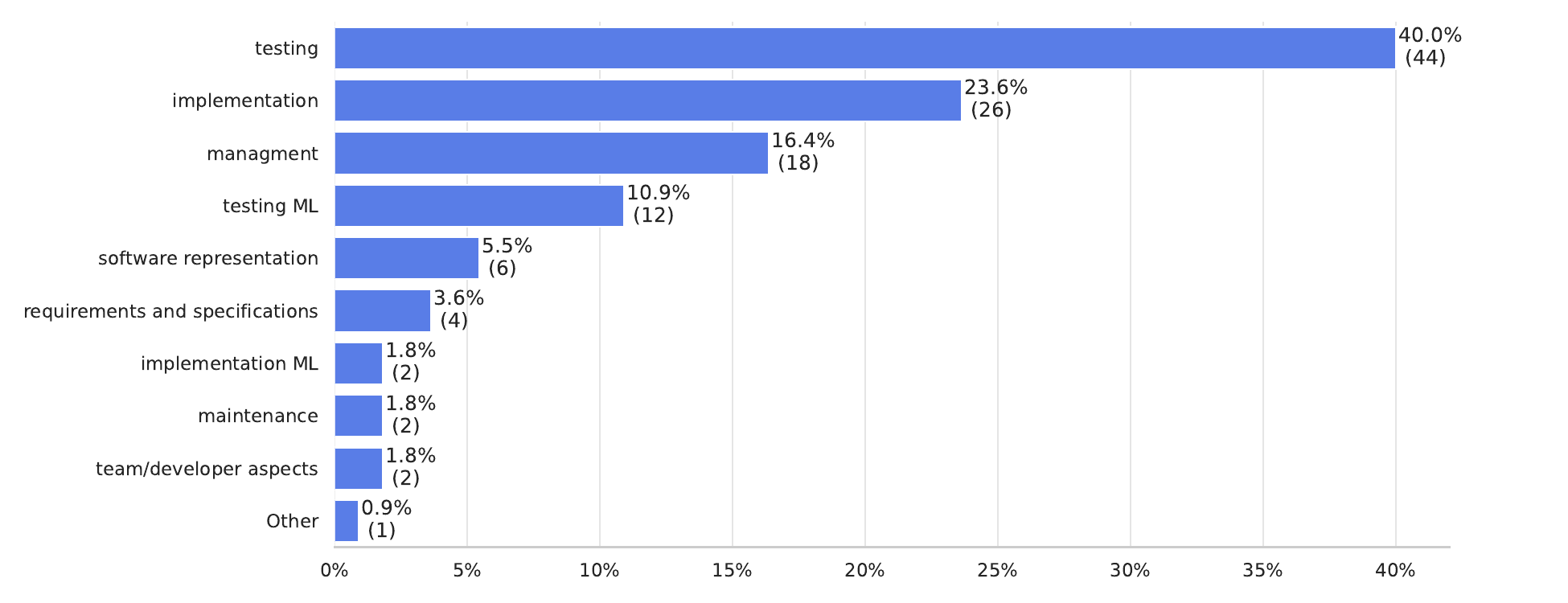}
    \caption{Percentage of papers in which each \textit{SE task category} appears.}
    \label{fig:SEtasksp}
\end{figure}

\FloatBarrier
\newpage
\subsection{Codes associated to the Quality attributes}

\begin{figure}[h]
    \centering
    \includegraphics[width=\textwidth]{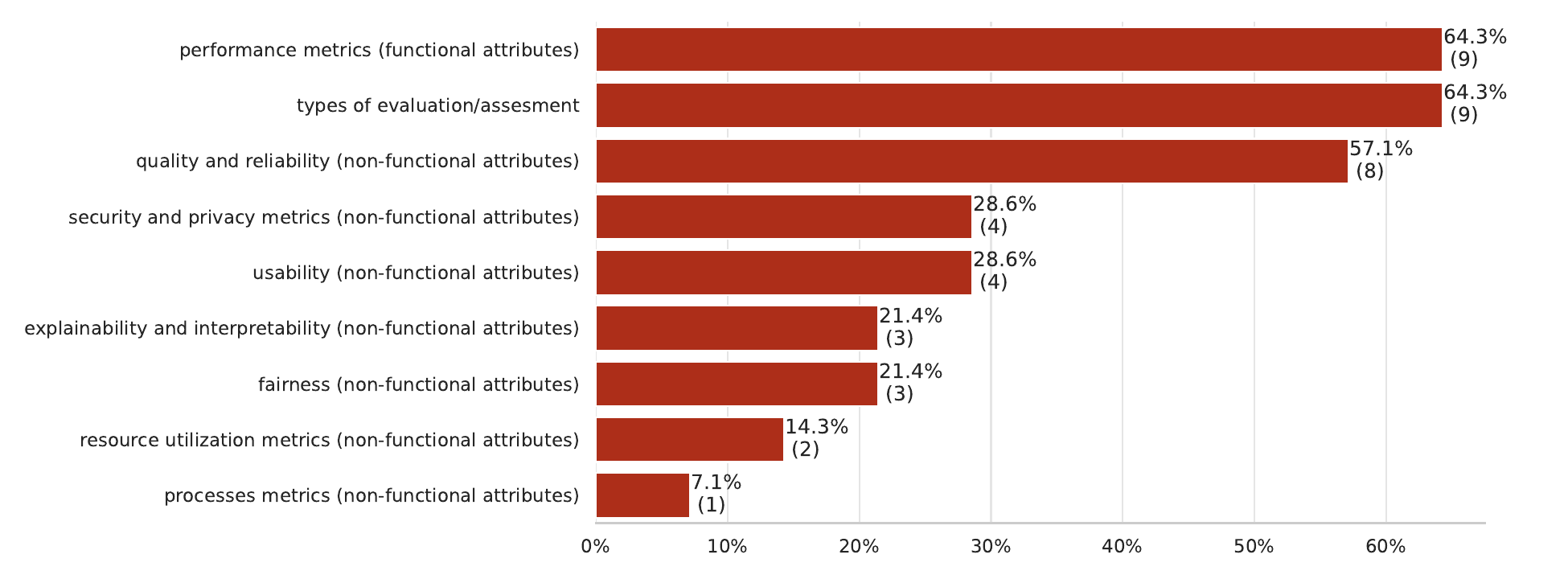}
    \caption{Percentage of interviews in which each \textit{quality attribute category} appears.}
    \label{fig:QA}
\end{figure}

\subsection{Codes associated to the ML Challenges}

\begin{figure}[h]
    \centering
    \includegraphics[width=\textwidth]{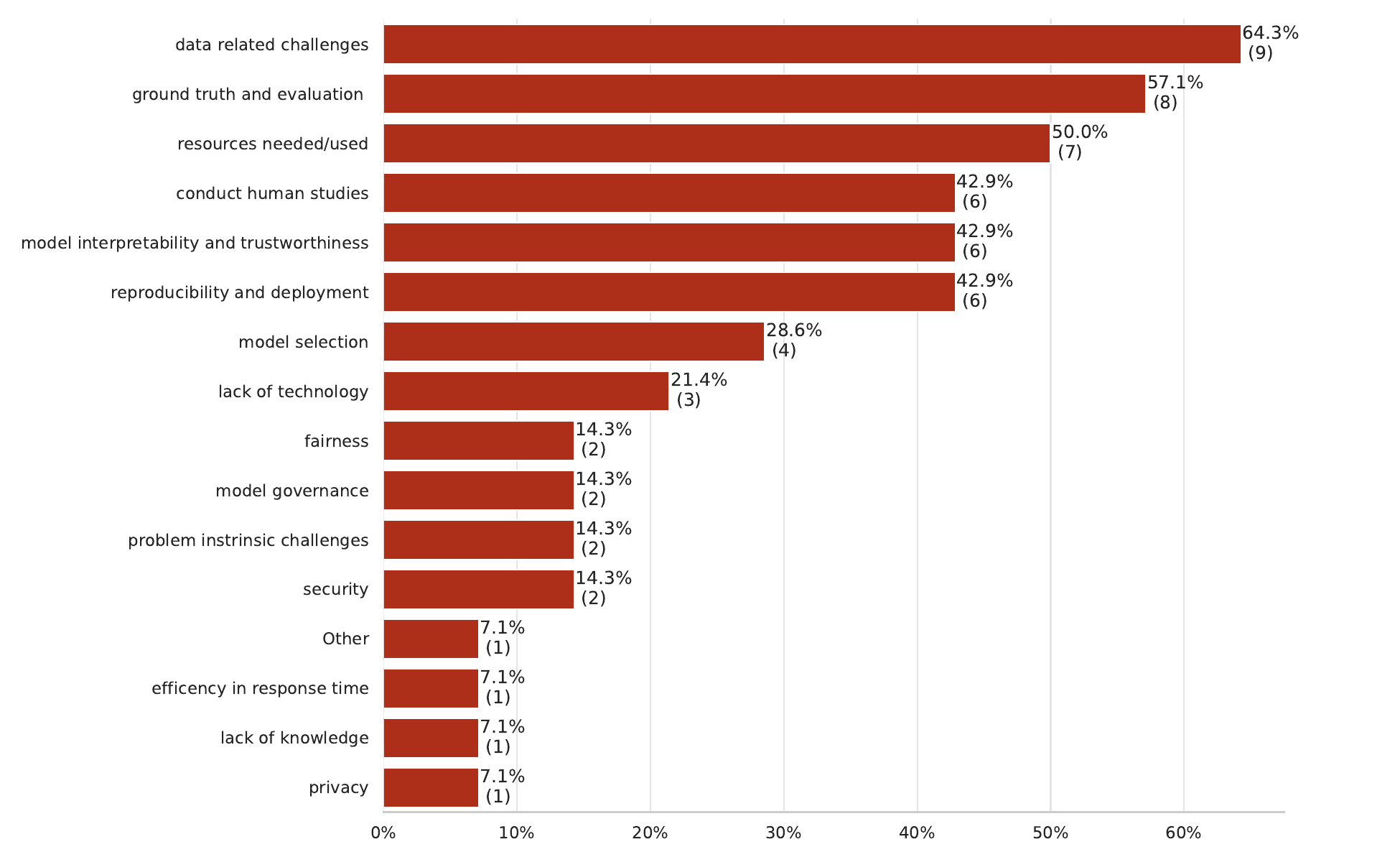}
    \caption{Percentage of interviews in which each \textit{challenge category} appears.}
    \label{fig:challenges}
\end{figure}

\FloatBarrier
\newpage
\subsection{Codes associated to the Reviewer's perspective}

\begin{figure}[h]
    \centering
    \includegraphics[width=\textwidth]{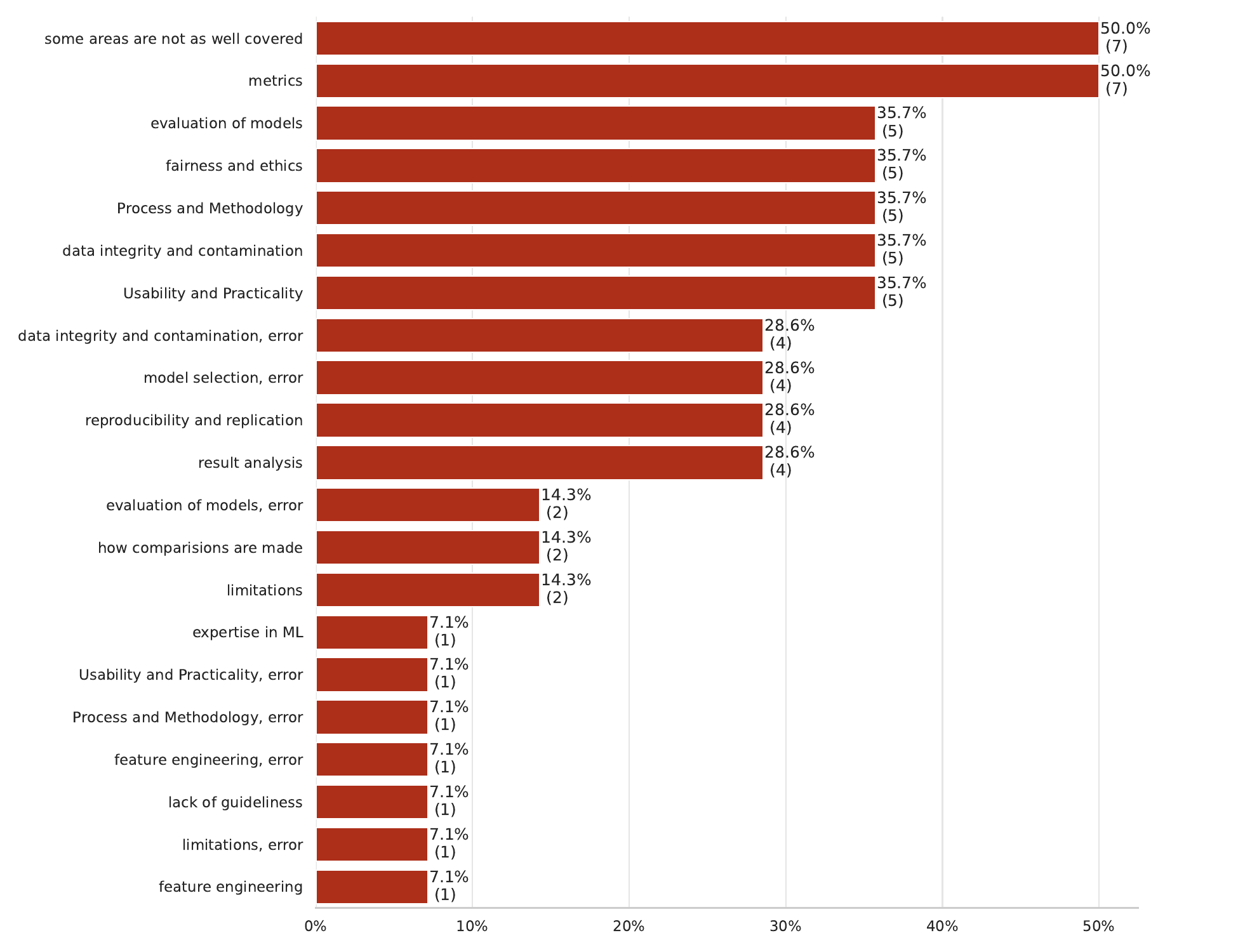}
    \caption{Percentage of interviews in which each \textit{reviewers' perspective category} appears.}
    \label{fig:reviewers}
\end{figure}

\FloatBarrier
\newpage
\subsection{Codes associated to the Educator's perspective}

\begin{figure}[h]
    \centering
    \includegraphics[width=\textwidth]{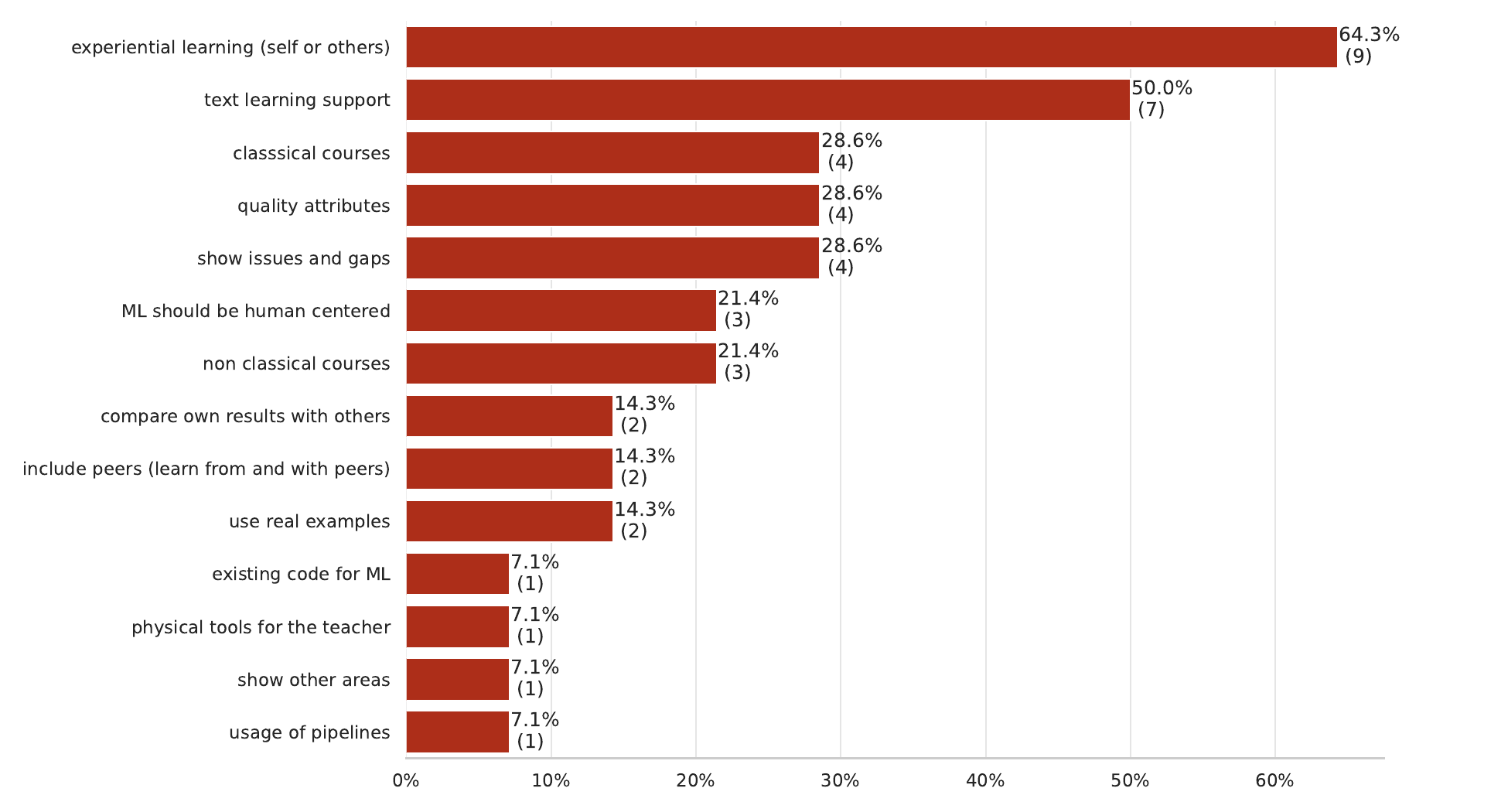}
    \caption{Percentage of interviews in which each \textit{educators' perspective category} appears.}
    \label{fig:educators}
\end{figure}

\FloatBarrier
\newpage
\section{Supplementary Material for the Results of the Open Coding, 2023-2025}

\subsection{Codes associated to the ML practices, 2023-2025}
\begin{figure}[h]
    \centering
    \includegraphics[width=\textwidth]{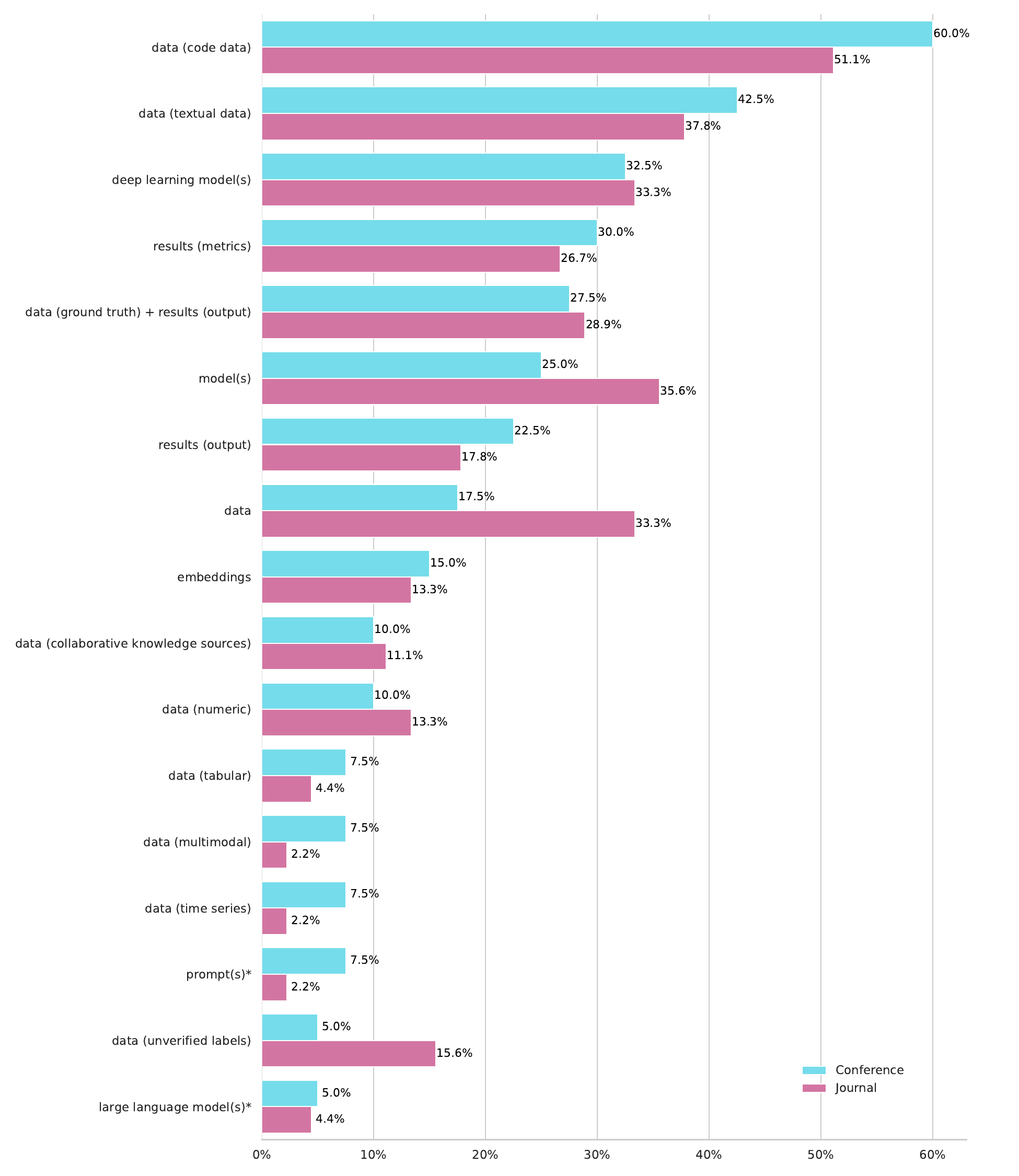}
    \caption{Percentage per venue (\ie conferences, journals) in which each \textit{input category} appears. The figure displays the first half of the \textit{input codes}, organized by the subject \textit{papers} appearance.}
    \label{fig:input_axial1-new}
\end{figure}

\begin{figure}[h]
    \centering
    \includegraphics[width=\textwidth]{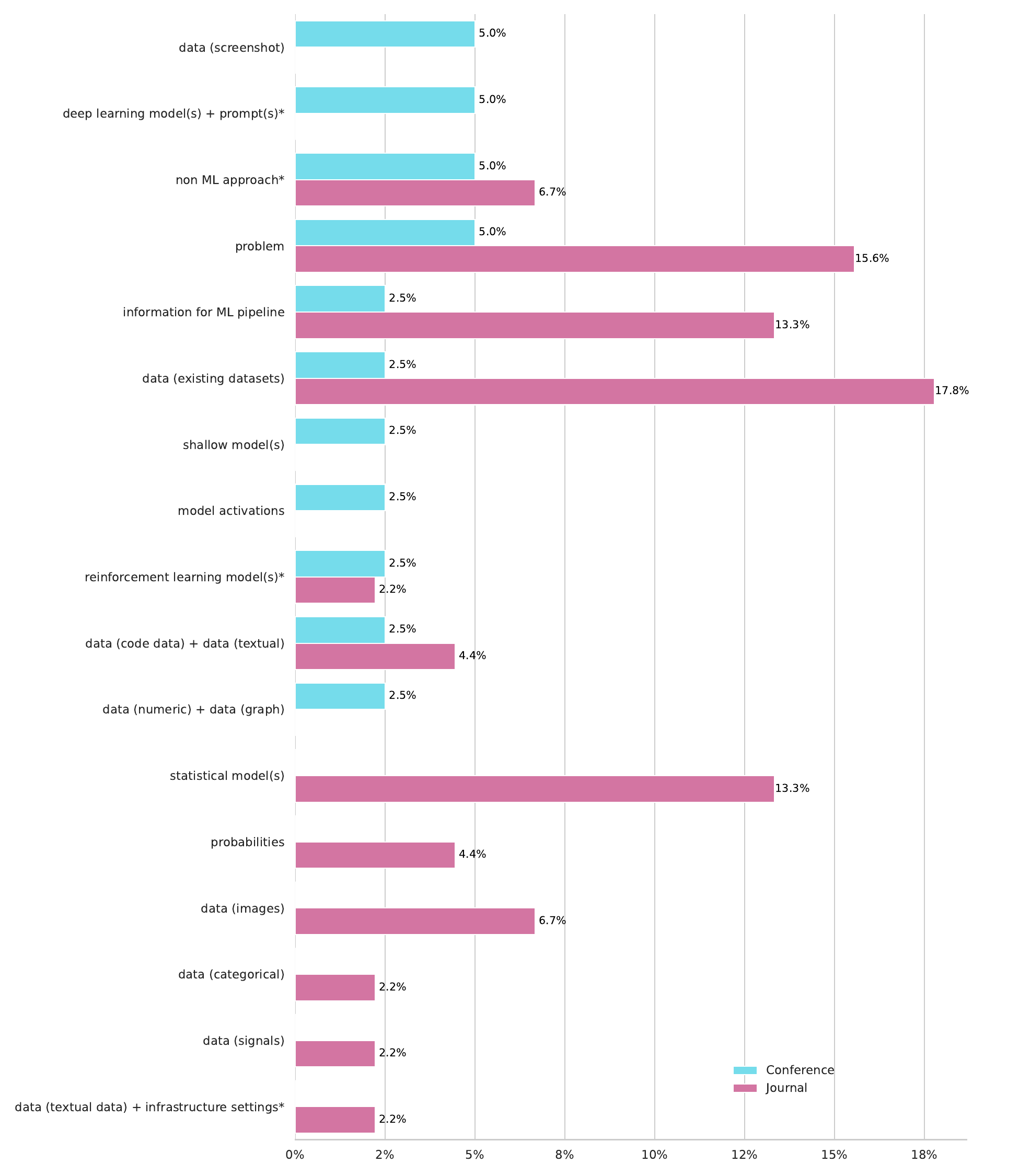}
    \caption{Percentage per venue (\ie conferences, journals) in which each \textit{input category} appears. The figure displays the second half of the \textit{input codes}, organized by the subject \textit{papers} appearance.}
    \label{fig:input_axial2-new}
\end{figure}

% ------------------------------------------------------ Techniques --------------------------------------

\begin{figure}[h]
    \centering
    \includegraphics[width=\textwidth]{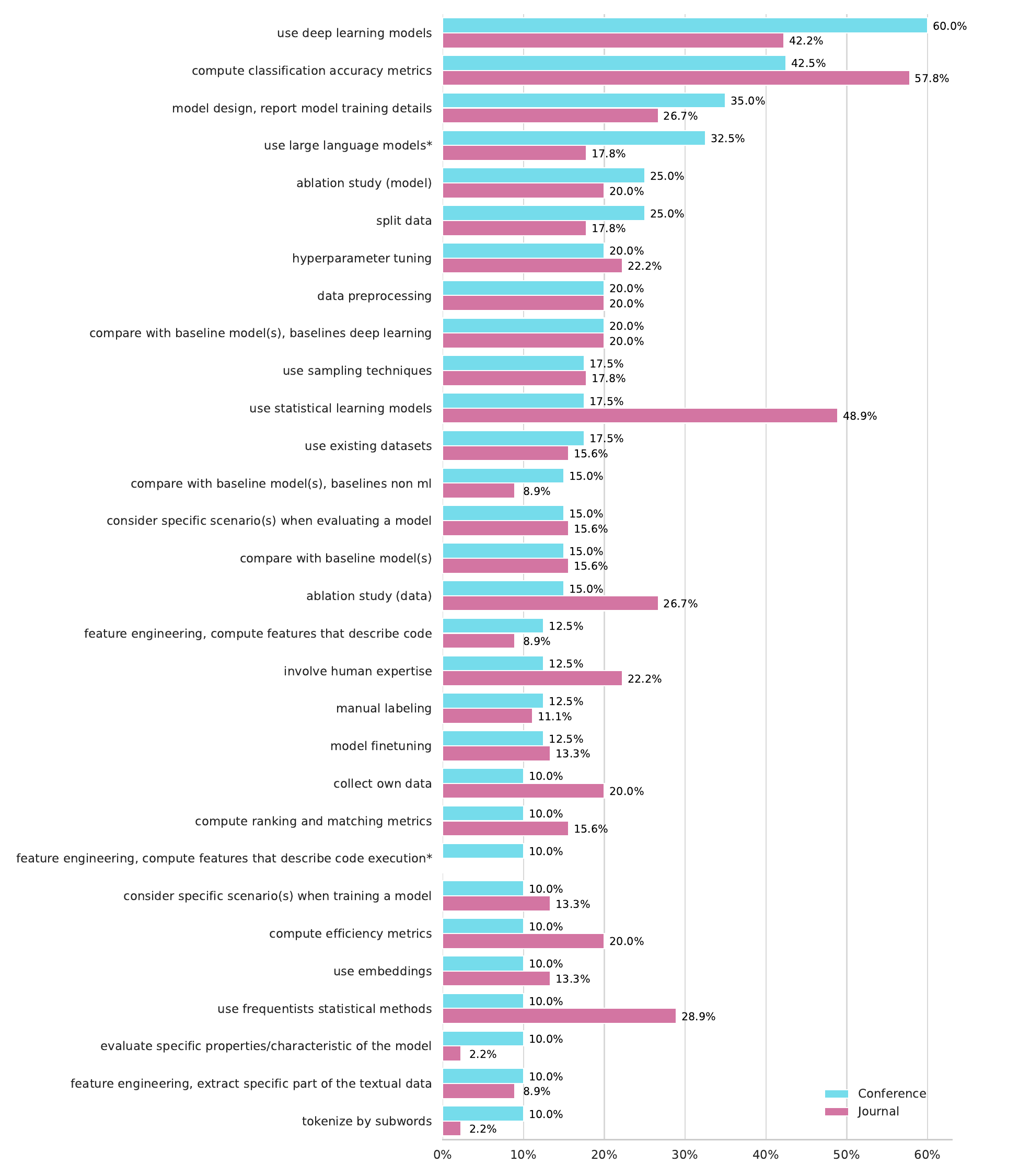}
    \caption{Percentage per venue (\ie conferences, journals) in which each \textit{technique category} appears. The figure displays approximately the first fifth  of the \textit{technique codes}, organized by the subject \textit{papers} appearance.}
    \label{fig:technique_axial1-new}
\end{figure}

\begin{figure}[h]
    \centering
    \includegraphics[width=\textwidth]{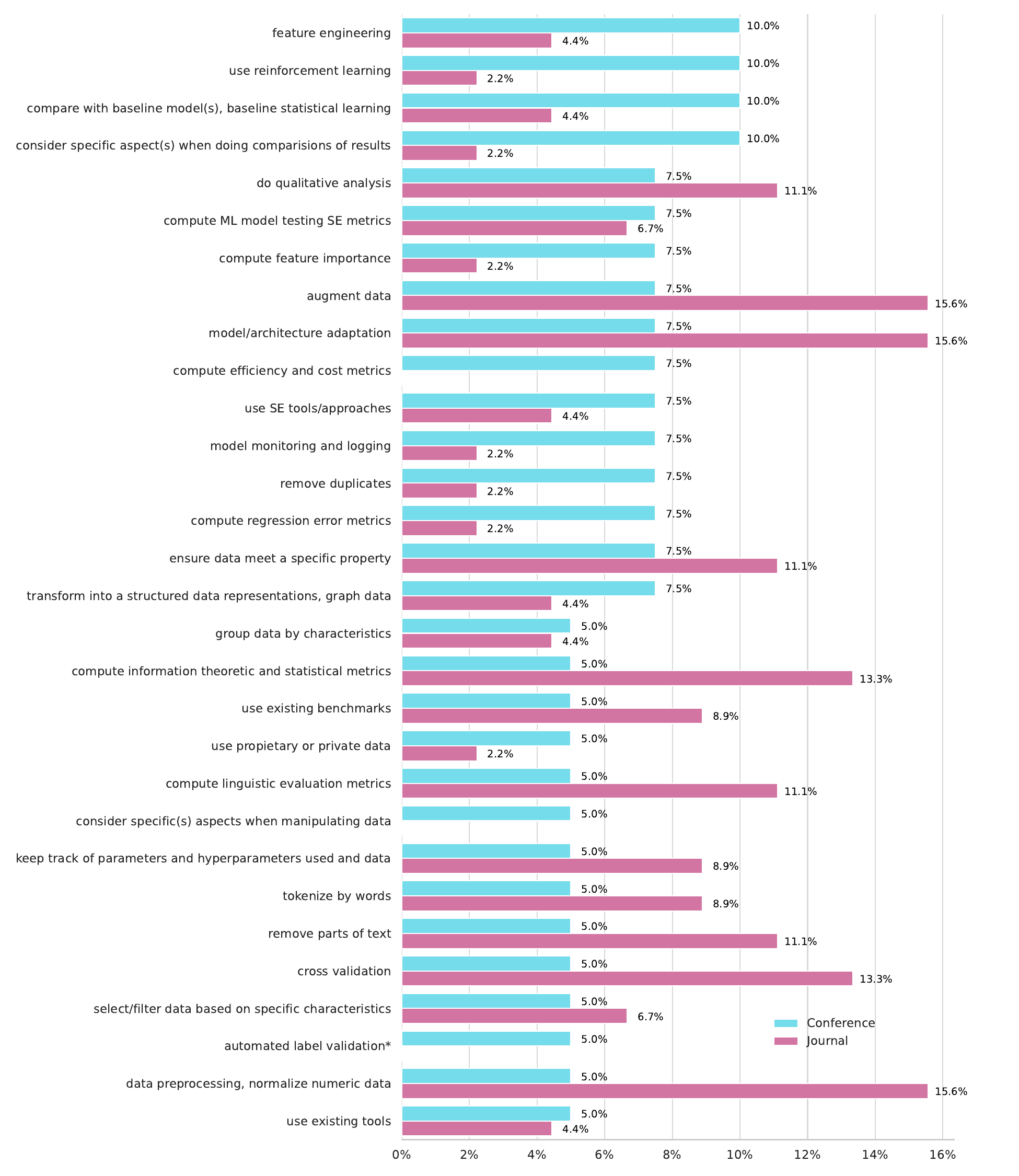}
    \caption{Percentage per venue (\ie conferences, journals) in which each \textit{technique category} appears. The figure displays approximately the second fifth of the \textit{technique codes},  organized by the subject \textit{papers} appearance.}
    \label{fig:technique_axial2-new}
\end{figure}

\begin{figure}[h]
    \centering
    \includegraphics[width=\textwidth]{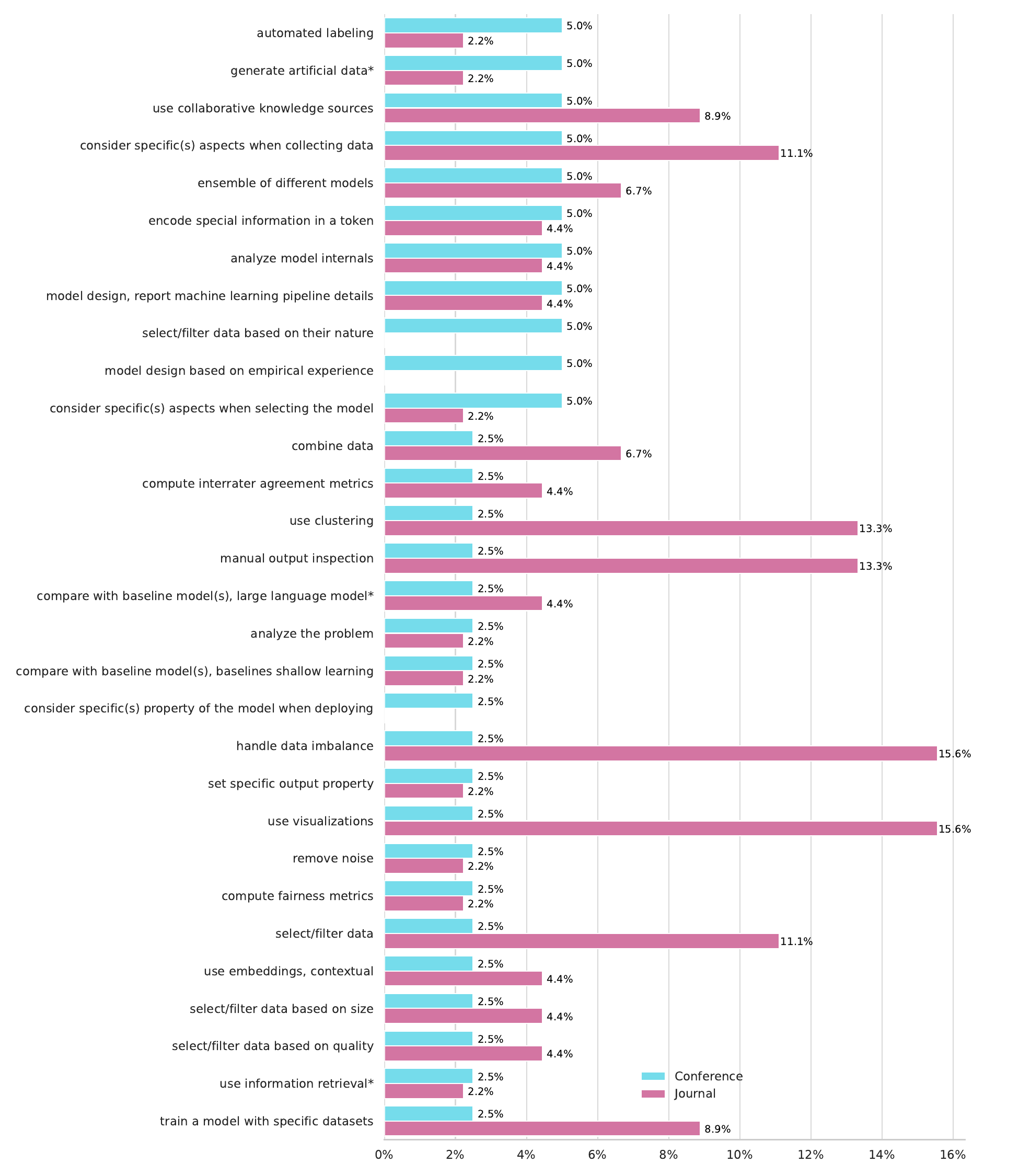}
    \caption{Percentage per venue (\ie conferences, journals) in which each \textit{technique category} appears. The figure displays approximately the third fifth of the \textit{technique codes}, organized by the subject \textit{papers} appearance.}
    \label{fig:technique_axial3-new}
\end{figure}

\begin{figure}[h]
    \centering
    \includegraphics[width=\textwidth]{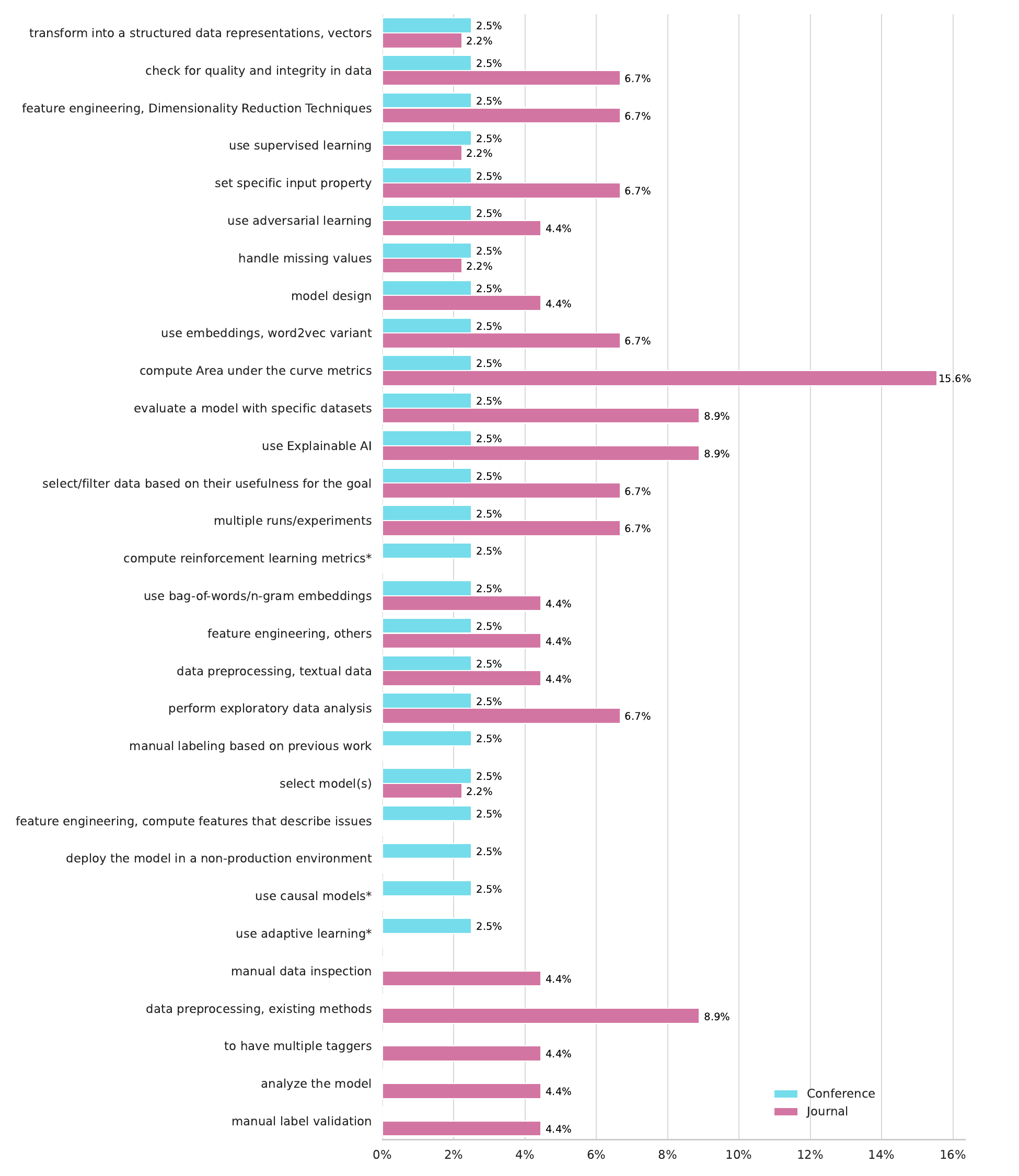}
    \caption{Percentage per venue (\ie conferences, journals) in which each \textit{technique category} appears. The figure displays approximately the fourth fifth of the \textit{technique codes},  organized by the subject \textit{papers} appearance.}
    \label{fig:technique_axial4-new}
\end{figure}

\begin{figure}[h]
    \centering
    \includegraphics[width=\textwidth]{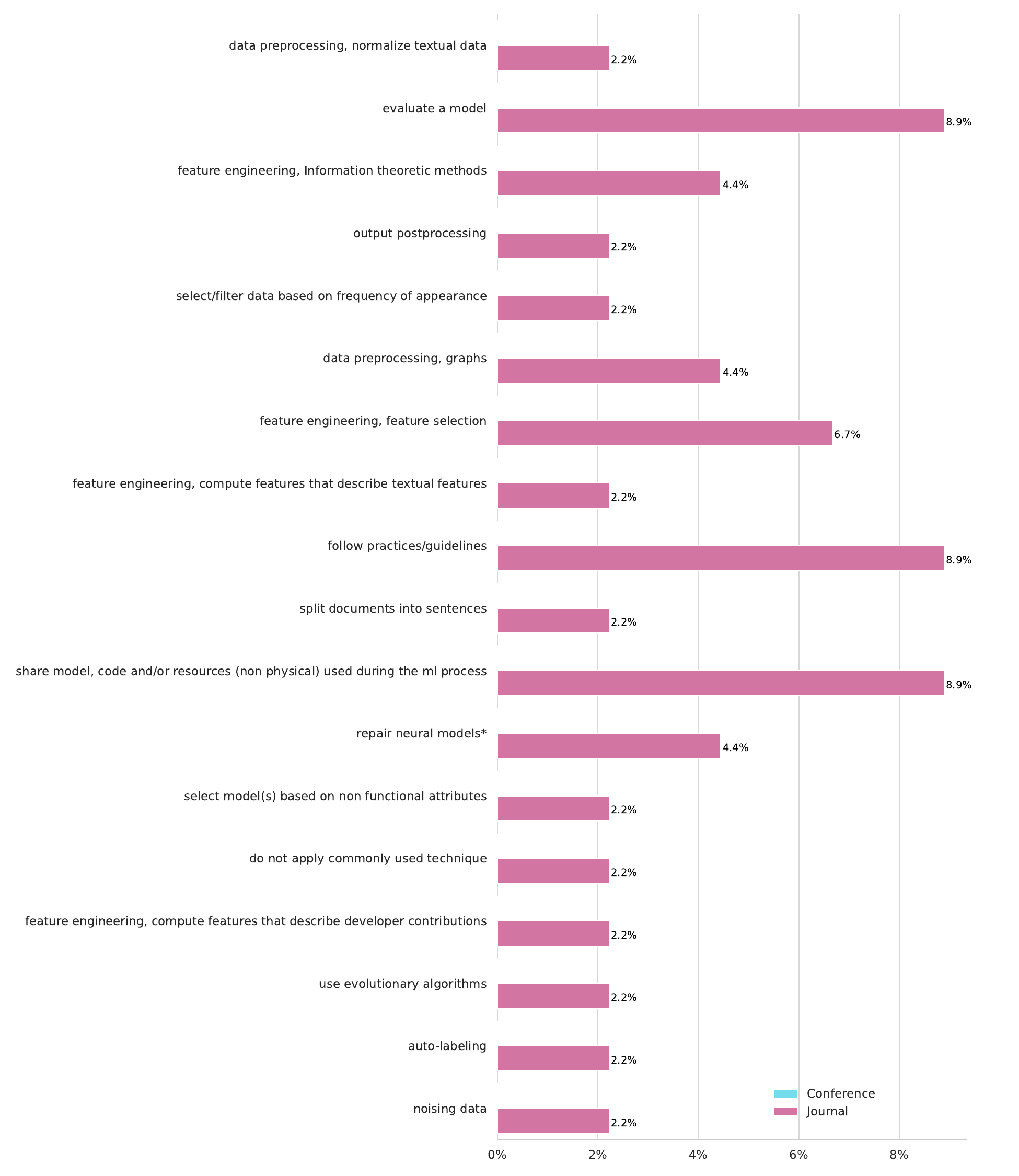}
    \caption{Percentage per venue (\ie conferences, journals) in which each \textit{technique category} appears. The figure displays approximately the last fifth of the \textit{technique codes},  organized by the subject \textit{papers} appearance.}
    \label{fig:technique_axial5-new}
\end{figure}

% ------------------------------------------------------ Purpose --------------------------------------

\begin{figure}[h]
    \centering
    \includegraphics[width=\textwidth]{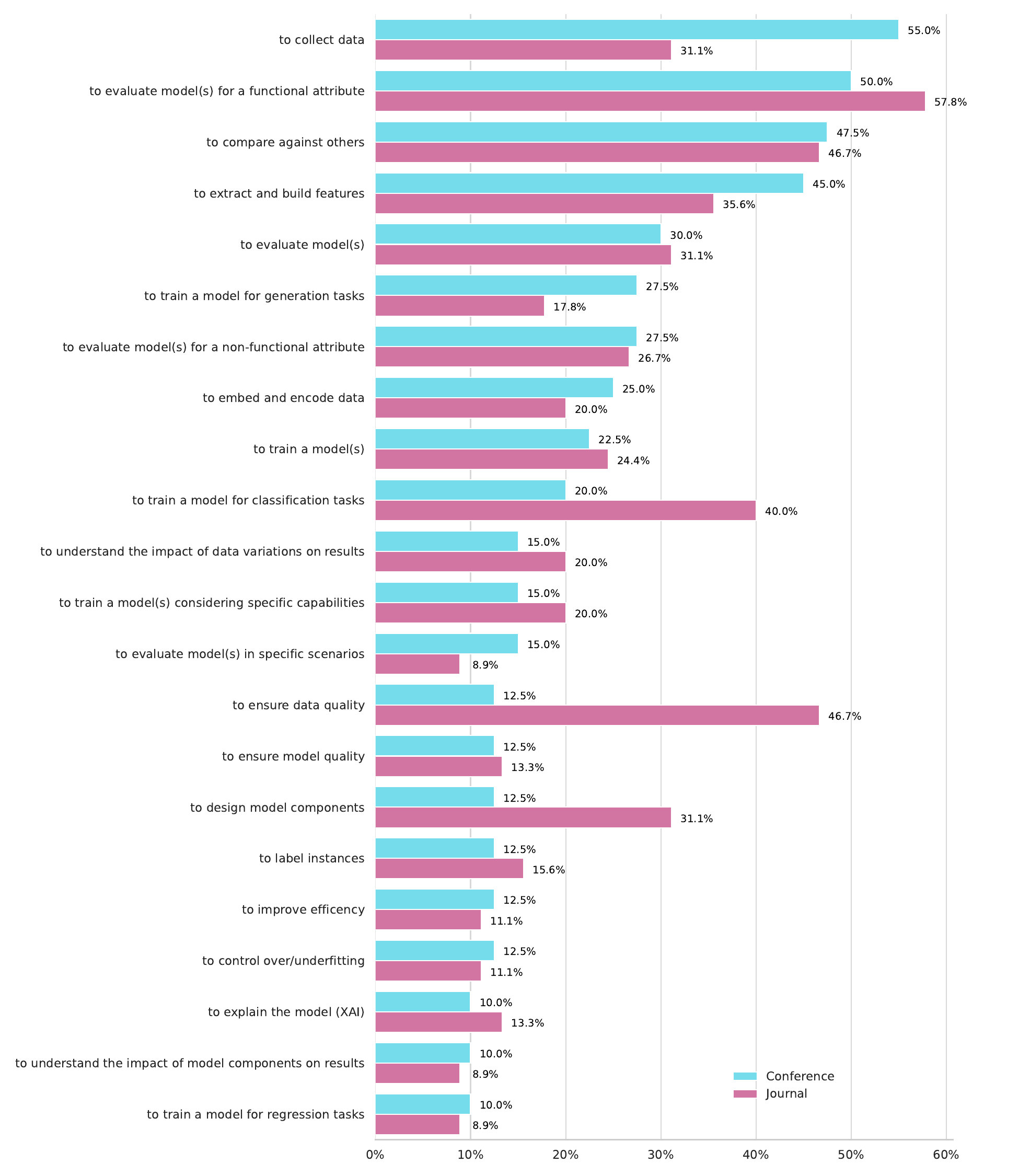}
    \caption{Percentage per venue (\ie conferences, journals) in which each \textit{purpose/output category} appears. The figure displays approximately the first third of the \textit{purpose/output codes}, organized by the subject \textit{papers} appearance.}
    \label{fig:output1-new}
\end{figure}

\begin{figure}[h]
    \centering
    \includegraphics[width=\textwidth]{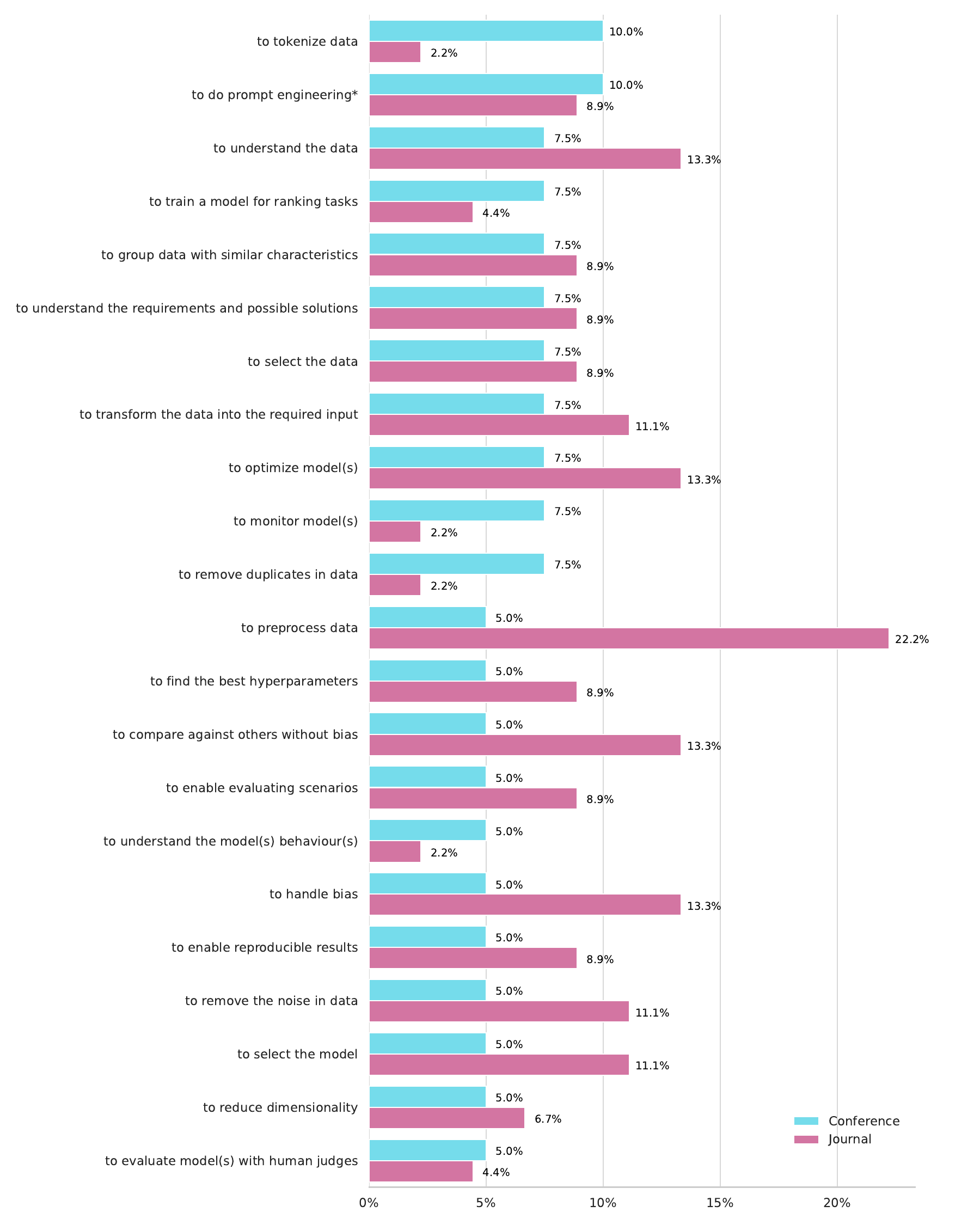}
    \caption{Percentage per venue (\ie conferences, journals) in which each \textit{purpose/output category} appears. The figure displays approximately the second third of the \textit{purpose/output codes},  organized by the subject \textit{papers} appearance.}
    \label{fig:output2-new}
\end{figure}

\begin{figure}[h]
    \centering
    \includegraphics[width=\textwidth]{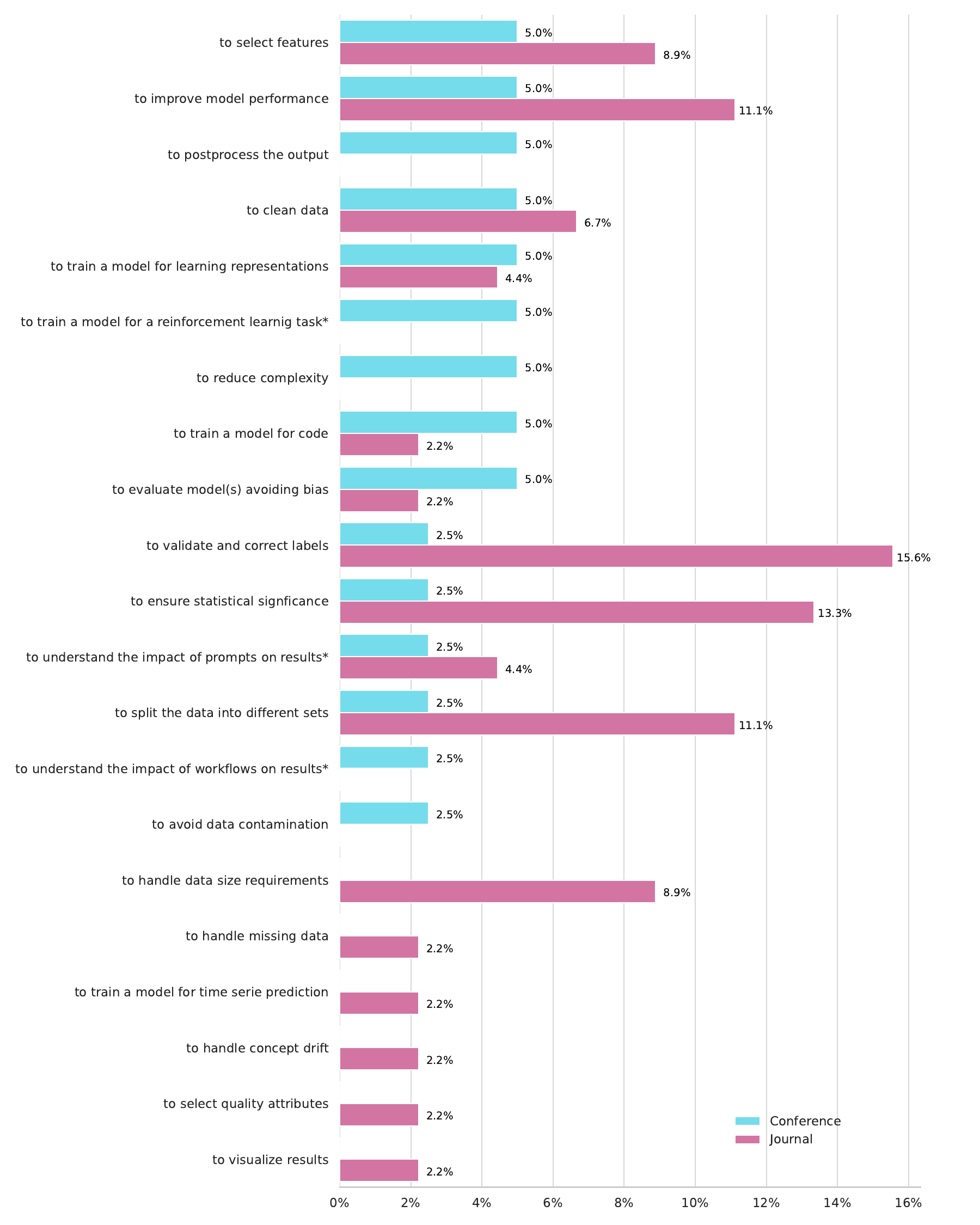}
    \caption{Percentage per venue (\ie conferences, journals) in which each \textit{purpose/output category} appears. The figure displays approximately the last third of the \textit{purpose/output codes},  organized by the subject \textit{papers} appearance.}
    \label{fig:output3-new}
\end{figure}

\FloatBarrier
\newpage

\subsection{Categories associated to the ML Stage, 2023-2025}

\begin{figure}[h]
    \centering
    \includegraphics[width=\textwidth]{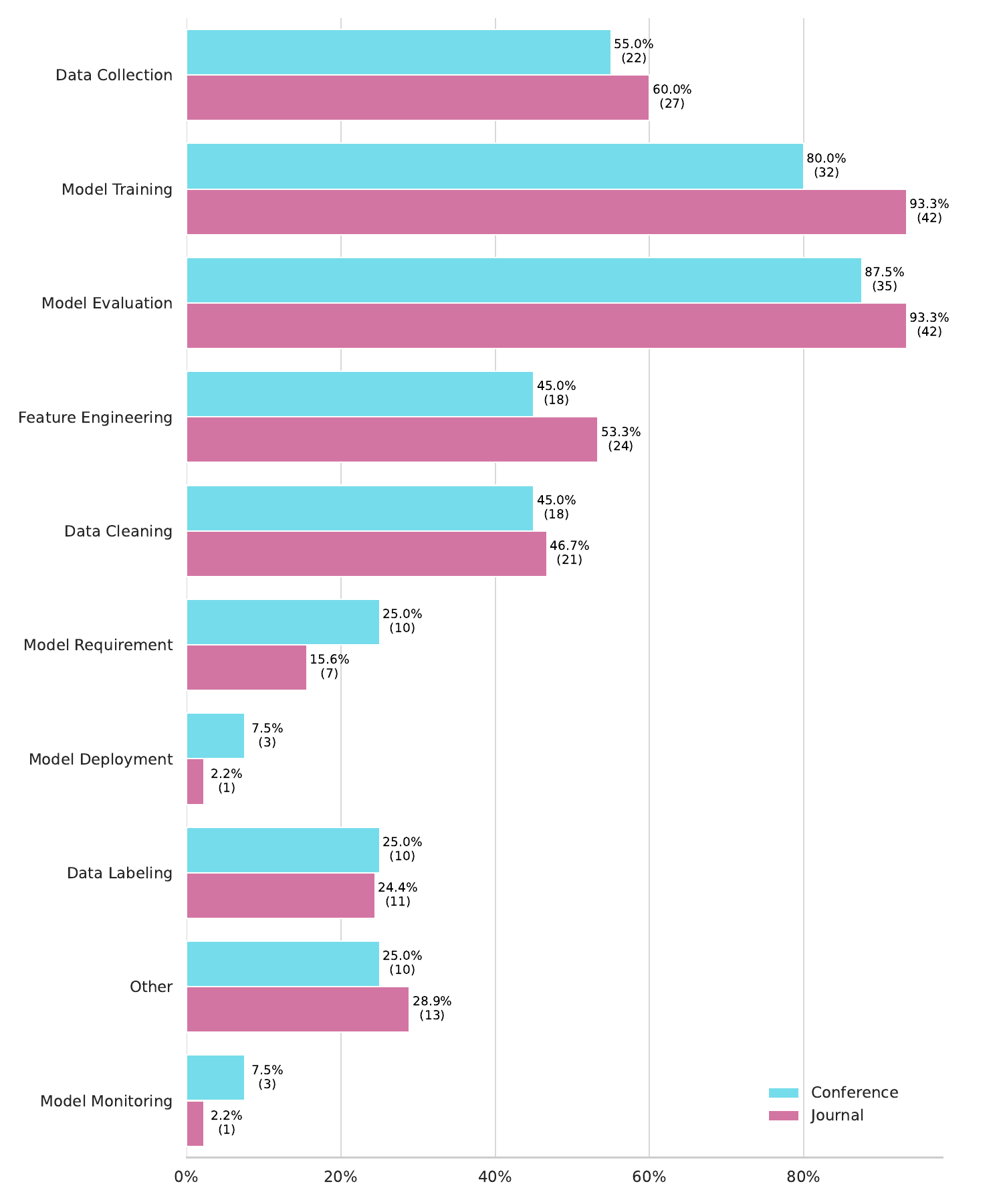}
    \caption{Percentage per venue (\ie conferences, journals) in which each \textit{ML stage category} appears.}
    \label{fig:stages-new}
\end{figure}

\FloatBarrier
\newpage
\subsection{Categories associated to the SE Tasks, 2023-2025}

\begin{figure}[h]
    \centering
    \includegraphics[width=\textwidth]{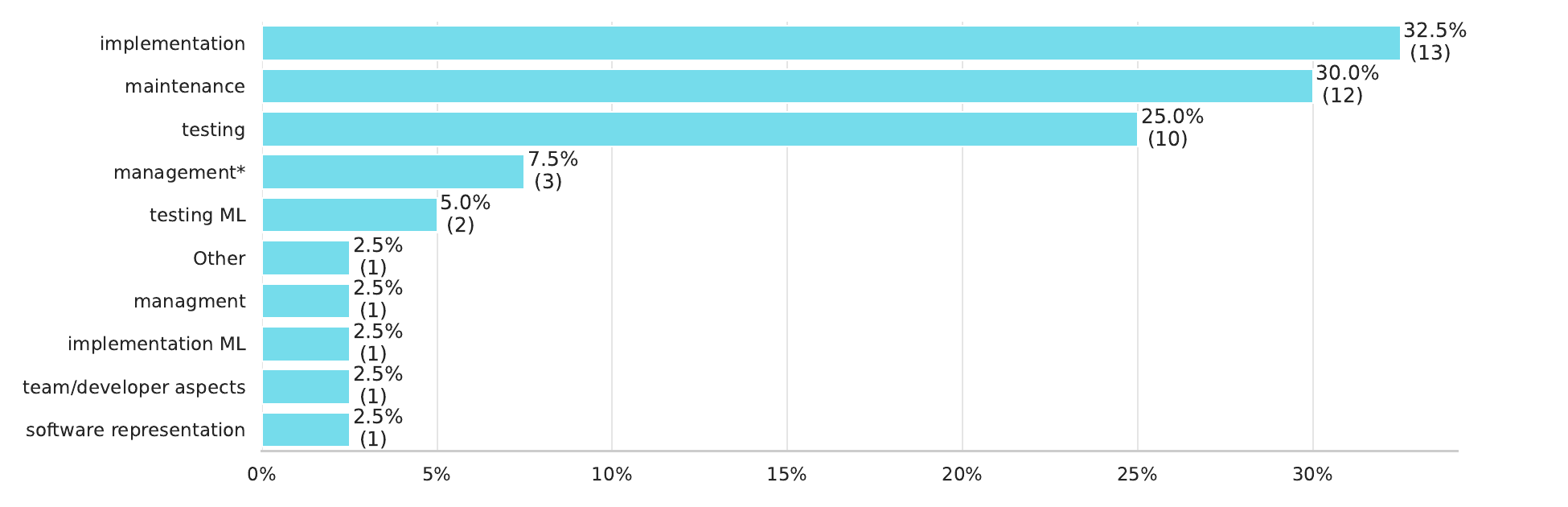}
    \caption{Percentage of journals in which each \textit{SE task category} appears.}
    \label{fig:SEtasksI-new}
\end{figure}

\begin{figure}[h]
    \centering
    \includegraphics[width=\textwidth]{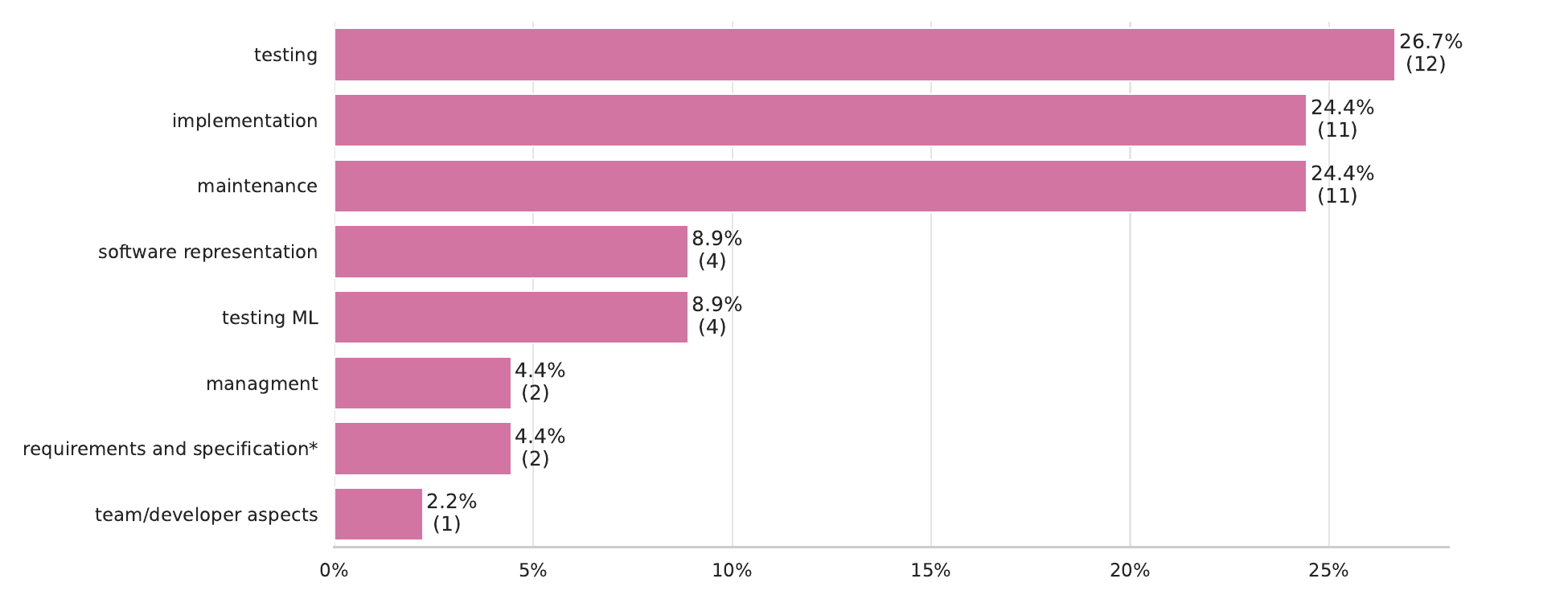}
    \caption{Percentage of conferences in which each \textit{SE task category} appears.}
    \label{fig:SEtasksp-new}
\end{figure}

\end{document}